\documentclass[12pt]{article}

\usepackage[margin=1in]{geometry}
\usepackage{amsmath}
\usepackage[T1]{fontenc}
\usepackage[utf8]{inputenc}
\usepackage{newtxtext,newtxmath}
\usepackage{booktabs}
\usepackage{array}
\usepackage{threeparttable}
\usepackage{graphicx}
\usepackage{caption}
\usepackage{xcolor}
\usepackage{setspace}
\usepackage{placeins}
\usepackage{chngcntr}
\usepackage[authoryear,round]{natbib}
\usepackage{xurl}
\usepackage[hidelinks]{hyperref}

\title{Priced In or Bet On? Cram-School Lessons, Jeonse Leases, and Expected Appreciation in Seoul}
\author{Samiha Tariq\thanks{PhD Candidate in Economics, College of Business and Analytics, Southern Illinois University Carbondale, Carbondale, IL 62901, USA.}}
\date{November 2026}

\newcommand{\ChkAccBaseKaptMHsSale}{0.046}
\newcommand{\ChkAccSizeCHagGap}{0.024}
\newcommand{\ChkAccSizeCellMHh}{\ensuremath{-}0.051}
\newcommand{\ChkAccSizeCellMHhSe}{0.004}
\newcommand{\ChkAccSizeCellMHs}{0.009}
\newcommand{\ChkAccSizeJHagGap}{\ensuremath{-}0.020}
\newcommand{\ChkAccSizeKaptMHsSale}{0.023}
\newcommand{\ChkAccSizeMDStJeonse}{\ensuremath{-}0.096}
\newcommand{\ChkAccSizeMDStSale}{\ensuremath{-}0.048}
\newcommand{\ChkAccSizeMHhJeonse}{0.039}
\newcommand{\ChkAccSizeMHhSale}{0.089}
\newcommand{\ChkAccSizeMHsGap}{0.002}
\newcommand{\ChkAccSizeMHsGapSe}{0.016}
\newcommand{\ChkAccSizeMHsJeonse}{0.029}
\newcommand{\ChkAccSizeMHsJeonseP}{\ensuremath{p=0.063}}
\newcommand{\ChkAccSizeMHsJeonsePct}{5.9}
\newcommand{\ChkAccSizeMHsSale}{0.027}
\newcommand{\ChkAccSizeMHsSaleP}{\ensuremath{p=0.023}}
\newcommand{\ChkAccSizeMHsSalePct}{5.6}
\newcommand{\ChkAccSizeMHsSaleSe}{0.012}
\newcommand{\ChkAuxHhOnHs}{0.224}
\newcommand{\ChkAuxHhOnHsSe}{0.069}
\newcommand{\ChkCellHalfEqualMP}{\ensuremath{p=0.009}}
\newcommand{\ChkCellHalfEqualNoTwentyTwoMP}{\ensuremath{p=0.063}}
\newcommand{\ChkCellHalfMedianCells}{9,980}
\newcommand{\ChkCellHalfTwentyTwoHOneM}{\ensuremath{-}0.048}
\newcommand{\ChkCellHalfTwentyTwoHOneMCells}{3,058}
\newcommand{\ChkCellHalfTwentyTwoHTwoM}{\ensuremath{-}0.028}
\newcommand{\ChkCellHalfTwentyTwoHTwoMCells}{1,527}
\newcommand{\ChkCellOldRecentN}{0.020}
\newcommand{\ChkCellPosAgeThirtyM}{\ensuremath{-}0.358}
\newcommand{\ChkCellPosAgeThirtyMPctAbs}{30}
\newcommand{\ChkCellPosAgeThirtyMSe}{0.032}
\newcommand{\ChkCellPosHsOldM}{\ensuremath{-}0.009}
\newcommand{\ChkCellPosHsYoungM}{0.004}
\newcommand{\ChkCellPosPowerShare}{13}
\newcommand{\ChkCellPosSeRatio}{32}
\newcommand{\ChkCellRecentNewM}{0.003}
\newcommand{\ChkCellRecentNewMSe}{0.012}
\newcommand{\ChkCellRegEqualMP}{\ensuremath{p=0.61}}
\newcommand{\ChkCellRegOneM}{0.003}
\newcommand{\ChkCellRegThreeM}{\ensuremath{-}0.002}
\newcommand{\ChkCellRegTwoM}{0.006}
\newcommand{\ChkCellShareRecentM}{0.010}
\newcommand{\ChkCellShareRecentMSe}{0.010}
\newcommand{\ChkCellTierEqualMP}{\ensuremath{p=0.68}}
\newcommand{\ChkCellTierHighM}{0.007}
\newcommand{\ChkCellTierHighMSe}{0.025}
\newcommand{\ChkCellTierLowM}{0.004}
\newcommand{\ChkCellTierMidM}{0.014}
\newcommand{\ChkCellYoungEarlyN}{\ensuremath{-}0.009}
\newcommand{\ChkCellYoungRecentDiffNP}{\ensuremath{p=0.11}}
\newcommand{\ChkCellYoungRecentM}{\ensuremath{-}0.023}
\newcommand{\ChkCellYoungRecentMP}{\ensuremath{p=0.044}}
\newcommand{\ChkCellYoungRecentN}{\ensuremath{-}0.032}
\newcommand{\ChkCellYoungRecentNCells}{43,528}
\newcommand{\ChkCellYoungRecentNP}{\ensuremath{p=0.005}}
\newcommand{\ChkCellYoungRecentNSe}{0.011}
\newcommand{\ChkCellYoungRecentRowsN}{\ensuremath{-}0.035}
\newcommand{\ChkDecBaseYoungMHsSale}{0.064}
\newcommand{\ChkDecBuilderMHsSale}{0.045}
\newcommand{\ChkDecHhMHsSale}{0.025}
\newcommand{\ChkDecSizeYoungMHsSale}{0.029}
\newcommand{\ChkEvRegAllSaleJumpNovP}{\ensuremath{p<0.001}}
\newcommand{\ChkEvRegAllSaleJumpNovPct}{1.4}
\newcommand{\ChkEvRegAllSalePreP}{\ensuremath{p<0.001}}
\newcommand{\ChkEvRegCappedJumpFebP}{\ensuremath{p=0.94}}
\newcommand{\ChkEvRegCappedJumpJunP}{\ensuremath{p=0.57}}
\newcommand{\ChkEvRegNewJJumpFebPct}{\ensuremath{-}0.4}
\newcommand{\ChkEvRegNewJJumpFebPctHi}{0.9}
\newcommand{\ChkEvRegNewJJumpFebPctLo}{\ensuremath{-}1.8}
\newcommand{\ChkEvRegNewJJumpFebZonePctHi}{1.2}
\newcommand{\ChkEvRegNewJJumpJunPct}{0.9}
\newcommand{\ChkEvRegNewJJumpJunPctHi}{1.8}
\newcommand{\ChkEvRegNewJJumpJunPctLo}{0.0}
\newcommand{\ChkEvRegNewJJumpJunZonePctHi}{1.7}
\newcommand{\ChkEvRegNewJObs}{151,443}
\newcommand{\ChkEvRegSalesJumpFebP}{\ensuremath{p=0.002}}
\newcommand{\ChkEvRegSalesJumpFebPct}{0.8}
\newcommand{\ChkEvRegSalesJumpJunP}{\ensuremath{p=0.036}}
\newcommand{\ChkEvRegSalesJumpJunPct}{0.8}
\newcommand{\ChkEvRegSalesObs}{91,348}
\newcommand{\ChkExpJeonseNSig}{14}
\newcommand{\ChkExpMCapHsJeonse}{0.029}
\newcommand{\ChkExpMCapHsSale}{0.045}
\newcommand{\ChkExpMKmHsJeonse}{0.095}
\newcommand{\ChkExpMKmHsSale}{0.074}
\newcommand{\ChkExpMNoEdgeHsJeonse}{0.043}
\newcommand{\ChkExpMNoEdgeHsSale}{0.055}
\newcommand{\ChkExpMPreEarlyHsJeonse}{0.042}
\newcommand{\ChkExpMPreEarlyHsSale}{0.037}
\newcommand{\ChkExpMPreHsJeonse}{0.040}
\newcommand{\ChkExpMPreHsSale}{0.042}
\newcommand{\ChkExpMTimeHsJeonse}{0.040}
\newcommand{\ChkExpMTimeHsSale}{0.052}
\newcommand{\ChkExpNKmHsGapP}{\ensuremath{p=0.19}}
\newcommand{\ChkExpNKmHsJeonse}{0.102}
\newcommand{\ChkExpNKmHsSale}{0.065}
\newcommand{\ChkExpNNoEdgeShopGap}{0.098}
\newcommand{\ChkExpNNoEdgeShopGapP}{\ensuremath{p<0.001}}
\newcommand{\ChkExpNNoEdgeShopJeonse}{0.006}
\newcommand{\ChkExpNSplitBrkJeonse}{0.085}
\newcommand{\ChkExpNSplitBrkSale}{0.093}
\newcommand{\ChkExpNSplitHealthJeonse}{0.029}
\newcommand{\ChkExpNSplitHealthSale}{0.053}
\newcommand{\ChkExpNSplitRepairGap}{0.169}
\newcommand{\ChkExpNSplitRepairGapP}{\ensuremath{p<0.001}}
\newcommand{\ChkExpNSplitRepairJeonse}{\ensuremath{-}0.044}
\newcommand{\ChkExpNSplitRepairSale}{\ensuremath{-}0.213}
\newcommand{\ChkExpSpecs}{15}
\newcommand{\ChkFamPassBH}{19}
\newcommand{\ChkFamPassBonf}{15}
\newcommand{\ChkFamPassFive}{28}
\newcommand{\ChkFamQCellYoungN}{0.014}
\newcommand{\ChkFamQEntryMin}{0.063}
\newcommand{\ChkFamQMHsJeonse}{0.079}
\newcommand{\ChkFamTests}{54}
\newcommand{\ChkFamThreshold}{0.00093}
\newcommand{\ChkGridFourCells}{51}
\newcommand{\ChkGridTwoCells}{153}
\newcommand{\ChkInfCHagCellPWild}{\ensuremath{p=0.12}}
\newcommand{\ChkInfCHagGapPWild}{\ensuremath{p=0.032}}
\newcommand{\ChkInfJHagGapPWild}{\ensuremath{p=0.004}}
\newcommand{\ChkInfMHsCellPWild}{\ensuremath{p=0.89}}
\newcommand{\ChkInfMHsJeonsePWild}{\ensuremath{p=0.095}}
\newcommand{\ChkInfMHsJeonsePZone}{\ensuremath{p=0.071}}
\newcommand{\ChkInfMHsSalePGridMax}{\ensuremath{p=0.006}}
\newcommand{\ChkInfMHsSalePWild}{\ensuremath{p=0.005}}
\newcommand{\ChkKaptTxnShare}{89.1}
\newcommand{\ChkLooDaechiMHsJeonse}{0.028}
\newcommand{\ChkLooThreeMHsJeonse}{0.022}
\newcommand{\ChkLooThreeMHsSale}{0.053}
\newcommand{\ChkOsterMJeonseHigh}{0.27}
\newcommand{\ChkOsterMJeonseLow}{1.07}
\newcommand{\ChkOsterMSaleHigh}{0.46}
\newcommand{\ChkOsterMSaleLow}{1.68}
\newcommand{\ChkWildDraws}{99,999}
\newcommand{\DatCorrCramEighteenFifteen}{0.96}
\newcommand{\DatCorrCramEighteenTwentyFive}{0.94}
\newcommand{\DatCorrCramNonfood}{0.29}
\newcommand{\DatCorrCramNonstudent}{0.43}
\newcommand{\DatCorrCramOtherEdu}{0.58}
\newcommand{\DatCorrCramShops}{0.29}
\newcommand{\DatCorrHsOther}{0.85}
\newcommand{\DatCorrHsSbizEighteen}{0.82}
\newcommand{\DatCorrNeisSbiz}{0.90}
\newcommand{\DatCorrNeisSbizEighteen}{0.87}
\newcommand{\DatDateFirst}{1 January 2015}
\newcommand{\DatDateLast}{12 October 2026}
\newcommand{\DatFlagBlank}{92,592}
\newcommand{\DatFlagBlankBeforePct}{96}
\newcommand{\DatFlagBlankPct}{12}
\newcommand{\DatFlagJeonse}{780,063}
\newcommand{\DatFlagNew}{400,730}
\newcommand{\DatFlagRenewal}{286,741}
\newcommand{\DatGeoBlockN}{3,687}
\newcommand{\DatGeoBlockPct}{36.4}
\newcommand{\DatGeoBlockSpread}{150}
\newcommand{\DatGeoBlockTxPct}{5.0}
\newcommand{\DatGeoComplexes}{10,116}
\newcommand{\DatGeoExactTxPct}{88.3}
\newcommand{\DatGeoLocated}{8,729}
\newcommand{\DatGeoLocatedGuMaxPct}{91}
\newcommand{\DatGeoLocatedJongnoPct}{71}
\newcommand{\DatGeoLocatedJungguPct}{66}
\newcommand{\DatGeoLocatedLargePct}{91}
\newcommand{\DatGeoLocatedMeanTx}{356}
\newcommand{\DatGeoLocatedMedianTx}{54}
\newcommand{\DatGeoLocatedNewPct}{11}
\newcommand{\DatGeoLocatedPct}{86}
\newcommand{\DatGeoLocatedSmallPct}{88}
\newcommand{\DatGeoLocatedTxPct}{93.3}
\newcommand{\DatGeoLotN}{4,688}
\newcommand{\DatGeoLotPct}{46.3}
\newcommand{\DatGeoLotPointsMedian}{6}
\newcommand{\DatGeoLotTxPct}{86.6}
\newcommand{\DatGeoNameN}{206}
\newcommand{\DatGeoNameSpread}{100}
\newcommand{\DatGeoPointBiz}{796,404}
\newcommand{\DatGeoPointFiles}{16}
\newcommand{\DatGeoPointRows}{1,466,068}
\newcommand{\DatGeoRebuildAgreePct}{100.0}
\newcommand{\DatGeoRoadN}{148}
\newcommand{\DatGeoUnmatchedMeanTx}{160}
\newcommand{\DatGeoUnmatchedMedianTx}{66}
\newcommand{\DatGeoUnmatchedN}{1,387}
\newcommand{\DatGeoUnmatchedNewPct}{11}
\newcommand{\DatGeoUnmatchedPct}{13.7}
\newcommand{\DatGeoUnmatchedTxPct}{6.7}
\newcommand{\DatGradeAnyCuePct}{46}
\newcommand{\DatGradeElemPct}{21}
\newcommand{\DatGradeHsAlsoLowerPct}{42}
\newcommand{\DatGradeHsN}{5,278}
\newcommand{\DatGradeHsPct}{37}
\newcommand{\DatGradeMidPct}{21}
\newcommand{\DatGradeNameCuePct}{12}
\newcommand{\DatGradeNoCueN}{7,595}
\newcommand{\DatGradeNoCuePct}{54}
\newcommand{\DatGradeOtherN}{8,868}
\newcommand{\DatGradeOtherPct}{63}
\newcommand{\DatGradeTextOnlyPct}{34}
\newcommand{\DatJeonseClean}{1,526,777}
\newcommand{\DatJeonseRatioMedianPct}{71}
\newcommand{\DatJeonseShareHOneTwenty}{71}
\newcommand{\DatJeonseShareHOneTwentySix}{50}
\newcommand{\DatJeonseShareHOneTwentyThree}{58}
\newcommand{\DatLeaseFiles}{24}
\newcommand{\DatLeasesRaw}{2,489,023}
\newcommand{\DatLeasesTwenty}{196,650}
\newcommand{\DatLeasesTwentyTwo}{259,616}
\newcommand{\DatMapBinHigh}{160}
\newcommand{\DatMaxCram}{496}
\newcommand{\DatMeanJeonseArea}{75}
\newcommand{\DatMeanSaleAge}{19}
\newcommand{\DatMeanSaleArea}{77}
\newcommand{\DatMedBiz}{942}
\newcommand{\DatMedCram}{16}
\newcommand{\DatMedNeisHs}{6}
\newcommand{\DatMedNeisOther}{14}
\newcommand{\DatMedNonedu}{876}
\newcommand{\DatMedNonfood}{325}
\newcommand{\DatMedNonstudent}{192}
\newcommand{\DatMedianJeonsePriceM}{6.0}
\newcommand{\DatMedianSaleNeisHs}{9}
\newcommand{\DatMedianSalePriceM}{8.5}
\newcommand{\DatNeisAcad}{14,146}
\newcommand{\DatNeisAcadHagwon}{8,477}
\newcommand{\DatNeisAcadPct}{55}
\newcommand{\DatNeisAcadRoom}{5,669}
\newcommand{\DatNeisCourseFilledPct}{75}
\newcommand{\DatNeisFeeFilledPct}{24}
\newcommand{\DatNeisHsOpenedAfterEighteenPct}{55}
\newcommand{\DatNeisKeyedPct}{99.9}
\newcommand{\DatNeisOpenedAfterEighteenPct}{53}
\newcommand{\DatNeisOpenedBeforeFifteenPct}{31}
\newcommand{\DatNeisOpenedKnownPct}{100}
\newcommand{\DatNeisOpenedMedianYear}{2019}
\newcommand{\DatNeisPlacedMinGuPct}{98}
\newcommand{\DatNeisPlacedN}{14,006}
\newcommand{\DatNeisPlacedPct}{99.0}
\newcommand{\DatNeisPlacedSpread}{0}
\newcommand{\DatNeisSeoulAll}{25,507}
\newcommand{\DatPNinetyCram}{43}
\newcommand{\DatPNinetyNeisHs}{24}
\newcommand{\DatRightMedianIncrease}{5}
\newcommand{\DatSaleFiles}{12}
\newcommand{\DatSalesCancelled}{17,521}
\newcommand{\DatSalesClean}{845,315}
\newcommand{\DatSalesDirect}{12,985}
\newcommand{\DatSalesDirectPct}{1.5}
\newcommand{\DatSalesRaw}{862,836}
\newcommand{\DatSampleBlockTxPct}{5.3}
\newcommand{\DatSampleComplexes}{8,427}
\newcommand{\DatSampleDongs}{332}
\newcommand{\DatSampleJeonse}{1,426,021}
\newcommand{\DatSampleLostFilterPct}{0.1}
\newcommand{\DatSampleLostLocatePct}{6.7}
\newcommand{\DatSampleMonths}{142}
\newcommand{\DatSampleN}{2,197,510}
\newcommand{\DatSampleNAll}{2,359,107}
\newcommand{\DatSampleSales}{771,489}
\newcommand{\DatSampleSharePct}{93.2}
\newcommand{\DatSampleZones}{46}
\newcommand{\DatSbizBizEighteen}{457,024}
\newcommand{\DatSbizBizTwentySix}{554,092}
\newcommand{\DatSbizCramDropPct}{11}
\newcommand{\DatSbizCramEighteen}{9,690}
\newcommand{\DatSbizCramFifteen}{6,579}
\newcommand{\DatSbizCramPointsEighteen}{21,702}
\newcommand{\DatSbizCramTwentyFive}{16,739}
\newcommand{\DatSbizCramTwentySix}{14,852}
\newcommand{\DatSbizEduDropPct}{2}
\newcommand{\DatSbizEduEighteen}{27,393}
\newcommand{\DatSbizIncompleteCramPct}{43}
\newcommand{\DatSbizIncompleteEduPct}{61}
\newcommand{\DatSnapshots}{14}
\newcommand{\DatValBlockAllN}{2,503}
\newcommand{\DatValBlockFarPct}{4}
\newcommand{\DatValBlockMedian}{66}
\newcommand{\DatValBlockN}{2,458}
\newcommand{\DatValBlockOverlapPct}{75}
\newcommand{\DatValBlockPNinety}{195}
\newcommand{\DatValLotNameN}{3,253}
\newcommand{\DatValLotNamePNinety}{1}
\newcommand{\DatValLotNameSamePct}{90}
\newcommand{\DatValLotRoadFarPct}{0.5}
\newcommand{\DatValLotRoadN}{4,580}
\newcommand{\DatValLotRoadSamePct}{90}
\newcommand{\DatValPairComplexes}{2,503}
\newcommand{\DatValPairMedian}{87}
\newcommand{\DatValPairN}{25,336}
\newcommand{\DatValPairOverlapPct}{62}
\newcommand{\DatValPairPNinety}{302}
\newcommand{\DatValSeoulSampleDistOne}{19}
\newcommand{\DatValSeoulSampleDistTwo}{103}
\newcommand{\DatValSeoulSampleN}{2}
\newcommand{\DatValSeoulSampleRows}{5}
\newcommand{\DatWithinZonePct}{81}
\newcommand{\DatWolseClean}{962,246}
\newcommand{\DatYearFirst}{2015}
\newcommand{\DatYearLast}{2026}
\newcommand{\DatZoneAreaMax}{31}
\newcommand{\DatZoneAreaMedian}{11}
\newcommand{\DatZoneAreaMin}{5}
\newcommand{\DatZoneAreaTotal}{601}
\newcommand{\DatZoneComplexesMax}{690}
\newcommand{\DatZoneComplexesMedian}{170}
\newcommand{\DatZoneComplexesMin}{38}
\newcommand{\DatZoneNearestN}{0}
\newcommand{\DatZoneUpdFirst}{August 2016}
\newcommand{\DatZoneUpdLast}{April 2023}
\newcommand{\DatZonesNational}{1,687}
\newcommand{\DatZonesSeoul}{46}
\newcommand{\FigSpecBaseMove}{6 to 45}
\newcommand{\FigSpecJeonseAbove}{21}
\newcommand{\FigSpecJeonseMax}{20.9}
\newcommand{\FigSpecJeonseMin}{4.2}
\newcommand{\FigSpecN}{30}
\newcommand{\FigSpecSaleMax}{15.6}
\newcommand{\FigSpecSaleMin}{4.8}
\newcommand{\FigZoneComplexes}{8,729}
\newcommand{\FigZones}{46}
\newcommand{\ResAHagJeonse}{0.047}
\newcommand{\ResAHagSale}{0.025}
\newcommand{\ResBHagJeonse}{0.066}
\newcommand{\ResBHagSale}{0.036}
\newcommand{\ResCHagGap}{0.025}
\newcommand{\ResCHagGapP}{\ensuremath{p=0.003}}
\newcommand{\ResCHagGapSe}{0.009}
\newcommand{\ResCHagJeonse}{0.051}
\newcommand{\ResCHagJeonseP}{\ensuremath{p<0.001}}
\newcommand{\ResCHagJeonsePct}{3.6}
\newcommand{\ResCHagJeonseSe}{0.014}
\newcommand{\ResCHagSale}{0.026}
\newcommand{\ResCHagSaleP}{\ensuremath{p=0.032}}
\newcommand{\ResCHagSalePct}{1.8}
\newcommand{\ResCHagSaleSe}{0.012}
\newcommand{\ResCellAllCHag}{0.014}
\newcommand{\ResCellAllCHagP}{\ensuremath{p=0.042}}
\newcommand{\ResCellAllCHagSe}{0.007}
\newcommand{\ResCellAllJHag}{\ensuremath{-}0.004}
\newcommand{\ResCellAllJHagSe}{0.008}
\newcommand{\ResCellAllJShop}{0.040}
\newcommand{\ResCellAllJShopSe}{0.009}
\newcommand{\ResCellAllMHs}{0.002}
\newcommand{\ResCellAllMHsHi}{0.024}
\newcommand{\ResCellAllMHsLo}{\ensuremath{-}0.020}
\newcommand{\ResCellAllMHsMaxApprShare}{35}
\newcommand{\ResCellAllMHsMinusSaleZ}{5.2}
\newcommand{\ResCellAllMHsP}{\ensuremath{p=0.85}}
\newcommand{\ResCellAllMHsPct}{0.4}
\newcommand{\ResCellAllMHsPctHi}{4.9}
\newcommand{\ResCellAllMHsPctLo}{\ensuremath{-}3.8}
\newcommand{\ResCellAllMHsSe}{0.011}
\newcommand{\ResCellAllNHs}{0.001}
\newcommand{\ResCellAllNHsSe}{0.011}
\newcommand{\ResCellAllNShop}{0.042}
\newcommand{\ResCellAllNShopP}{\ensuremath{p<0.001}}
\newcommand{\ResCellAllNShopPct}{3.0}
\newcommand{\ResCellAllNShopPctHi}{4.4}
\newcommand{\ResCellAllNShopPctLo}{1.6}
\newcommand{\ResCellAllNShopSe}{0.010}
\newcommand{\ResCellAvgGap}{\ensuremath{-}0.579}
\newcommand{\ResCellComplexShare}{61}
\newcommand{\ResCellExactMHs}{0.003}
\newcommand{\ResCellExactNShop}{0.038}
\newcommand{\ResCellFromBanCHag}{0.008}
\newcommand{\ResCellFromBanMHs}{0.002}
\newcommand{\ResCellFromBanNShop}{0.048}
\newcommand{\ResCellObsCells}{222,400}
\newcommand{\ResCellObsComplexes}{5,123}
\newcommand{\ResCellPreBanCHag}{0.022}
\newcommand{\ResCellPreBanCHagP}{\ensuremath{p=0.025}}
\newcommand{\ResCellPreBanMHs}{0.009}
\newcommand{\ResCellPreBanNShop}{0.039}
\newcommand{\ResCellRatioPct}{56}
\newcommand{\ResDHagJeonse}{0.100}
\newcommand{\ResDHagSale}{0.038}
\newcommand{\ResDaechiCount}{733}
\newcommand{\ResDaechiScale}{4.4}
\newcommand{\ResEFAllEarlyMarkup}{0.181}
\newcommand{\ResEFAllEarlyMarkupPct}{19.9}
\newcommand{\ResEFAllEarlyNHs}{0.0107}
\newcommand{\ResEFAllEarlyNHsP}{\ensuremath{p=0.042}}
\newcommand{\ResEFAllEarlyNHsPct}{2.1}
\newcommand{\ResEFAllFullCHag}{0.0005}
\newcommand{\ResEFAllFullCHagSe}{0.0012}
\newcommand{\ResEFAllFullCells}{74,443}
\newcommand{\ResEFAllFullContracts}{349,424}
\newcommand{\ResEFAllFullExactNHs}{0.0046}
\newcommand{\ResEFAllFullExactNHsP}{\ensuremath{p=0.099}}
\newcommand{\ResEFAllFullMOth}{\ensuremath{-}0.0035}
\newcommand{\ResEFAllFullMarkup}{0.086}
\newcommand{\ResEFAllFullMarkupPct}{9.0}
\newcommand{\ResEFAllFullMedHs}{9}
\newcommand{\ResEFAllFullNHs}{0.0049}
\newcommand{\ResEFAllFullNHsP}{\ensuremath{p=0.068}}
\newcommand{\ResEFAllFullNHsPct}{1.0}
\newcommand{\ResEFAllFullNHsPctMed}{0.8}
\newcommand{\ResEFAllFullNHsSe}{0.0027}
\newcommand{\ResEFAllFullNShop}{\ensuremath{-}0.0030}
\newcommand{\ResEFAllFullNShopSe}{0.0022}
\newcommand{\ResEFAllLateMarkup}{0.053}
\newcommand{\ResEFAllLateMarkupPct}{5.4}
\newcommand{\ResEFAllLateNHs}{0.0025}
\newcommand{\ResEFAllLateNHsP}{\ensuremath{p=0.35}}
\newcommand{\ResEFBHMinQ}{0.19}
\newcommand{\ResEFBHTests}{9}
\newcommand{\ResEFCapEarlyMarkup}{0.212}
\newcommand{\ResEFCapEarlyMarkupPct}{23.6}
\newcommand{\ResEFCapEarlyNHs}{0.0131}
\newcommand{\ResEFCapEarlyNHsP}{\ensuremath{p=0.025}}
\newcommand{\ResEFCapEarlyNHsPct}{2.6}
\newcommand{\ResEFCapEarlyNShop}{\ensuremath{-}0.0112}
\newcommand{\ResEFCapEarlyNShopP}{\ensuremath{p=0.013}}
\newcommand{\ResEFCapFullMarkup}{0.112}
\newcommand{\ResEFCapFullMarkupPct}{11.9}
\newcommand{\ResEFCapFullNHs}{0.0048}
\newcommand{\ResEFCapLateNHs}{0.0005}
\newcommand{\ResEFCapLateNHsP}{\ensuremath{p=0.85}}
\newcommand{\ResEFCapped}{141,768}
\newcommand{\ResEFLocated}{644,100}
\newcommand{\ResEFMinP}{0.025}
\newcommand{\ResEFNew}{374,741}
\newcommand{\ResEFVolEarlyNHs}{0.0092}
\newcommand{\ResEFVolEarlyNHsP}{\ensuremath{p=0.091}}
\newcommand{\ResEFVolFullMarkup}{0.058}
\newcommand{\ResEFVolFullMarkupPct}{6.0}
\newcommand{\ResEFVolFullNHs}{0.0048}
\newcommand{\ResEFVolFullNHsP}{\ensuremath{p=0.14}}
\newcommand{\ResEFVolLateNHs}{0.0035}
\newcommand{\ResEFVolLateNHsP}{\ensuremath{p=0.32}}
\newcommand{\ResEFVoluntary}{127,591}
\newcommand{\ResEHagJeonse}{0.050}
\newcommand{\ResEHagSale}{0.024}
\newcommand{\ResEvCappedPostP}{\ensuremath{p=0.13}}
\newcommand{\ResEvCappedPreP}{\ensuremath{p=0.93}}
\newcommand{\ResEvNewJCoefTwentyOneHTwo}{0.0147}
\newcommand{\ResEvNewJCoefTwentyOneHTwoSe}{0.0050}
\newcommand{\ResEvNewJFixedEighteenJumpJun}{0.0012}
\newcommand{\ResEvNewJFixedTwentyTwoJumpJun}{0.0022}
\newcommand{\ResEvNewJJumpFeb}{0.0013}
\newcommand{\ResEvNewJJumpFebPct}{0.6}
\newcommand{\ResEvNewJJumpFebPctHi}{4.4}
\newcommand{\ResEvNewJJumpFebPctLo}{\ensuremath{-}3.1}
\newcommand{\ResEvNewJJumpFebSe}{0.0043}
\newcommand{\ResEvNewJJumpJun}{0.0027}
\newcommand{\ResEvNewJJumpJunPct}{1.2}
\newcommand{\ResEvNewJJumpJunPctHi}{3.8}
\newcommand{\ResEvNewJJumpJunPctLo}{\ensuremath{-}1.4}
\newcommand{\ResEvNewJJumpJunSe}{0.0030}
\newcommand{\ResEvNewJPostF}{1.20}
\newcommand{\ResEvNewJPostP}{\ensuremath{p=0.30}}
\newcommand{\ResEvNewJPreF}{3.81}
\newcommand{\ResEvNewJPreP}{\ensuremath{p=0.005}}
\newcommand{\ResEvNewJWinter}{\ensuremath{-}0.0018}
\newcommand{\ResEvNewJWinterSe}{0.0019}
\newcommand{\ResEvNewJYoungPreP}{\ensuremath{p=0.25}}
\newcommand{\ResEvSalesCoefTwentyFiveHTwo}{0.0121}
\newcommand{\ResEvSalesCoefTwentyFiveHTwoSe}{0.0055}
\newcommand{\ResEvSalesJumpFeb}{0.0022}
\newcommand{\ResEvSalesJumpFebP}{\ensuremath{p=0.10}}
\newcommand{\ResEvSalesJumpJun}{0.0027}
\newcommand{\ResEvSalesJumpJunP}{\ensuremath{p=0.13}}
\newcommand{\ResEvSalesPostP}{\ensuremath{p=0.23}}
\newcommand{\ResEvSalesYoungCoefTwentyFiveHTwo}{0.0059}
\newcommand{\ResEvSalesYoungJumpFeb}{0.0029}
\newcommand{\ResEvSalesYoungJumpFebP}{\ensuremath{p=0.078}}
\newcommand{\ResFHagJeonse}{0.064}
\newcommand{\ResFHagSale}{0.032}
\newcommand{\ResGGHundredJeonsePct}{25.0}
\newcommand{\ResGGHundredSalePct}{15.2}
\newcommand{\ResGGTenJeonsePct}{5.1}
\newcommand{\ResGGTenSalePct}{1.7}
\newcommand{\ResGGThirtyJeonsePct}{12.8}
\newcommand{\ResGGThirtySalePct}{5.4}
\newcommand{\ResHHagGap}{\ensuremath{-}0.022}
\newcommand{\ResHHagGapP}{\ensuremath{p=0.008}}
\newcommand{\ResHHagGapSe}{0.008}
\newcommand{\ResHHagJeonse}{0.031}
\newcommand{\ResHHagSale}{0.053}
\newcommand{\ResHShopJeonse}{0.043}
\newcommand{\ResHShopSale}{\ensuremath{-}0.064}
\newcommand{\ResHsPSeventyFive}{12}
\newcommand{\ResHsPTwentyFive}{3}
\newcommand{\ResHsPctileTarget}{97}
\newcommand{\ResHsTarget}{50}
\newcommand{\ResIHagJeonse}{0.027}
\newcommand{\ResIHagSale}{0.067}
\newcommand{\ResIOthEduJeonse}{0.042}
\newcommand{\ResIOthEduSale}{\ensuremath{-}0.077}
\newcommand{\ResJHagGap}{\ensuremath{-}0.036}
\newcommand{\ResJHagGapP}{\ensuremath{p<0.001}}
\newcommand{\ResJHagJeonse}{0.025}
\newcommand{\ResJHagJeonsePct}{1.8}
\newcommand{\ResJHagJeonseSe}{0.012}
\newcommand{\ResJHagSale}{0.061}
\newcommand{\ResJHagSalePct}{4.3}
\newcommand{\ResJHagSaleSe}{0.011}
\newcommand{\ResJShopJeonseP}{\ensuremath{p=0.004}}
\newcommand{\ResJShopJeonsePct}{3.7}
\newcommand{\ResJShopSalePct}{\ensuremath{-}5.4}
\newcommand{\ResKHagGap}{\ensuremath{-}0.020}
\newcommand{\ResKShopJeonse}{0.040}
\newcommand{\ResKShopSale}{\ensuremath{-}0.072}
\newcommand{\ResMHsGap}{\ensuremath{-}0.020}
\newcommand{\ResMHsGapP}{\ensuremath{p=0.19}}
\newcommand{\ResMHsGapSe}{0.016}
\newcommand{\ResMHsJeonse}{0.036}
\newcommand{\ResMHsJeonseKRW}{34}
\newcommand{\ResMHsJeonseP}{\ensuremath{p=0.034}}
\newcommand{\ResMHsJeonsePct}{7.3}
\newcommand{\ResMHsJeonsePctHi}{14.6}
\newcommand{\ResMHsJeonsePctIqr}{4.3}
\newcommand{\ResMHsJeonsePctLo}{0.5}
\newcommand{\ResMHsJeonseSe}{0.017}
\newcommand{\ResMHsSale}{0.056}
\newcommand{\ResMHsSaleKRW}{83}
\newcommand{\ResMHsSaleP}{\ensuremath{p<0.001}}
\newcommand{\ResMHsSalePct}{11.8}
\newcommand{\ResMHsSalePctHi}{18.7}
\newcommand{\ResMHsSalePctIqr}{6.8}
\newcommand{\ResMHsSalePctLo}{5.2}
\newcommand{\ResMHsSalePctTxn}{9.6}
\newcommand{\ResMHsSaleSe}{0.016}
\newcommand{\ResMOthJeonse}{0.005}
\newcommand{\ResMOthJeonseHi}{0.040}
\newcommand{\ResMOthJeonseLo}{\ensuremath{-}0.030}
\newcommand{\ResMOthJeonseSe}{0.018}
\newcommand{\ResMOthSale}{\ensuremath{-}0.016}
\newcommand{\ResMOthSaleHi}{0.016}
\newcommand{\ResMOthSaleLo}{\ensuremath{-}0.049}
\newcommand{\ResMOthSaleSe}{0.017}
\newcommand{\ResMZeroRegGap}{0.001}
\newcommand{\ResMZeroRegGapP}{\ensuremath{p=0.86}}
\newcommand{\ResMZeroRegGapSe}{0.008}
\newcommand{\ResMZeroRegJeonse}{0.037}
\newcommand{\ResMZeroRegJeonseSe}{0.010}
\newcommand{\ResMZeroRegSale}{0.036}
\newcommand{\ResMZeroRegSaleSe}{0.010}
\newcommand{\ResMedDongCount}{8}
\newcommand{\ResMedJeonseArea}{77}
\newcommand{\ResMedJeonsePriceM}{6.0}
\newcommand{\ResMedSaleArea}{83}
\newcommand{\ResMedSalePriceM}{8.5}
\newcommand{\ResNHsJeonse}{0.034}
\newcommand{\ResNHsSale}{0.055}
\newcommand{\ResNOthJeonse}{\ensuremath{-}0.015}
\newcommand{\ResNOthSale}{0.007}
\newcommand{\ResNShopJeonse}{0.061}
\newcommand{\ResNShopSale}{\ensuremath{-}0.070}
\newcommand{\ResOldJeonseTwentyHOne}{0.0030}
\newcommand{\ResOldJeonseTwentyHOnePct}{1.3}
\newcommand{\ResOldJeonseTwentyHOneSe}{0.0035}
\newcommand{\ResOldSaleNineteenHTwo}{0.0048}
\newcommand{\ResOldSaleNineteenHTwoP}{\ensuremath{p=0.044}}
\newcommand{\ResOldSaleTwentyHOne}{0.0022}
\newcommand{\ResOldSaleTwentyHOneSe}{0.0033}
\newcommand{\ResPerBanJHagGap}{\ensuremath{-}0.059}
\newcommand{\ResPerBanMHsGap}{\ensuremath{-}0.029}
\newcommand{\ResPerBanMHsGapP}{\ensuremath{p=0.15}}
\newcommand{\ResPerBanMHsSale}{0.058}
\newcommand{\ResPerBanNShopGap}{0.161}
\newcommand{\ResPerBanNShopJeonse}{0.070}
\newcommand{\ResPerBanNShopSale}{\ensuremath{-}0.090}
\newcommand{\ResPerNoPermitMHsGap}{\ensuremath{-}0.030}
\newcommand{\ResPerNoPermitMHsGapP}{\ensuremath{p=0.044}}
\newcommand{\ResPerNoPermitMHsSale}{0.064}
\newcommand{\ResPerPostBanMHsGap}{\ensuremath{-}0.025}
\newcommand{\ResPerPostBanMHsGapP}{\ensuremath{p=0.16}}
\newcommand{\ResPerPreBanMHsGap}{\ensuremath{-}0.006}
\newcommand{\ResPerPreBanMHsGapP}{\ensuremath{p=0.68}}
\newcommand{\ResPerPreBanMHsJeonse}{0.040}
\newcommand{\ResPerPreBanMHsJeonsePct}{8.4}
\newcommand{\ResPerPreBanMHsSale}{0.046}
\newcommand{\ResPerPreBanMHsSalePct}{9.6}
\newcommand{\ResPerPreCurbsCHagGap}{0.026}
\newcommand{\ResPerPreCurbsCHagGapP}{\ensuremath{p=0.022}}
\newcommand{\ResPerPreCurbsJHagGap}{\ensuremath{-}0.008}
\newcommand{\ResPerPreCurbsMHsGap}{0.005}
\newcommand{\ResPerPreCurbsMHsGapP}{\ensuremath{p=0.78}}
\newcommand{\ResPerPreCurbsMHsJeonse}{0.042}
\newcommand{\ResPerPreCurbsMHsSale}{0.038}
\newcommand{\ResPerPreCurbsNShopGap}{0.081}
\newcommand{\ResPerPreCurbsNShopJeonse}{0.028}
\newcommand{\ResPerPreCurbsNShopSale}{\ensuremath{-}0.054}
\newcommand{\ResPermitJeonseShare}{10.9}
\newcommand{\ResPermitSaleShare}{7.8}
\newcommand{\ResRecentAllJMHsGap}{\ensuremath{-}0.029}
\newcommand{\ResRecentAllJMHsGapP}{\ensuremath{p=0.073}}
\newcommand{\ResRecentAllJYoungMHsGap}{\ensuremath{-}0.046}
\newcommand{\ResRecentAllJYoungMHsGapP}{\ensuremath{p=0.019}}
\newcommand{\ResRecentAllJYoungMHsSale}{0.086}
\newcommand{\ResRecentAllJYoungNHsGap}{\ensuremath{-}0.057}
\newcommand{\ResRecentAllJYoungNHsGapP}{\ensuremath{p<0.001}}
\newcommand{\ResRecentCapJCHagGap}{0.003}
\newcommand{\ResRecentCapJCHagGapP}{\ensuremath{p=0.68}}
\newcommand{\ResRecentCapJMHsGap}{\ensuremath{-}0.025}
\newcommand{\ResRecentCapJMHsGapP}{\ensuremath{p=0.10}}
\newcommand{\ResRecentCapJNShopJeonse}{0.003}
\newcommand{\ResRecentNewJCHagGap}{0.026}
\newcommand{\ResRecentNewJMHsGap}{\ensuremath{-}0.030}
\newcommand{\ResRecentNewJMHsGapP}{\ensuremath{p=0.066}}
\newcommand{\ResRecentNewJNShopJeonse}{0.060}
\newcommand{\ResRecentNewJYoungMHsGap}{\ensuremath{-}0.047}
\newcommand{\ResRecentNewJYoungMHsGapP}{\ensuremath{p=0.017}}
\newcommand{\ResRecentNewJYoungNHsGap}{\ensuremath{-}0.058}
\newcommand{\ResRecentNewJYoungNHsGapP}{\ensuremath{p<0.001}}
\newcommand{\ResRowsAllMHs}{\ensuremath{-}0.004}
\newcommand{\ResRowsAllMHsSe}{0.013}
\newcommand{\ResRowsAllNShop}{0.050}
\newcommand{\ResScaleHs}{1.99}
\newcommand{\ResScaleHsTxn}{1.63}
\newcommand{\ResWithinZoneShareHs}{78}

\begin{document}

\maketitle

\begin{abstract}
South Korean families spend heavily on hagwons, private cram schools that prepare students for the university entrance exam. This paper asks what access to them is worth in housing and whether the premium includes expected appreciation. Using \DatSampleN{} Seoul apartment sales and jeonse (lump-sum deposit lease) contracts from 2015 to 2026, located with a new business-address geocoder, the analysis compares flats within the same middle-school assignment zone and month. Flats near hagwons preparing high-school students for the exam sell and lease for more: moving from the median complex to the \ResHsPctileTarget{}th percentile of such hagwons within 500 meters is associated with sale prices \ResMHsSalePct\% and jeonse deposits \ResMHsJeonsePct\% higher, or \ChkAccSizeMHsSalePct\% and \ChkAccSizeMHsJeonsePct\% with complex size held fixed. Hagwons teaching younger children add no sale premium once these are held fixed. A larger premium in deposits than in prices across complexes reflects convenience: buyers pay less than renters to live near shops and services, above all repair and personal services. For the same flat type, complex and month, hagwon access produces no detectable change in the jeonse-to-price ratio, although the test detects the reconstruction value of old buildings; nearby shops raise it. Hagwon access is thus capitalized in proportion into prices and deposits, as expected of a current-flow amenity without extra expected appreciation. New tenants pay no lasting premium over renewing tenants, and deposits did not respond to major education-policy announcements. The housing premium is a cost of the education race that measures of tutoring spending omit.
\end{abstract}



\clearpage
\onehalfspacing

\section{Introduction}\label{sec:intro}

South Korean families spend more on private tutoring than almost any others. In 2025 they spent KRW~27.5 trillion on private education for school-age children. In Seoul, 82.6\% of students took part, and participating high-school students spent about KRW~1.05 million a month \citep{mods2026results}. Most of this money goes to \emph{hagwons}, private cram schools that prepare students for school examinations and, above all, for the national university entrance exam. A growing literature links this ``education fever'' to Korea's very low fertility: when parents compete for their children's relative standing, the cost of raising each child rises \citep{anderson2013education,kim2024status}. Studies of these costs measure tutoring fees. Yet hagwons cluster in a few districts, so families may also pay for them through housing, by living within reach of the best cram schools. Little is known about how large that housing cost is, what kind of tutoring it buys, or whether owners and renters pay it alike.

This paper measures what Seoul families pay in the housing market for access to hagwons. Seoul offers three advantages for this question. First, hagwons are dense and highly concentrated, in districts such as Daechi-dong, Mok-dong and Junggye-dong, but they are also spread across the city, so exposure varies widely. Second, public schools are assigned by residence: within a middle-school assignment zone, every flat gives its children the same lottery over the same middle schools, whether the family owns or rents. Comparing flats within a zone separates the premium for private tutoring from the premium for public school assignment, which earlier work shows is itself priced in Seoul \citep{lee2015school,chung2015school,kim2026parental}. Third, Seoul has two prices for the same housing stock. Besides sale prices, a large share of flats are let on \emph{jeonse}, a lease in which the tenant pays a lump-sum deposit, often half the value of the flat or more, instead of monthly rent, and recovers it at the end of the lease \citep{ambrose2003modeling}. Because the deposit is itself a capitalized value, the jeonse-to-price ratio of the same flat is a direct, market-based measure of how an amenity is priced in the asset market relative to the lease market.

The data cover \DatSampleN{} brokered sales and jeonse contracts in \DatSampleComplexes{} Seoul apartment complexes between 2015 and 2026 from the real-transaction records of the Ministry of Land, Infrastructure and Transport, \DatSampleSharePct\% of all such contracts in the city. Because the official geocoders require Korean identity verification, each complex is located from the addresses and coordinates of the businesses registered inside it, and the result is validated (Appendix~\ref{app:geocoding}). Hagwon access is the number of hagwons within 500~m of the complex, measured in two ways: from the national business register in December 2018, and from the official registry of hagwons, which records the courses each one teaches and makes it possible to separate hagwons that prepare high-school students for university entrance from those that teach younger children. Flats are compared only with flats in the same middle-school assignment zone and month.

The paper reports five findings. First, hagwon access is priced beyond school assignment, in both markets, and the premium is carried by high-school and exam-preparation hagwons. Moving a complex from the median of \DatMedNeisHs{} such hagwons within 500~m to \ResHsTarget{}, about the \ResHsPctileTarget{}th percentile, is associated with sale prices \ResMHsSalePct\% higher and jeonse deposits \ResMHsJeonsePct\% higher; a move across the interquartile range is associated with \ResMHsSalePctIqr\% and \ResMHsJeonsePctIqr\%. For a flat of median size and price, the larger move corresponds to about KRW~\ResMHsSaleKRW{} million in price. Exam hagwons cluster near large complexes, and holding complex size fixed roughly halves the sale premium, to \ChkAccSizeMHsSalePct\%, while the deposit premium, at \ChkAccSizeMHsJeonsePct\%, changes little; distances to schools and subway stations explain neither. The sale premium survives clustering by school zone, dropping the three densest hagwon districts and bounds on selection on unobservables; the deposit premium is estimated less precisely. Holding exam hagwons fixed, hagwons for younger children carry no premium in sale prices; in deposits the contrast is imprecise. Because those hagwons serve many of the same families and would load on a generic ``good neighborhood'' factor, the grade contrast is a partial falsification check, and it ties the premium to the competition for university places rather than to childcare.

Second, an apparent tenure difference turns out to be about convenience. In cross-complex comparisons with the business-register count, the hagwon premium is larger in jeonse deposits than in sale prices (a gap of \ResCHagGap{} log points, \ResCHagGapP). But hagwon districts are also busy commercial districts. Buyers pay less than renters to live near businesses that have no reason to follow students: doubling the number of estate agents, clinics and repair and personal services within 500~m is associated with changes of \ResJShopJeonsePct\% in deposits and \ResJShopSalePct\% in prices. The gap comes from everyday services such as hair salons, laundries and repair shops; estate agents and clinics are priced about alike in the two markets. Once any measure of commercial density is held fixed, or once hagwons are measured from the registry, the jeonse premium is no longer larger.

Third, and best identified, hagwon access is capitalized in proportion into prices and deposits. For the same flat size in the same complex and month, in \ResCellObsCells{} such cells, high-school hagwons produce no detectable change in the jeonse-to-price ratio: the coefficient is \ResCellAllMHs, with a 95\% interval of \ResCellAllMHsLo{} to \ResCellAllMHsHi. The test has power: in the same cells it detects that buildings of 30 years or more, whose prices carry reconstruction value, have ratios about \ChkCellPosAgeThirtyMPctAbs\% lower. Nearby shops, by contrast, raise the ratio, by \ResCellAllNShopPct\% for a doubling. Because this comparison holds fixed everything the two prices of a flat share, it does not require hagwons to be randomly located, only that nothing else that moves the ratio varies with hagwon access. In a user-cost framework with investor landlords (Section~\ref{sec:framework}), the proportional response is what one expects of a current-flow amenity; if no offsetting wedge is at work, it bounds the share of the sale premium that could reflect expected appreciation at about \ResCellAllMHsMaxApprShare\%. One pattern qualifies this result: in buildings under 25 years after mid-2021, the cross-complex sale premium exceeds the jeonse premium once shops are controlled, and a smaller gap of the same sign appears within complexes, which does not survive the correction for multiple testing.

Fourth, newcomers pay no lasting ``entry fee.'' Seoul's 2020 lease reform capped deposit increases for renewing tenants, so new contracts carry a markup over renewals in the same complex and month (\ResEFAllFullMarkupPct\% on average). The markup is barely larger near exam hagwons (\ResEFAllFullNHsPct\% for the median-to-\ResHsTarget{} move), appears only in 2021--22, and vanishes once renewals caught up with the market. Fifth, the premium is stable. It is present before and during Seoul's mortgage and land-permit rules of 2017--2026, which do not explain the main patterns, and the deposit premium did not respond measurably to three windows of national education-policy news that changed the stakes of exam preparation: the 2019 plan to abolish elite high schools, the 2023 removal of ``killer questions'' from the entrance exam, and the 2024 expansion of medical-school places. For the 2023 and 2024 windows, the upper 95\% bounds on net jumps in new-contract deposits are \ResEvNewJJumpJunPctHi\% and \ResEvNewJJumpFebPctHi\% for Daechi-dong relative to a median neighborhood. Sale premia rose gradually from 2017 rather than at the announcements.

Care is needed about what these estimates identify. The level premia are capitalization premia conditional on school assignment, flat characteristics and local commercial density. They are not the causal effect of opening a hagwon: hagwons locate where demand for them is high, and sorting on unobserved neighborhood traits remains possible, especially because school zones are large. The grade placebo, the commercial-density controls, controls for schools, transport and complex size, bounds on selection on unobservables and the stability across periods and policy regimes narrow this concern; Section~\ref{sec:strategy-identification} discusses what remains. The within-complex comparison of the two prices, and the comparison of new and renewal contracts, rest on weaker assumptions.

The paper contributes to three literatures. First, it extends the study of how education is capitalized into housing, which has focused on public schools and the rules that ration them \citep{black1999better,bayer2007unified,lee2015school,chung2015school}, to private supplementary education, which no assignment rule rations. A literature, mostly in Korean, relates hagwon supply to apartment prices, using district aggregates \citep{jin2005education,kim2007impacts,sung2023influence} or cross-sections of complexes within one or two districts \citep{youn2011impact,lee2020effects,lee2024analysis}; most, though not all, of these studies find higher prices where hagwons are more numerous. This paper appears to be the first to estimate the hagwon premium across an entire city while holding public-school assignment fixed, the first to separate hagwons that prepare high-school students for university entrance from those that teach younger children, and the first to compare the premium in sale prices and jeonse deposits for the same type of flat in the same complex and month. It complements evidence from China, where restricting tutoring moved competition into school-district housing \citep{jia2025unintended,chen2023pay}, by measuring the premium for tutoring access itself where tutoring is legal and dense.

Second, the paper adds to work on price--rent ratios and on how amenities are priced in rents and prices \citep{himmelberg2005assessing,hattapoglu2014dependency,zheng2016much,gabe2022relationship}. Jeonse makes the rent side a capitalized value, observed for the same flat type in the same complex and month, so the comparison does not depend on assumed rent-to-price conversions. The results also carry a methodological lesson: a cross-complex difference between rent and price capitalization can reflect correlated amenities and the composition of what is sold and what is rented, and it disappears in the within-complex comparison. This appears to be the first evidence that commercial convenience raises deposits relative to prices within the same complex and month. Third, the paper uses the new and renewal flags recorded since the 2020 lease reform \citep{koo2025putting,hur2025right} to show that its capped renewals produced a temporary newcomer markup, not a lasting entry fee to hagwon districts. Finally, the estimates add a cost that measures of tutoring spending omit to the study of education costs and fertility in East Asia \citep{anderson2013education,kim2024status,kim2026impact}; the paper does not estimate effects on fertility.

The rest of the paper is organized as follows. Section~\ref{sec:background} describes hagwons, school assignment, jeonse and Seoul's housing rules, and Section~\ref{sec:literature} reviews related work. Section~\ref{sec:data} presents the data and the geocoding, and Section~\ref{sec:strategy} the methodology. Section~\ref{sec:results} reports the main results, Section~\ref{sec:robust} robustness checks and the policy shocks, and Section~\ref{sec:conclusion} concludes.
\section{Institutional background}\label{sec:background}

This section describes the tutoring market and the exam it serves, public school assignment in Seoul, the jeonse lease, Seoul's housing rules, and the education-policy announcements used later as tests.

\subsection{Hagwons and the university-entrance race}\label{sec:background-hagwon}

Spending on private tutoring is measured by the annual Private Education Expenditures Survey (\emph{sagyoyukbi josa}), which the national statistics office and the Ministry of Education run on about 74,000 students in about 3,000 schools \citep{mods2026results}. In 2025, families spent KRW~27.5 trillion on private education for elementary, middle and high-school students, 5.7\% less than in 2024 as student numbers fell. Of all students, 75.7\% took part, and spending averaged KRW~458,000 per student per month, or KRW~604,000 per participating student \citep{mods2026results}. Most of this money goes to \emph{hagwons} (private cram schools): of the KRW~336,000 spent per student per month on academic subjects, KRW~267,000 paid for hagwon classes. High-school students take part less often (63.0\%) but spend the most when they do, KRW~793,000 a month. Seoul is the outlier. Its participation rate was 82.6\%, average spending was KRW~663,000 per student per month, and participating Seoul high-school students spent KRW~1.05 million a month \citep{mods2026results}.

Much of this spending targets university entrance. The College Scholastic Ability Test (CSAT, \emph{suneung}) is a national exam set by the Korea Institute for Curriculum and Evaluation and sat on a single day, usually in November. The latest sitting, on 13 November 2025, drew 554,174 applicants \citep{joongang2025half,joongang2025korea}. Universities admit students in two rounds. Early admission (\emph{susi}) relies mainly on high-school records, although many programs also require minimum CSAT grades. Regular admission (\emph{jeongsi}) relies mainly on CSAT scores \citep{kcue2025university,joongang2025half}. Nationwide, regular admission fills 19.7\% of places for the 2027 intake, and 92.7\% of those places are allocated mainly on CSAT scores \citep{kcue2025university}. The exam matters more at selective Seoul universities: in November 2019 the government required 16 of them to fill at least 40\% of places through regular admission from the 2023 intake \citep{newsis2019expansion}.

Hagwons cluster in a few hagwon districts (\emph{hagwon-ga}). The best known are Daechi-dong in Gangnam-gu, Mok-dong in Yangcheon-gu and Junggye-dong in Nowon-gu, often called Seoul's three main school districts; in Junggye-dong the hagwons sit around the Eunhaeng intersection \citep{newsis2025house}. Figure~\ref{fig:map} shows the same three clusters in the data (Section~\ref{sec:data}). These neighborhoods also have sought-after public schools, which is why the assignment rules in Section~\ref{sec:background-schools} matter.

Hagwons are regulated under the Act on the Establishment and Operation of Private Teaching Institutes and Extracurricular Lessons. Anyone opening one must meet the Act's facility standards and register with the superintendent of the city or provincial office of education \citep[Art.~6]{korea2018act}. The resulting registry underlies the grade classification (Appendix~\ref{app:grades}). The Act also lets the superintendent set teaching hours for school-subject hagwons, small teaching centers and private tutors, within limits set by city or provincial ordinance \citep[Art.~16]{korea2018act}. Seoul's ordinance confines such teaching to 5~a.m.--10~p.m. for all grades, a curfew formally introduced in 2008 \citep{joongang2025ordinance}. In 12 of the 17 cities and provinces, high-school students may attend hagwons later than in Seoul, in eight of them until midnight \citep{segye2025extending}. A 2025 proposal in the Seoul city council to extend high-school hours to midnight was held in committee and lapsed at the end of the council's term on 1 July 2026 \citep{smc2026bill}.

\subsection{Public school assignment in Seoul}\label{sec:background-schools}

Public schools in Seoul are assigned by residence at every level. For elementary school, the local \emph{dong} office lists the children living in its area, the head of the district office of education draws school catchments (\emph{tonghak guyeok}), and the \emph{dong} office designates the school that each child will attend \citep[Arts.~15--17]{korea2017enforcement}. Catchment boundaries are not observed (Section~\ref{sec:data}).

Middle schools are assigned by lottery. The head of the district office of education allocates applicants to schools by lottery within districts and school groups, and the superintendent sets these middle-school assignment zones (\emph{hakgyogun}) with the approval of the city council \citep[Art.~68]{korea2017enforcement}. Seoul has 46 such zones, which subdivide its 25 districts (\emph{gu}), and a child cannot normally attend a middle school outside the zone of residence without moving \citep{ytn2020fact}. Within a zone, schools are allocated by a computerized lottery that may take transport and the balance of enrollment across schools into account \citep{metro2024seoul,naeil2023middle}. Proximity is not guaranteed: a child can be sent to a distant school in the zone when nearby ones are full \citep{seoulshinmun2022middle}.

Assignment at both levels follows where the family lives, not whether it owns. District offices treat as the residence the address where the whole family actually lives, as shown in the resident registration (\emph{jumin deungnok}), and they investigate false registrations \citep{naeil2023middle,seongdong2026guide}. The rules draw no distinction between owners and tenants; one office even lists protecting a jeonse deposit among the documented reasons why not all family members need to move \citep{seongdong2026guide}. An owner-occupier and a jeonse tenant of the same flat therefore face the same assignment, and two flats in the same school zone give their children the same lottery over the same set of middle schools. Comparing flats within a zone and month (Section~\ref{sec:strategy}) holds those middle-school options fixed, but not the elementary catchment.

High schools were equalized first. Under the high-school equalization policy (\emph{pyeong\-jun\-hwa}), from March 1974 Seoul and Busan stopped letting high schools select their own students; instead, students sat a common qualifying exam and were assigned to schools within school districts by lottery \citep{seoulshinmun2003reviewing,segye2025seoul}. Seoul's 2010 high-school choice reform put stages of choice before the lottery. Students first list two schools anywhere in the city, and 20\% of each school's places are filled from these applications; they then list two schools in their own school group, which fills a further 40\%. The remaining students are assigned by lottery within a wider integrated group \citep{hankyung2008apply,seoulshinmun2010seoul,newsis2026seoul}. The staged procedure still governs assignment to Seoul's 212 general high schools for 2026 \citep{newsis2026seoul}. \citet{chung2015school} studies how the 2010 reform affected house prices (Section~\ref{sec:literature}).

A separate tier of schools stands outside the lottery. Autonomous private high schools (\emph{jasago}), foreign-language high schools (\emph{oego}) and international high schools (\emph{gukjego}) run their own admissions, and their admits are excluded from the general assignment \citep{newsis2024retention,newsis2026seoul}. In 2019 there were 79 such schools nationwide: 42 autonomous private, 30 foreign-language and 7 international \citep{kyunghyang2025abolishing}.

\subsection{Jeonse}\label{sec:background-jeonse}

Under a \emph{jeonse} lease (lump-sum deposit lease), the tenant pays the landlord a large deposit, typically 40--80\% of the property's value, pays no periodic rent, and gets the nominal deposit back when the lease ends. The landlord need not hold the deposit in escrow, so the tenant bears the risk of default \citep{ambrose2003modeling}. Older English-language studies spell the term \emph{chonsei} \citep{ambrose2003modeling,kim2013financial}. A lease agreed for less than two years is deemed to last two years \citep[Art.~4]{korea2023housing}. The jeonse-to-price ratio (\emph{jeonsega-yul}) expresses the deposit as a share of the sale price. For Seoul apartments, the KB Kookmin Bank series stood at 75.1\% in June 2016 \citep{newsis2017february}, 50.87\% in May 2023 and 50.92\% in January 2026 \citep{heraldeconomy2026seoul}.

Monthly-rent leases (\emph{wolse}) have gained ground. In January--August 2025, 62.2\% of housing leases nationwide and 46.8\% of apartment leases were monthly-rent leases, up from 57.4\% and 43.7\% a year earlier \citep{newsis2025monthly}. Deposits and rents are linked by the conversion rate (\emph{jeonwolse jeonhwanyul}), the annual rent charged per won of deposit given up. When a landlord converts part of a deposit into monthly rent, the law caps this rate at the Bank of Korea base rate plus a margin set by decree, or at a decree ceiling if that is lower \citep[Art.~7-2]{korea2023housing}. The margin fell from 3.5 to 2 percentage points on 29 September 2020 \citep{moleg2021timing}. Market conversion rates move less than the base rate. In the third quarter of 2026 the rate on Seoul apartment leases was 4.5\%, against a base rate of 3\%, and since 2022 it has stayed between 3.9\% and 4.9\% while the base rate ranged from 1.25\% to 3.5\% \citep{seoul2026conversion}.

Amendments to the Housing Lease Protection Act took effect on promulgation on 31 July 2020 and also applied to leases already running \citep{korea2023housing}. They gave tenants a renewal right (\emph{gyeyak gaengsin cheonggu-gwon}): a tenant may demand one renewal, for a further two years, which the landlord may refuse only on listed grounds such as rent arrears \citep[Art.~6-3]{korea2023housing}. A rent or deposit increase may not exceed one-twentieth (5\%), and local ordinances may set a lower cap \citep[Art.~7]{korea2023housing}. The cap governs increases within an existing tenancy, not the terms a landlord agrees with a new tenant. Since mid-2020 the same flat can therefore carry a market deposit for newcomers and a capped deposit for sitting tenants, a gap that Section~\ref{sec:robust} examines.

Mandatory lease reporting (\emph{jeonwolse singoje}) began on 1 June 2021 under the Real Estate Transaction Reporting Act. In the capital region, the metropolitan cities and other cities, leases with a deposit above KRW~60 million or a monthly rent above KRW~300,000 must be reported, including renewals, which also report the previous rent \citep{newsis2021molit}. These reports are why the lease records identify new and renewal contracts, the previous deposit and use of the renewal right only from mid-2021 (Section~\ref{sec:data}).

A wave of jeonse fraud, in which tenants lost their deposits, led the National Assembly to pass a special act for victims on 25 May 2023 \citep{newsis2023national}. By September 2026, 41,559 victims had been recognized under it. Apartments accounted for 13.3\% of the affected homes, against 28.3\% for multi-unit villas and 21.0\% for officetels \citep{heraldeconomy2026jeonse}, so the fraud wave was concentrated outside the apartment market studied here.

\subsection{Seoul housing regulation, 2017--2026}\label{sec:background-housing}

Seoul's housing market was tightly regulated for most of the sample period (Table~\ref{tab:policy}). The August 2017 measures designated all 25 districts as overheated speculation districts (\emph{tugi gwayeol jigu}) and 11 of them, including Gangnam, Yangcheon and Nowon, as speculation zones (\emph{tugi jiyeok}); in both, loan-to-value and debt-to-income limits fell to 40\% \citep{kyunghyang2017seoul,kyunghyang2017real}. From 17 December 2019, banks could not lend for the purchase of homes priced above KRW~1.5 billion in these areas \citep{dailian2019mortgage}. The ban was lifted on 1 December 2022 \citep{newsis2022fsc}. Land-transaction permit zones (\emph{toji georae heoga guyeok}) go further: a purchase needs prior permission, and a buyer of housing must live in it for two years \citep{fnnews2020land}. Daechi-dong entered a permit zone in June 2020, and the 14 complexes of the Mok-dong new town followed in April 2021. Coverage widened to all apartments in four districts in March 2025 and to all apartments in Seoul in October 2025 (Table~\ref{tab:policy}). Both the loan ban and the permit zones bore on expensive flats in the main hagwon districts, so Section~\ref{sec:robust} re-estimates the premium by regulatory period and without the permit-zone neighborhoods.

\begin{table}[tbp]
\centering
\caption{Seoul housing regulations, 2017--2026}\label{tab:policy}
\footnotesize
\begin{tabular}{@{}p{0.12\textwidth}p{0.57\textwidth}p{0.25\textwidth}@{}}
\toprule
Effective date & Rule & Source \\
\midrule
3 Aug 2017 & August 2 measures: all 25 districts become overheated speculation districts and 11 become speculation zones; LTV and DTI capped at 40\% & \citet{kyunghyang2017seoul} \\
17 Dec 2019 & December 16 measures: no purchase mortgages for homes priced above KRW~1.5 billion in regulated areas & \citet{dailian2019mortgage} \\
23 Jun 2020 & Permit zone: Samseong-dong, Daechi-dong and Cheongdam-dong (Gangnam-gu) and Jamsil-dong (Songpa-gu); two-year owner occupancy & \citet{fnnews2020land} \\
27 Apr 2021 & Permit zones: Apgujeong (24 complexes), Yeouido (16), Mok-dong new town (14) and the Seongsu redevelopment area & \citet{etoday2021land} \\
1 Dec 2022 & KRW~1.5 billion mortgage ban lifted; LTV 50\% for non-owners and one-home owners (announced 10 Nov 2022) & \citet{newsis2022fsc} \\
5 Jan 2023 & Regulated-area designations lifted in all districts except Gangnam, Seocho, Songpa and Yongsan & \citet{bizwatch2023regulated} \\
13 Feb 2025 & Permit zone lifted for 291 of 305 apartment complexes in Jamsil, Samseong, Daechi and Cheongdam; 14 reconstruction complexes, including Eunma, kept & \citet{newsis2025land} \\
24 Mar 2025 & Permit zone: all apartments in Gangnam, Seocho, Songpa and Yongsan (about 2,200 complexes), initially to 30 Sep 2025 & \citet{newsis2025seoul} \\
16 Oct 2025 & October 15 measures: all 25 districts again overheated speculation districts and adjustment areas & \citet{heraldeconomy2025owner} \\
20 Oct 2025 & Permit zone: apartments in all 25 districts, until 31 Dec 2026 & \citet{heraldeconomy2025owner} \\
\bottomrule
\end{tabular}

\smallskip
\begin{minipage}{0.96\textwidth}\footnotesize
\emph{Notes:} LTV = loan-to-value ratio; DTI = debt-to-income ratio. Permit zones (land-transaction permit zones) require official permission before a purchase. Dates are the dates on which each rule took effect, from the sources listed.
\end{minipage}
\end{table}

\subsection{Education-policy shocks}\label{sec:background-shocks}

Section~\ref{sec:robust} tests whether the hagwon premium responded to three national announcements (Table~\ref{tab:events}). Each changed, on a known date, the expected payoff to exam preparation for families across Seoul.

\paragraph{Elite high schools (November 2019).} On 7 November 2019 the Ministry of Education announced that autonomous private, foreign-language and international high schools would become general high schools by 2025 \citep{kyunghyang2025abolishing,newsis2019yoo}. The decree amendment was finalized on 28 February 2020, with conversion due at the end of February 2025 \citep{newsis2024retention}. On 28 November 2019 the government also ordered the regular-admission expansion at 16 Seoul universities described above \citep{newsis2019expansion}. In principle, abolishing the schools that do not admit by residence sends every student into residence-based assignment, which raises the value of living near strong general high schools and the hagwons that prepare students to compete in them. Net transfers of elementary students into schools in Gangnam and Seocho, Gangnam's ``Eighth School District,'' rose from 1,064 in 2018 to 1,849 in 2020, although the ministry attributed the rise to transport, amenities, housing supply and hagwon districts \citep{newsis2021transfers}. The plan was reversed in steps. On 29 July 2022 the new government said it would keep autonomous private high schools \citep{newsis2022plan}. On 21 June 2023 it confirmed that all three types would stay \citep{newsis2023high}, and on 16 January 2024 the cabinet approved the decree amendment \citep{newsis2024retention}.

\paragraph{``Killer questions'' (June 2023).} On 15 June 2023 the education minister relayed President Yoon Suk Yeol's instruction that the CSAT exclude material not covered in school classes \citep{ytn2023president}. On 22 June the ministry opened a center for reports of a ``private-education cartel'' and other misconduct, and the vice minister accused large CSAT-specialist hagwons of exploiting the hardest questions \citep{newsis2023education}. On 26 June it published examples of these ``killer questions'' from the past three CSATs and the June 2023 mock exam, promised to remove such items, and announced action against large exam hagwons together with the Fair Trade Commission, the police and the Seoul and Gyeonggi education offices \citep{newsis2023killer}. In principle, an exam confined to the school curriculum lowers the payoff to intensive exam preparation, and the crackdown raised the cost of supplying it; both should, if anything, reduce the value of living near exam-prep hagwons. The decision to keep elite high schools came in the same week (21 June 2023), so this window bundles two announcements.

\paragraph{Medical-school places (February 2024).} On 6 February 2024 the health minister announced that medical-school admissions, frozen at 3,058 since 2006, would rise by 2,000 to 5,058 from the 2025 intake \citep{newsis2024medical}. The final 2025 intake was 4,567, or 1,509 above the old quota \citep{newsis2024admission}. On 17 April 2025 the education minister returned the 2026 intake to 3,058 \citep{metro2025medical}. On 10 February 2026 the government set the 2027 intake at 3,548, with increases averaging 668 a year through 2031. All the added places go to 32 medical schools outside Seoul and are filled through a regional-doctor track whose graduates must work in the region for ten years \citep{joongang2026govt,koreaherald2026education}. Within weeks of the announcement, hagwons launched medical-school courses, including night classes for working adults \citep{koreatimes2024private}. In principle, more places raise the payoff to top-end exam preparation and so the value of living near exam-prep hagwons. The later adjustments reversed the 2026 increase and steered new places toward students from outside Seoul, which should, if anything, lower that value for Seoul families.
\section{Related literature}\label{sec:literature}

The paper draws on five literatures: school capitalization, shadow education, the price--rent relationship, jeonse, and education costs and fertility. Each strand closes with what this paper adds.

\subsection{School quality, school assignment and house prices}\label{sec:literature-schools}

Where public school places are allocated by residence, families buy access to schools through the housing market. Comparing houses on opposite sides of attendance-zone boundaries, \citet{black1999better} finds that parents pay 2.5\% more for a 5\% increase in test scores. \citet{bayer2007unified} embed the boundary design in a sorting model and find a willingness to pay of less than 1\% for the same improvement. \citet{gibbons2013valuing} find that a one-standard-deviation increase in either school value-added or pupils' prior achievement raises prices by about 3\%, and \citet{fack2010better} find 1.4--2.4\% per standard deviation in Paris, with smaller premia where private schools offer an alternative. Reviews find capitalization across many countries and designs, mostly below 4\% per standard deviation of test scores and smaller in fixed-effects designs \citep{black2011housing,nguyenhoang2011capitalization}; a meta-analysis finds that boundary and neighborhood fixed effects absorb much of the influence of other amenities \citep{turnbull2021meta}. \citet{zahirovicherbert2008school} relate school quality to prices and selling time, \citet{sah2016estimating} show that school-proximity estimates depend on spatial controls, and \citet{huang2021proximity} find that distance to the school is still priced within Auckland's enrollment zones. \citet{kuminoff2014capitalization} add a caveat that applies to any capitalization estimate, including those here: in their boundary designs, capitalization understates parents' willingness to pay for school improvements by as much as 75\%.

Korean evidence centers on the rules that replaced selective entrance exams. \citet{lee2015school} studies Seoul's 1974 switch from exam-based high-school admission to district-based random assignment and finds that residential land prices rose by about 13 percentage points more in the better school district (about 26 points across district boundaries). \citet{hahn2012influence} use two cities that moved from self-selection to geographic assignment and find that good high schools raised nearby house prices and rents. For Seoul's later high-school choice reform, \citet{chung2015school} uses boundary and panel fixed effects with sale, rental and asking prices and finds that it lowered prices in high-performing districts by roughly 10--27\% relative to low-performing ones, and \citet{park2021school} find declines of 4.7\% within 0.5 km of affected schools after the announcement and a further 5.3\% after admission. Using boundary fixed effects and about half a million transactions, \citet{kim2026parental} show that Seoul flats near middle schools with stronger high-school admission records command higher prices.

These papers value \emph{public} schooling, which assignment rules ration by residence. This paper holds school assignment fixed, comparing flats within the same middle-school assignment zone and month, and asks what families pay for proximity to \emph{private} supplementary schooling that no assignment rule rations. Since Seoul replaced exam-based high-school admission with district-based assignment in 1974 \citep{lee2015school}, much of the competition for university entrance has moved into private tutoring \citep{kim2010private}, which is where the premium estimated here lies. \citet{kim2026parental} also mark a limit of this design: middle-school assignment within a zone depends on residential proximity, so flats near hagwon clusters may also be closer to sought-after middle schools, which the zone-by-month comparison does not absorb. Section~\ref{sec:robust-access} therefore controls for distance to the nearest middle school; school outcomes are not observed.

\subsection{Shadow education and housing}\label{sec:literature-shadow}

\citet{bray1999shadow} examines private supplementary tutoring, the ``shadow education system,'' and its implications for education planners; later surveys weigh its effects on achievement against its costs to households and to equity \citep{dang2008growing,park2016learning}. Korea is the leading case. \citet{kim2010private} report that Korean households spent 2.57\% of GDP on private tutoring in 2006 and argue that this demand is a market response to an equalized, heavily regulated secondary system that feeds a steep university hierarchy. \citet{lee2011impact} show that students' desire for prestigious university entry predicts parental spending. Restricting supply has had little effect: \citet{choi2016regulating} find that Korea's hagwon curfews did not reduce the hours or money spent on tutoring, least of all for high-school students.

When tutoring is restricted, competition can move into housing. Using a boundary discontinuity design in one Chinese city, \citet{jia2025unintended} find that the 2021 ``double reduction'' restrictions on private tutoring raised property values in high-quality school districts. In Chengdu, \citet{chen2023pay} find that a lottery for private-school places raised school-district premia by 2.8\% for key-school districts and 7.8\% for elite-school districts. In both cases spending shifted toward the channel of educational competition that remained open.

A literature, mostly in Korean, relates hagwon supply to apartment prices directly. One group works with district aggregates. \citet{jin2005education} find that education variables affect both sale prices and jeonse deposits across Seoul's submarkets in 2001--2004, and \citet{eom2006effects} estimate that a district whose educational environment is twice as good has apartment prices about 64\% higher. \citet{kim2007impacts} relate 2004--2006 price growth in Seoul's 25 districts and nearby new towns to growth in hagwons per secondary student, using spatial autoregressive models, and conclude that the price surge did not stem from education factors. \citet{lee2007study} find the private education environment significant in a cross-section of Seoul's districts, and \citet{sung2023influence} find private academies positively associated with district-level prices for every flat size. A second group estimates hedonic or spatial models for apartments within one district or a few neighboring districts. For jeonse deposits in Seoul's three Gangnam districts, \citet{youn2011impact} find the number of exam-preparation cram schools significant in a hedonic model but not in a hierarchical linear model. \citet{ha2013relationship} find educational environment as important as physical characteristics in Yangcheon-gu; \citet{lee2020effects} find academies within 2 km associated with higher 2017 prices in Daegu's Suseong-gu but not in Seoul's Gangnam-gu; \citet{jo2023study} rank the private education environment as the strongest price factor in Bongseon-dong, Gwangju; and \citet{lee2024analysis} find, with spatial error and spatial lag models for Suseong-gu in 2022, that academy density explains prices better than distance to the nearest academy.

Most of these studies, though not all, find hagwon access associated with higher apartment prices, and several also examine jeonse. None compares flats assigned to the same public schools across a whole city, none estimates separate premia for hagwons teaching high-school students and those teaching younger children, and none compares the sale and jeonse premia for the same type of flat in the same complex and month. The grade split also serves as a partial falsification check: hagwons for younger children follow many of the same families, yet carry no premium in sale prices once exam-preparation hagwons are held fixed. That points to the university-entrance competition described by \citet{kim2010private}, not to childcare. Unlike the Chinese studies, this paper has no policy that closes hagwons, so its level estimates are capitalization premia conditional on school assignment and local commercial density, not causal effects of hagwon supply.

\subsection{Amenities, tenure and the price--rent relationship}\label{sec:literature-pricerent}

\citet{himmelberg2005assessing} argue that the price-to-rent ratio is a poor guide to housing costs unless it is compared with the user cost of owning. In this framework, rent--price ratios differ across units with maintenance costs, vacancy rates, risk premia and expected growth \citep{bracke2015house}. \citet{campbell2009moves} decompose the aggregate rent--price ratio into expected rent growth, real interest rates and a housing premium, and \citet{hattapoglu2014dependency} show that appreciation expectations depend on past price growth and on locational and structural fundamentals. A location attribute that also raises expected appreciation should therefore lower the rent--price ratio, while one that only delivers current services should leave it unchanged. Section~\ref{sec:framework} builds on this logic.

Rented and owned units differ, which complicates comparisons. \citet{bracke2015house} finds lower rent--price ratios for bigger and more central units in central London, \citet{hill2016hedonic} show that adjusting for the quality of sold and rented houses lowers Sydney's observed price--rent ratio by about 18\%, and \citet{halket2020housing} find that structure and unobserved quality, but not location, govern which English dwellings are rented. In Hong Kong, \citet{huang2018accounts} find that estates with higher turnover have lower rental yields. In Seoul, \citet{lee2005dwelling} find that expected redevelopment raises the prices of apartments older than about 15--19 years, an appreciation component that sale prices carry and lease deposits need not. These results motivate the comparison of the two prices within the same complex, flat size and month.

Some studies ask whether school quality is capitalized in rents as well as prices. \citet{zheng2016much} exploit a Beijing rule under which only owners can enroll children in the local school: rent differences between paired flats in adjacent buildings control for unobserved neighborhood quality, and the price premium inside a key primary school's zone is about RMB 2,266 per square meter. \citet{song2022capitalization} find school quality capitalized in Beijing rents too, and \citet{gabe2022relationship} find it capitalized in US multi-family rents, with renters valuing primary schools most. \citet{hilber2017economic} reviews evidence that capitalization is stronger where supply is constrained. In Seoul, school assignment follows registered residence rather than ownership (Section~\ref{sec:background}), and renters and owners have the same access to hagwons, so a gap between the two prices cannot reflect access rights.

Consumption amenities are also capitalized, and they increasingly draw residents to cities \citep{glaeser2001consumer,couture2020urban}. The number and quality of nearby restaurants raise housing values \citep{kuang2017quality}, a new Walmart raises prices within half a mile by 2--3\% \citep{pope2015walmart}, and online grocery delivery raises prices by about 2.5\% \citep{chung2026grocery}. \citet{koster2012impact} find a willingness to pay of about 2.5\% for mixed land use, and only among apartment occupiers.

This paper adds a rent--price comparison for a private education amenity, holding flat type, complex and month fixed. Hagwon access is capitalized in the same proportion into sale prices and jeonse deposits, consistent with a current-flow amenity that carries no extra expected appreciation. The renter--buyer gap seen in cross-complex comparisons belongs instead to shops and services of every kind: renters pay more, and buyers less, to live near them. This appears to be the first evidence on this tenure difference in the pricing of convenience within the same complex and month.

\subsection{Jeonse}\label{sec:literature-jeonse}

\citet{ambrose2003modeling} model the jeonse contract, in which the tenant pays a large up-front deposit, no periodic rent is due, and the landlord returns the nominal deposit at maturity. Because landlords need not hold the deposit in escrow, the tenant bears the default risk. Ambrose and Kim value the contract's embedded compound options in a contingent-claims model, confirm its predictions on monthly data for 1986--2000, and interpret jeonse as an indigenous market response to Korean conditions. \citet{kim2013financial} explains jeonse as a response to financial repression that lets landlords finance housing investment without heavy mortgage borrowing, and predicts its decline once the arbitrage gains disappear. \citet{cho2005interest} shows that the ratio of sale prices to jeonse deposits depends on the ratio of inflation to the real interest rate. \citet{hwang2006dividend} find support for the dividend pricing model in Korean condominium prices and rents before and after the 1997--98 crisis, \citet{oh2019role} argue that jeonse supports prices in downturns through sellers' loss aversion, and \citet{lee2009analysis} find that the jeonse-to-price ratio is lower in Seoul's high-price areas.

The 2020 lease laws introduced a two-year renewal right, a 5\% cap on increases at renewal, and lease reporting. \citet{koo2025putting} find that they raised average rents in Greater Seoul by 17.7\% over two years and shifted contracts from pure jeonse toward mixed deposit-and-rent leases; \citet{yun2025circumventing} find that landlords raised maintenance fees; and \citet{hur2025right} find that rent increases were 12.29 percentage points lower when tenants exercised the renewal right, most of all for jeonse. The spread of deposit fraud has prompted work on jeonse risk. \citet{kim2022relationship} links high jeonse ratios to unreturned deposits in Seoul in 2020--2022, and \citet{kim2023physical} find high-risk jeonse (ratios of 80\% or more) more likely in row houses and multi-household buildings and less likely where schools are more accessible.

This literature treats the jeonse ratio as a market-level object or a risk indicator. Here the jeonse deposit serves as a second price of the same flat, and the within-complex jeonse ratio as a test of whether a local amenity carries expected appreciation beyond its current services. The new-contract and renewal flags recorded since the 2020 reform also make it possible to test whether newcomers pay an ``entry fee'' near hagwons; there is no lasting one.

\subsection{Education costs and fertility in East Asia}\label{sec:literature-fertility}

\citet{anderson2013education} argue that ``education fever'' has made child-rearing so expensive in Korea that couples stop at one or two children. In the quantitative model of \citet{kim2024status}, Korean fertility would be 28\% higher without status externalities in education. \citet{kim2026impact} estimates from a provincial panel that a 1\% increase in shadow-education spending lowers the total fertility rate by 0.18--0.26\%. In China, the 2021 tutoring ban raised expected total fertility by 7--8\% \citep{meng2026education}, and rising house prices delay marriage and first births \citep{qiu2026housing}. These studies, surveyed with the wider economics of fertility by \citet{doepke2023economics}, measure the cost of education through tutoring spending. The estimates here add a cost they omit, the housing premium families pay, as buyers or as renters, to live near exam-preparation hagwons. The paper does not estimate effects on fertility.
\section{Data}\label{sec:data}

The data combine four administrative sources: transaction records for every Seoul apartment sale and lease, two sources on hagwon locations, and the boundaries of the middle-school assignment zones. Table~\ref{tab:summary} summarizes the analysis sample.

\subsection{Housing transactions}\label{sec:data-transactions}

Transactions come from the Real Transaction Price Disclosure System of the Ministry of Land, Infrastructure and Transport (MOLIT): all Seoul apartment sales and leases with contract dates from \DatDateFirst{} to \DatDateLast{} \citep{molit2026real}, \DatSalesRaw{} sales in \DatSaleFiles{} annual files and \DatLeasesRaw{} leases in \DatLeaseFiles{} half-year files. Each record gives the district, the legal dong (\textit{beopjeong-dong}, the statutory neighborhood unit), the lot number, the road address, the complex name, the exclusive floor area, the floor, the year of construction and the contract date. Sales report the price; leases report the deposit, any monthly rent and the lease term. Unit numbers are withheld. Records are organized by contract date, so late reports and cancellations can still revise the last few months.

\paragraph{Sales.} The \DatSalesCancelled{} sales that carry a cancellation date are dropped, leaving \DatSalesClean{}. A further \DatSalesDirect{} sales (\DatSalesDirectPct\%) were concluded directly between the parties rather than through a licensed broker. They are kept in the cleaned file but excluded from the analysis, because unbrokered deals include transfers between relatives and other transactions that may not take place at market prices.

\paragraph{Leases.} All \DatLeasesRaw{} lease records are usable: \DatJeonseClean{} are jeonse contracts, in which the tenant pays only a lump-sum deposit, and \DatWolseClean{} are monthly-rent contracts (\textit{wolse}), which combine a smaller deposit with a monthly payment. The analysis uses jeonse only. The jeonse deposit is the whole price of the lease, whereas a monthly-rent contract can be compared with it only through a conversion rate between deposits and rents, which varies over time and across districts.

\paragraph{The June 2021 filing change.} The lease records change source in June 2021, when reporting of lease contracts became mandatory (Section~\ref{sec:background}). Earlier records come from fixed-date confirmations (\textit{hwakjeong ilja}), which tenants obtain to protect their deposits. Such filings probably under-record monthly-rent contracts, whose tenants have smaller deposits to protect. Recorded apartment leases rose from \DatLeasesTwenty{} in 2020 to \DatLeasesTwentyTwo{} in 2022. The jeonse share of recorded leases fell from \DatJeonseShareHOneTwenty\% in the first half of 2020 to \DatJeonseShareHOneTwentyThree\% in the first half of 2023 and \DatJeonseShareHOneTwentySix\% in the first half of 2026; part of the early level reflects coverage rather than behavior. Because every specification compares flats within the same zone and month, a citywide change in recording is absorbed by the fixed effects. Section~\ref{sec:robust} also splits the sample by period.

\paragraph{New and renewal contracts.} The new reports also record whether a lease is new or a renewal, and the previous deposit. Of the \DatFlagJeonse{} jeonse contracts signed from June 2021, \DatFlagNew{} are new, \DatFlagRenewal{} are renewals and \DatFlagBlank{} (\DatFlagBlankPct\%) carry no flag. Before June 2021 the flag is blank for \DatFlagBlankBeforePct\% of jeonse contracts; this information was never collected and cannot be recovered. Renewals that used the statutory renewal right raised the deposit by a median of exactly \DatRightMedianIncrease\%, the legal cap (Section~\ref{sec:background}). The new-contract comparisons in Section~\ref{sec:robust} rely on this flag.

\subsection{Hagwon locations}\label{sec:data-hagwons}

\paragraph{Business data.} The first hagwon measure comes from the commercial-district store data of the Small Enterprise and Market Service (SEMAS) \citep{semas2026commercial}. These quarterly snapshots list operating stores and small businesses nationwide, with an industry category, the lot and road addresses, and longitude and latitude. The business data comprise \DatSnapshots{} snapshots of education businesses in Seoul and Gyeonggi between December 2015 and June 2026, and the two complete snapshots of all Seoul businesses for December 2018 (\DatSbizBizEighteen{} businesses) and June 2026 (\DatSbizBizTwentySix{}). A cram hagwon is a business in the subcategory of entrance-exam and school-subject academies (\textit{ipsi$\cdot$gyogwa hagwon}). In December 2018 Seoul had \DatSbizCramEighteen{} such businesses among \DatSbizEduEighteen{} education businesses. Counts around each complex also include Gyeonggi businesses (\DatSbizCramPointsEighteen{} cram hagwons in Seoul and Gyeonggi together), so complexes near the city boundary are not undercounted.

This measure is held fixed at December 2018, for three reasons. First, the register's coverage widened over time. Its count of Seoul cram hagwons rose from \DatSbizCramFifteen{} in December 2015 to \DatSbizCramTwentyFive{} in December 2025, and the 2015 count is less than half the number of academic hagwons in the official registry today (\DatNeisAcad{}). Changes in a complex's count over time therefore mix hagwon entry with changes in coverage. Second, December 2018 is one of the two dates for which the complete snapshot of all businesses was obtained, which the density controls in Section~\ref{sec:data-density} require. It is the earlier of the two and predates the education-policy shocks of 2019--2024 studied in Section~\ref{sec:robust}. Third, the spatial ranking of complexes is very stable: the correlation of log counts within 500~m across complexes is \DatCorrCramEighteenFifteen{} between December 2018 and December 2015, and \DatCorrCramEighteenTwentyFive{} between December 2018 and December 2025. A fixed measure thus captures persistent differences in access. The estimates use these cross-sectional differences within school zones, not the opening of new hagwons.

The later releases have known defects. The September 2023, December 2023 and March 2024 releases contain only \DatSbizIncompleteEduPct\% of the Seoul education businesses in the December 2022 release (\DatSbizIncompleteCramPct\% of cram hagwons); the missing businesses are absent, not reclassified. The June 2023 release could not be obtained. The June 2026 release lists \DatSbizCramDropPct\% fewer cram hagwons than December 2025, while education businesses as a whole fell by only \DatSbizEduDropPct\%, which suggests a partial reclassification. None of these releases enters the December 2018 measure. Two further caveats apply to the business data in general: they record businesses, not licensed hagwons, and business names rarely state the grade level taught.

\paragraph{Hagwon registry.} The second source is the official registry of hagwons and tutoring rooms kept by the offices of education and published on the NEIS Education Information Open Portal \citep{neis2026hagwon}. The Seoul snapshot of 12 October 2026 has \DatNeisSeoulAll{} records, of which \DatNeisAcad{} are in the academic field (exam preparation and supplementary study): \DatNeisAcadHagwon{} hagwons and \DatNeisAcadRoom{} tutoring rooms (\textit{gyoseupso}), which are smaller teaching establishments. Records include the name, opening date, capacity, course list, posted fees and road address, but no coordinates. Each academic hagwon is placed at the median position of the business-data records at the same road address, which succeeds for \DatNeisPlacedPct\% of them. Text cues in the name, course list and fees then sort hagwons by the grade level they teach: \DatGradeHsPct\% carry a high-school or exam cue (Appendix~\ref{app:grades}).

The registry has one important limitation: survivorship. It lists only hagwons open on the snapshot date and records no closures. Of the academic hagwons in the 2026 snapshot, \DatNeisOpenedAfterEighteenPct\% opened after December 2018 and only \DatNeisOpenedBeforeFifteenPct\% opened before 2015. Applied to transactions from 2015 onward, the registry measures early exposure with survivors and later entrants. If hagwons entered where demand was rising, part of the measured premium could reflect that later demand. Two facts limit the concern. The spatial pattern is stable: registry counts within 500~m correlate at \DatCorrNeisSbizEighteen{} with the December 2018 business counts and at \DatCorrNeisSbiz{} with the December 2025 business counts. And the period splits in Section~\ref{sec:robust} show the high-school premium already in 2015--2019.

\subsection{Locating apartment complexes}\label{sec:data-geocoding}

The MOLIT records identify a complex by its address and name but give no coordinates. The official address-coordinate files and geocoding services issue data only after Korean identity verification by mobile phone or digital certificate, so a geocoder was built from public business addresses instead. Shops, clinics and tutoring rooms inside apartment complexes are registered at the complex's lot number and road address, with coordinates. The geocoder pools \DatGeoPointRows{} business records from \DatGeoPointFiles{} business-data snapshots and places each complex at the median position of the businesses registered at its address. Four match tiers are tried in order. In the lot, road and name tiers, the businesses' address matches the complex's lot number, road address or building name. In the block tier, the businesses share the complex's main lot number but not its sub-number, and the position is approximate.

The geocoder locates \DatGeoLocated{} of the \DatGeoComplexes{} complexes (\DatGeoLocatedPct\%), which account for \DatGeoLocatedTxPct\% of all transactions; the three exact tiers alone cover \DatGeoExactTxPct\%. Positions from the lot and road addresses coincide for \DatValLotRoadSamePct\% of the \DatValLotRoadN{} complexes that match both ways. When a complex is placed from the other sub-lots of its main lot, as the block tier would place it, the median error is \DatValBlockMedian{}~m and the 90th percentile \DatValBlockPNinety{}~m. Seoul's open-data portal publishes a public sample of \DatValSeoulSampleRows{} complexes from its apartment-complex dataset; the \DatValSeoulSampleN{} that can be matched lie \DatValSeoulSampleDistOne{}~m and \DatValSeoulSampleDistTwo{}~m from Seoul's coordinates. Table~\ref{tab:geocode} reports the tiers and checks, and Appendix~\ref{app:geocoding} gives the full method and its limitations. Section~\ref{sec:robust} reports estimates without block-tier locations.

\subsection{School assignment zones}\label{sec:data-zones}

Seoul assigns pupils to public middle schools within middle-school assignment zones (Section~\ref{sec:background}). The zone boundaries come from a shapefile published by the Korea Institute of Educational Facilities Safety, dated 20 September 2026 \citep{kiefs2026middle}. It contains \DatZonesNational{} zones nationwide, of which \DatZonesSeoul{} are in Seoul. They cover \DatZoneAreaTotal{}~km$^2$, and each zone measures between \DatZoneAreaMin{} and \DatZoneAreaMax{}~km$^2$ (median \DatZoneAreaMedian{}~km$^2$). Each located complex is assigned to the zone that contains its position (EPSG:5186), using GeoPandas and Shapely \citep{fleischmann2026geopandas,gillies2026shapely}. Every located complex falls inside a zone polygon, so the fallback rule, which assigns the nearest zone within 300~m, was not needed (\DatZoneNearestN{} complexes). The median school zone contains \DatZoneComplexesMedian{} located complexes (range \DatZoneComplexesMin{}--\DatZoneComplexesMax{}).

The boundaries are current. The Seoul zones were last revised between \DatZoneUpdFirst{} and \DatZoneUpdLast{}, and earlier boundaries are not available. Complexes near a boundary that moved during the sample may therefore be assigned to the wrong zone in the early years. The zones are also large, so comparisons within a zone still span several neighborhoods; elementary-school catchments would allow a sharper comparison. Within-zone comparison nonetheless retains most of the variation in hagwon access: \DatWithinZonePct\% of the cross-complex variance in log cram hagwons within 500~m lies within zones.

\subsection{Business density measures}\label{sec:data-density}

The hagwon measures are counts within 500~m of each complex (1~km as a check), using straight-line distances in EPSG:5186 and the k-d tree in SciPy \citep{virtanen2020scipy}. The business-data measure counts cram hagwons in December 2018. The registry measures count high-school/exam and other academic hagwons in the 2026 snapshot.

To separate hagwon access from general commercial convenience, the analysis counts businesses within 500~m in the complete December 2018 snapshot. Four groups are used:
(i) all businesses (median complex: \DatMedBiz{});
(ii) non-education businesses (\DatMedNonedu{}), a proxy for commercial density, busy streets and transit;
(iii) ``non-student'' businesses (\DatMedNonstudent{}): estate agents, clinics and other health services, and repair and personal services such as laundries and hair salons, the categories least likely to locate in order to serve students;
(iv) non-education businesses other than food, retail, and arts, sports and entertainment (\DatMedNonfood{}). Group (iv) drops personal computer rooms, karaoke rooms and study cafés.
A further measure counts other education businesses: all education businesses within 500~m except cram hagwons, such as art, music, sports and language academies. It uses the December 2018 education snapshot.

These controls are correlated with hagwon access but far from collinear. Across complexes, log cram hagwons within 500~m correlate at \DatCorrCramShops{} with log non-education businesses, \DatCorrCramNonstudent{} with non-student businesses (clinics often share buildings with hagwons), \DatCorrCramNonfood{} with group (iv), and \DatCorrCramOtherEdu{} with other education businesses. The full business snapshots cover Seoul only, so shop counts near the city boundary miss businesses on the Gyeonggi side, whereas the hagwon counts include them.

\paragraph{Schools, stations and complex size.} For the robustness checks in Section~\ref{sec:robust-access}, straight-line distances are measured from each complex to the nearest middle school and high school in the national school-location data (reference date October 2026) \citep{kiefs2026elementary}, with each high school typed as general, autonomous, special-purpose or vocational from the Ministry of Education's 2018 school list \citep{moe2019addresses}, and to the nearest urban-rail station of any operator \citep{knr2026urban}. Complex size (households) and the original builder come from the K-apt register of complexes that must publish their management fees \citep{molit2026kapt}, which is matched to \ChkKaptTxnShare\% of the analysis transactions; smaller complexes are not in the register. These sources describe the schools, lines and complexes of 2026, so schools and lines opened during the sample are treated as always present.

\subsection{Analysis sample and descriptive patterns}\label{sec:data-sample}

The analysis sample contains brokered sales and jeonse contracts in located complexes assigned to a school zone, with positive value, area of 10--300~m$^2$ and building age of 0--60 years. It has \DatSampleN{} transactions: \DatSampleSales{} sales and \DatSampleJeonse{} jeonse contracts. That is \DatSampleSharePct\% of the \DatSampleNAll{} brokered sales and jeonse contracts in Seoul. Nearly all the loss comes from complexes that could not be located (\DatSampleLostLocatePct\%); the value, area and age filters remove a further \DatSampleLostFilterPct\%. The sample spans \DatSampleComplexes{} complexes, \DatSampleDongs{} legal dongs, \DatSampleZones{} school zones and \DatSampleMonths{} contract months. Block-tier locations account for \DatSampleBlockTxPct\% of the sample.

\begin{table}[!htbp]
\centering
\begin{threeparttable}
\caption{Summary statistics for the analysis sample}
\label{tab:summary}
\small
\setlength{\tabcolsep}{4pt}
\begin{tabular}{@{}l rrr rrr@{}}
\toprule
 & \multicolumn{3}{c}{Sales} & \multicolumn{3}{c}{Jeonse} \\
\cmidrule(lr){2-4}\cmidrule(l){5-7}
 & Mean & SD & Median & Mean & SD & Median \\
\midrule
\multicolumn{7}{@{}l}{\textit{Panel A. Transaction}} \\
\quad Price or deposit per m$^2$ (KRW million) & 10.44 & 6.82 & 8.48 & 6.61 & 3.17 & 5.98 \\
\quad Exclusive floor area (m$^2$) & 77.4 & 29.7 & 83.0 & 75.5 & 27.5 & 77.1 \\
\quad Floor & 9.3 & 6.3 & 8 & 9.2 & 6.3 & 8 \\
\quad Building age (years) & 19.1 & 10.2 & 18 & 19.5 & 11.7 & 19 \\
\addlinespace
\multicolumn{7}{@{}l}{\textit{Panel B. Within 500 m of the complex}} \\
\quad Cram hagwons, Dec 2018 (business data) & 26.4 & 35.7 & 18 & 27.1 & 39.8 & 17 \\
\quad High-school/exam hagwons, 2026 registry & 15.3 & 27.4 & 9 & 16.9 & 33.1 & 9 \\
\quad Other academic hagwons, 2026 registry & 24.9 & 28.1 & 18 & 25.7 & 30.1 & 19 \\
\quad All businesses, Dec 2018 & 778.9 & 561.5 & 666 & 745.5 & 570.3 & 630 \\
\quad Non-student businesses, Dec 2018 & 159.2 & 94.3 & 148 & 152.4 & 97.2 & 138 \\
\addlinespace
\multicolumn{7}{@{}l}{\textit{Panel C. Sample size}} \\
\quad Transactions & \multicolumn{3}{c}{771,489} & \multicolumn{3}{c}{1,426,021} \\
\quad Complexes & \multicolumn{3}{c}{7,757} & \multicolumn{3}{c}{8,158} \\
\quad Middle-school assignment zones & \multicolumn{3}{c}{46} & \multicolumn{3}{c}{46} \\
\bottomrule
\end{tabular}
\begin{tablenotes}[flushleft]\footnotesize
\item \textit{Notes:} One observation is one transaction. Sales are brokered apartment sales; jeonse are lump-sum deposit leases (monthly-rent contracts excluded), both from MOLIT, 1 January 2015 to 12 October 2026. The sample keeps transactions in complexes located by the business-address geocoding and assigned to a middle-school assignment zone, with area 10--300 m$^2$ and building age 0--60 years, as in the main specifications. Prices and deposits are nominal. Panel B counts are per complex and are weighted here by transactions. Cram hagwons are academic cram schools in the Small Enterprise and Market Service business data (Seoul and Gyeonggi points); registry counts are academic hagwons and tutoring rooms in the NEIS registry snapshot of 12 October 2026, split by the grade cues of Appendix~\ref{app:grades}. Businesses are all Seoul businesses in the December 2018 snapshot; non-student businesses are estate agents, clinics, and repair and personal services.
\end{tablenotes}
\end{threeparttable}
\end{table}

Table~\ref{tab:summary} shows that the flats sold and the flats leased on jeonse are similar. Mean areas are \DatMeanSaleArea{} and \DatMeanJeonseArea{}~m$^2$, and buildings are about \DatMeanSaleAge{} years old on average in both markets. The median jeonse deposit is \DatMedianJeonsePriceM{} million KRW per m$^2$, \DatJeonseRatioMedianPct\% of the median sale price of \DatMedianSalePriceM{} million KRW per m$^2$.

Hagwon access is highly skewed. In December 2018 the median located complex had \DatMedCram{} cram hagwons within 500~m, the 90th percentile \DatPNinetyCram{} and the maximum \DatMaxCram{}, in Daechi-dong. In the registry, the median complex has \DatMedNeisHs{} high-school/exam hagwons within 500~m (90th percentile \DatPNinetyNeisHs{}) and \DatMedNeisOther{} other academic hagwons. Medians across transactions are higher than across complexes (Table~\ref{tab:summary}), so complexes with more hagwons nearby trade more often. The high-school and other registry counts correlate at \DatCorrHsOther{} across complexes, so the grade split in Section~\ref{sec:results} relies on limited independent variation (Appendix~\ref{app:grades}).

\begin{figure}[!htbp]
\centering
\includegraphics[width=\textwidth]{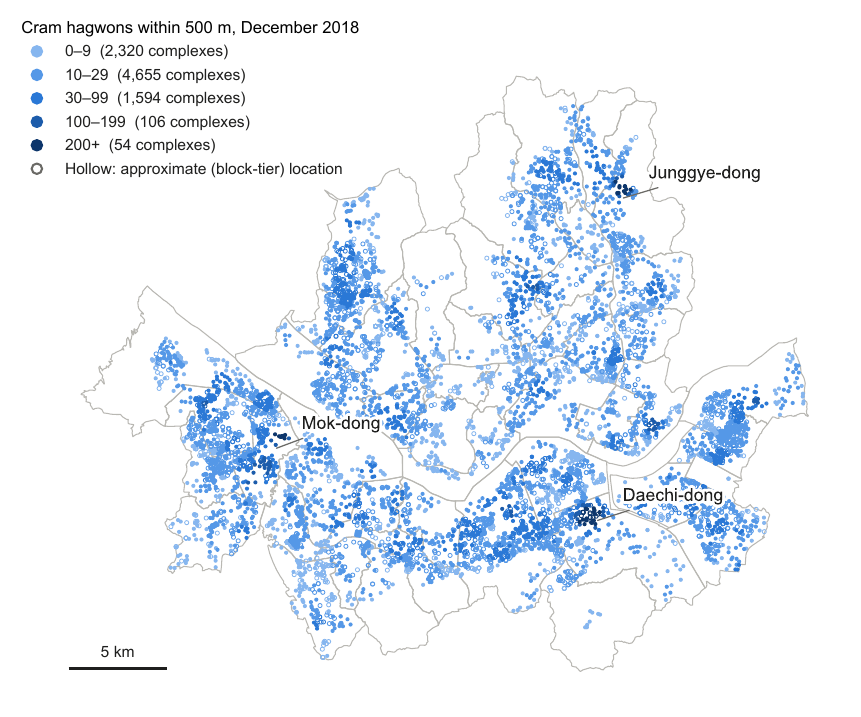}
\caption{Apartment complexes and cram-school access in Seoul. Each dot is one of the \DatGeoLocated{} located complexes, shaded by the number of academic cram hagwons within 500~m in December 2018 (business data). Hollow dots are approximate (block-tier) locations. Gray lines are the \DatZonesSeoul{} middle-school assignment zones.}
\label{fig:map}
\end{figure}

\begin{figure}[!htbp]
\centering
\includegraphics[width=\textwidth]{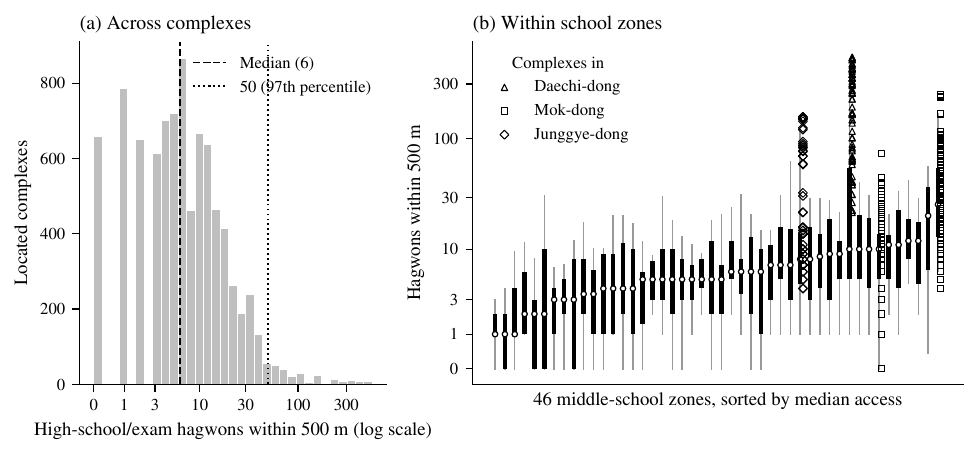}
\caption{High-school hagwon access across complexes and within school zones}
\label{fig:withinzone}
\begin{minipage}{0.95\textwidth}\footnotesize
\textit{Notes:} High-school/exam hagwons in the 2026 registry within 500~m of each of the \FigZoneComplexes{} located complexes, on a log scale. Panel (a): distribution across complexes; the dashed line marks the median and the dotted line 50 hagwons, the upper end of the headline comparison. Panel (b): one column per middle-school zone (\FigZones{} zones), sorted by median access. Thick bars show the interquartile range across the zone's complexes, thin lines the 5th to 95th percentiles and white dots the median. Open markers show the complexes of Daechi-dong, Mok-dong (with Sinjeong-dong) and Junggye-dong.
\end{minipage}
\end{figure}

Figure~\ref{fig:map} maps the located complexes. Hagwon access is concentrated in a few districts: Daechi-dong in Gangnam, Mok-dong in Yangcheon and Junggye-dong in Nowon. In all, \DatMapBinHigh{} complexes have 100 or more cram hagwons within 500~m. These clusters lie inside school zones, and complexes a short distance away in the same zone often have far fewer hagwons nearby. This within-zone contrast is the variation that the methodology in Section~\ref{sec:strategy} uses. Figure~\ref{fig:withinzone} shows it for the registry measure. Across complexes, high-school/exam hagwons within 500~m range from none to several hundred (panel a). Within most zones, complexes differ widely in access, and the clusters of Daechi-dong, Mok-dong and Junggye-dong sit at the top of zones that also contain complexes with few such hagwons nearby (panel b). In all, \ResWithinZoneShareHs\% of the variance of log high-school hagwons lies within zones.

\section{Methodology}\label{sec:strategy}

The empirical strategy measures how hagwon access is priced in two markets for the same housing stock: the market for ownership and the market for jeonse leases. Four designs answer different questions and rest on different assumptions (Table~\ref{tab:designs}). The first estimates the level of the premium among flats assigned to the same schools. The second compares the two prices of the same flat type, which asks whether the premium contains an asset-market component beyond the current service. The third asks whether newcomers pay more than sitting tenants, and the fourth whether the premium responded to education-policy news. Section~\ref{sec:framework} sets out what the jeonse-to-price ratio measures, and Section~\ref{sec:strategy-identification} discusses what each design can and cannot identify.

\begin{table}[tbp]
\centering
\caption{Four designs}\label{tab:designs}
\footnotesize
\begin{tabular}{@{}>{\raggedright\arraybackslash}p{0.17\textwidth}>{\raggedright\arraybackslash}p{0.23\textwidth}>{\raggedright\arraybackslash}p{0.23\textwidth}>{\raggedright\arraybackslash}p{0.27\textwidth}@{}}
\toprule
Design & Comparison & Question & Main assumption \\
\midrule
1. Hedonic within school zones (Section~\ref{sec:strategy-hedonic}) & Flats in the same middle-school assignment zone and month & Is hagwon access priced beyond school assignment? & Conditional on the controls, no unobserved trait that raises prices varies with hagwon access within a zone \\
\addlinespace
2. Same flat, two prices (Section~\ref{sec:strategy-withincell}) & Sale and jeonse contracts in the same complex, flat size and month & Is hagwon access priced in proportion into prices and deposits? & Within a zone, no determinant of the jeonse-to-price ratio other than the channels in Equation~\eqref{eq:ratio} varies with hagwon access \\
\addlinespace
3. New versus renewal contracts (Section~\ref{sec:strategy-entryfee}) & New and renewal jeonse in the same complex, flat size and month & Do newcomers near hagwons pay an extra premium? & Renewal deposits reflect the market of the previous contract \\
\addlinespace
4. Education-policy windows (Section~\ref{sec:strategy-events}) & The hagwon gradient across half-years, within complexes & Did the premium respond to news about exam stakes? & No other change in the gradient coincides with the windows \\
\bottomrule
\end{tabular}
\end{table}

\subsection{Conceptual framework: what the jeonse-to-price ratio measures}\label{sec:framework}

Within a complex, the same type of flat can be bought by an owner-occupier or by an investor who lets it on jeonse. Under jeonse the tenant pays no rent but lends the landlord a deposit $D$, returned at the end of the lease; the landlord's income is the return on the deposit \citep{ambrose2003modeling}. Let $c$ be the rate at which the market converts deposits into monthly rent (Section~\ref{sec:background-jeonse}), $g$ expected appreciation, and $\rho$ the required return plus holding costs (taxes, maintenance and depreciation) as shares of value. An investor who buys at price $P$ and lets the flat earns $cD$ from the deposit and $gP$ from appreciation, and requires $\rho P$. If investors are marginal, $cD = (\rho - g)P$.

Owner-occupiers and tenants value the flat's housing services. Write $R^{O}$ for the imputed rent of an owner-occupier and $R^{J} = cD$ for the implicit rent of a jeonse tenant. Under the user-cost condition \citep{poterba1984tax,himmelberg2005assessing}, $P = R^{O}/(\rho - g)$. Combining the two valuations gives the jeonse-to-price ratio
\begin{equation}
\ln\frac{D}{P} \;=\; \underbrace{\bigl(\ln R^{J} - \ln R^{O}\bigr)}_{\text{clientele}} \;+\; \ln(\rho - g) \;-\; \ln c .
\label{eq:ratio}
\end{equation}
If flats move freely between owner-occupation and letting, the investor condition holds, the marginal owner and tenant value the flat alike, and the clientele term is zero. The ratio is then pinned down by expected appreciation, holding costs and the conversion rate, as in rent--price decompositions \citep{campbell2009moves,hattapoglu2014dependency}. A difference in how renters and owners value an amenity shows up in which flats are rented, not in the ratio \citep{bracke2015house,halket2020housing}. The clientele term can differ from zero only where tenure is segmented, for example where buyers must occupy the flat themselves, as in Seoul's land-transaction permit zones (Section~\ref{sec:background-housing}).

Consider a local amenity $A$, such as hagwon access. Equation~\eqref{eq:ratio} separates three cases.
\begin{enumerate}
\item \textit{A current-flow amenity.} If $A$ raises the value of housing services but leaves $g$, $\rho$ and $c$ unchanged, it is capitalized in proportion into deposits and prices, and $D/P$ does not move.
\item \textit{An amenity valued partly for expected appreciation.} If buyers expect demand for $A$ to keep growing, so that $\partial g/\partial A > 0$, the price rises by more than the deposit and the ratio falls. Persistent growth differences of this kind across high-demand locations are documented by \citet{gyourko2013superstar}.
\item \textit{An amenity tied to a wedge that raises the ratio.} The ratio rises if, near $A$, renters value the flat more and tenure is segmented; if deposits are perceived as safer, so that $c$ is lower; or if holding costs or credit limits that tighten with value raise $\rho$ for the flats concerned.
\end{enumerate}
The level premia in each market mix these channels with every unobserved trait of the location. Their difference, measured for the same flat type in the same complex and month, removes what the two prices share and isolates the net of the three channels. This is the logic of the second design, which is close in spirit to the paired resale and rental comparison of \citet{zheng2016much}. A zero response fits the first case, but also offsetting forces in the second and third.

\subsection{Hedonic specification within school zones}\label{sec:strategy-hedonic}

For transaction $i$ in complex $c$, middle-school assignment zone $z(c)$ and month $t$, the model, estimated separately for sales and jeonse ($k \in \{S, J\}$), is
\begin{equation}
\ln V^{k}_{ict} \;=\; \beta^{k} h_{c} \;+\; \gamma^{k\prime} s_{c} \;+\; \psi^{k\prime} X_{i} \;+\; \mu^{k}_{z(c)t} \;+\; \alpha^{k}_{b(i)} \;+\; \varepsilon^{k}_{ict}.
\label{eq:hedonic}
\end{equation}
$V^{S}$ is the brokered sale price and $V^{J}$ the jeonse deposit, both per m$^2$ of exclusive area. Hagwon exposure is $h_{c} = \ln(1 + H_{c})$, where $H_{c}$ counts hagwons within 500~m of the complex. The vector $s_{c}$ holds log counts of other nearby businesses when they are controlled for, and $X_{i}$ contains log exclusive area and floor. The fixed effects $\mu^{k}_{zt}$ are school zone $\times$ month cells, and $\alpha^{k}_{b}$ are five-year building-age bins. In the tradition of hedonic pricing \citep{rosen1974hedonic} and of within-zone designs \citep{black1999better,huang2021proximity}, the zone $\times$ month effects compare each flat only with flats assigned to the same middle schools and traded in the same month. The test is therefore whether hagwon access is priced beyond school assignment, which is itself priced in Seoul \citep{chung2015school}. The fixed effects leave most of the variation intact: \DatWithinZonePct\% of the variance of log hagwon exposure across located complexes lies within zones (\ResWithinZoneShareHs\% for the registry high-school count).

To test whether the two premia differ, the two markets are stacked and every regressor and fixed effect is interacted with a jeonse indicator. The coefficient on $h_{c} \times J_{i}$ equals $\beta^{J} - \beta^{S}$, and its standard error accounts for the covariance between the two estimates. It is called the jeonse $-$ sale gap; a positive gap means a larger premium in deposits.

\paragraph{Exposure measures.} Two hagwon counts are used (Section~\ref{sec:data}). The \textit{SEMAS count} is the number of academic cram hagwons within 500~m in the SEMAS commercial-district data of December 2018 \citep{semas2026commercial}, four years into the sample; versions within 1~km and within the legal dong serve as checks. The \textit{registry counts} come from the NEIS hagwon registry of October 2026 \citep{neis2026hagwon}, which is split into high-school/exam hagwons, whose name, course list or fees mention high school, the College Scholastic Ability Test or exam preparation, and all other academic hagwons (Appendix~\ref{app:grades}). Specifications A--G use the SEMAS count, with C (500~m, zone $\times$ month effects) as the reference. H--K add controls for other businesses (Table~\ref{tab:app-shopdefs}). M0 uses the registry total, M splits it by grade, and N adds non-student shops to M.

\paragraph{Magnitudes.} Coefficients on log counts are elasticities with respect to one plus the count. For the high-school measure, two moves are reported: across the interquartile range of complexes, and from the median located complex, with \DatMedNeisHs{} high-school/exam hagwons within 500~m, to \ResHsTarget{}, about the \ResHsPctileTarget{}th percentile. The second is a move of $\ln(1+\ResHsTarget) - \ln(1+\DatMedNeisHs) = \ResScaleHs$ in exposure, and the implied change is $100\,[\exp(\ResScaleHs\,\hat\beta) - 1]$ percent. Measured from the median \textit{transaction}, which has \DatMedianSaleNeisHs{} such hagwons nearby, the same move is \ResScaleHsTxn{} log points and the magnitudes are smaller (for example, \ResMHsSalePctTxn\% rather than \ResMHsSalePct\% for the high-school sale premium). For other counts, the tables report the change implied by doubling.

\paragraph{Inference.} Standard errors are clustered by legal dong (\DatSampleDongs{} clusters), which allows errors to be correlated across complexes and months within a neighborhood \citep{cameron2015practitioner}. All models are estimated with pyfixest \citep{pyfixest2025pyfixest}, which absorbs the fixed effects by alternating projections \citep{gaure2013ols}.

\subsection{Same flat, two prices: the within-complex comparison}\label{sec:strategy-withincell}

The second design compares the deposit and the price of the same flat type. A cell $g$ is a complex $\times$ exclusive area (rounded to 1~m$^2$) $\times$ contract month. For every cell with at least one sale and one jeonse contract, the gap $\Delta_{g} = \bar{v}^{J}_{g} - \bar{v}^{S}_{g}$ is the mean log deposit per m$^2$ minus the mean log price per m$^2$, the cell's log jeonse-to-price ratio. The estimating equation is
\begin{equation}
\Delta_{g} \;=\; \theta\, h_{c(g)} \;+\; \gamma' s_{c(g)} \;+\; \psi' \Delta X_{g} \;+\; \mu_{z(g)t(g)} \;+\; \alpha_{b(g)} \;+\; u_{g},
\label{eq:withincell}
\end{equation}
where $\Delta X_{g}$ holds the within-cell differences in mean floor and mean log area, and the zone $\times$ month and age-bin effects let the average ratio differ across zones, periods and building ages. Each cell is weighted by $n^{J}_{g} n^{S}_{g} / (n^{J}_{g} + n^{S}_{g})$, the inverse of the sampling variance of a difference between two means.\footnote{Equation~\eqref{eq:withincell} is the cell-level form of a stacked transaction-level regression with cell fixed effects and zone $\times$ month $\times$ market effects, in which the coefficient of interest is exposure interacted with the jeonse indicator. Both use the same within-cell variation. The cell-level form is used because the demeaning of the stacked version, with about \ResCellObsCells{} cell effects, did not converge within 10,000 iterations.}

It is important to be precise about what the cells achieve. Exposure is measured per complex, so $\theta$ is still estimated from differences across complexes within a zone and month. What the cell comparison removes is the \textit{level} of each complex's prices: anything about a location, its schools, its buildings or its units that raises deposits and prices in the same proportion. In terms of Equation~\eqref{eq:ratio}, $\theta$ is the response of $\ln(D/P)$ to hagwon exposure, and it is biased only by determinants of the ratio itself that are correlated with exposure within a zone. That is a weaker requirement than in the level design, but not a trivial one (Section~\ref{sec:strategy-identification}).

\subsection{New versus renewal contracts}\label{sec:strategy-entryfee}

Since June 2021, reported leases state whether they are new contracts or renewals, and whether a renewing tenant used the statutory renewal right, which caps the increase at 5\% (Section~\ref{sec:background-jeonse}). The flag is blank for \DatFlagBlankPct\% of jeonse contracts in this period. If sitting tenants near hagwons hold below-market deposits, newcomers there pay an \textit{entry fee}. This is tested with the same cell design. For each complex $\times$ size $\times$ month cell with at least one new contract and one renewal, the markup $m_{g}$ is the mean log deposit per m$^2$ of new contracts minus that of renewals, and $m_{g}$ is regressed on exposure as in Equation~\eqref{eq:withincell}, with weights $n^{new}_{g} n^{ren}_{g}/(n^{new}_{g}+n^{ren}_{g})$.

A renewal deposit is anchored to the contract of two years earlier, so the markup mixes market rent growth since then with any newcomer premium. The two are separated by comparing new contracts with capped and with voluntary renewals, and by splitting the window at January 2023: before it, renewals were anchored to 2019--20 deposits in a rising market. A lasting entry fee would show up near hagwons in every comparison and period.

\subsection{Education-policy windows}\label{sec:strategy-events}

The fourth design asks whether the premium responded to news about the stakes of exam preparation. Three windows are studied: November 2019, June 2023 and February 2024 (Section~\ref{sec:background-shocks}). None is a single clean shock. The November 2019 window combines the plan to abolish elite high schools with a push toward admission by test scores. The June 2023 window combines the removal of ``killer questions'' from the College Scholastic Ability Test and a crackdown on hagwons with the decision to keep elite high schools. The February 2024 window is the medical-school expansion. Some components should raise demand for exam preparation and others lower it, so each window measures a net response.

Capped renewals cannot respond to news, so the main sample is new jeonse contracts for 60--85~m$^2$ flats from June 2021 to October 2026. The estimating equation is
\begin{equation}
\ln V_{ict} \;=\; \sum_{\tau \neq \text{2023H1}} \lambda_{\tau}\, e_{d(c)t}\, \mathbf{1}[t \in \tau] \;+\; \kappa\, e_{d(c)t} \;+\; \psi' X_{i} \;+\; \eta_{c} \;+\; \mu_{r(c)t} \;+\; \varepsilon_{ict},
\label{eq:events}
\end{equation}
where $\tau$ indexes half-years, $\eta_{c}$ are complex fixed effects and $\mu_{rt}$ are district $\times$ month effects. Each $\lambda_{\tau}$ is the change in the hagwon gradient relative to the first half of 2023, the last half-year before the June 2023 window. The tests cover jumps at June 2023 ($\lambda_{\text{2023H2}}$) and February 2024 ($\lambda_{\text{2024H1}} - \lambda_{\text{2023H2}}$), joint significance of the pre- and post-period coefficients, and a winter moving-season effect. Comparison samples are 60--85~m$^2$ sales, new contracts plus voluntary renewals, buildings under 25 years, and capped renewals as a placebo. The November 2019 window predates the new-contract sample, so it is assessed with an earlier specification on all contracts (Section~\ref{sec:robust-events}).

This design uses a coarser exposure measure than the rest of the paper: $e_{dt}$ is the log of one plus the SEMAS count of cram hagwons in the \textit{legal dong}, adjusted for the database's growing coverage and interpolated between snapshots, so it varies over time; fixed December 2018 and December 2022 counts serve as checks. Section~\ref{sec:robust-events} also re-estimates it with the 500~m registry measure, complex effects and zone $\times$ month effects. To express the coefficients in percent, the comparison is between Daechi-dong, with \ResDaechiCount{} academic cram hagwons in December 2018, and the median apartment dong, with \ResMedDongCount: a difference of \ResDaechiScale{} log points.

\subsection{Identification: what the designs can and cannot show}\label{sec:strategy-identification}

The level estimates from Equation~\eqref{eq:hedonic} are capitalization premia conditional on school assignment, flat attributes and local commercial density. They are not the causal effect of opening a hagwon, and under sorting they may partly price what comes with hagwons rather than hagwons themselves. The main threats are discussed in turn.

\paragraph{Sorting and unobserved neighborhood quality.} Hagwons locate where demand for them is high, near affluent, education-focused families, whose presence may itself be valued through peers, safety or amenities not observed in the data \citep{bayer2007unified,kuminoff2013new}. Zone $\times$ month effects remove what is common to a school zone, but zones are large (Section~\ref{sec:data-zones}), and the flat controls are limited to area, floor and building age. Section~\ref{sec:robust-access} adds distance to the nearest station, complex size and the builder, all of which may co-locate with hagwon clusters. Controls for shops, non-student services and other education businesses absorb part of general commercial density. The grade contrast is a partial falsification check: hagwons for younger children serve many of the same families, so a generic ``good neighborhood'' factor should load on them too. It is not a clean placebo, because the two counts are correlated at \DatCorrHsOther{} across complexes and high-school hagwons tend to be larger and more concentrated. Section~\ref{sec:robust-inference} reports how strong selection on unobservables would have to be to explain the level premia \citep{altonji2005selection,oster2019unobservable}.

\paragraph{Nearby schools.} A school zone is not a single catchment. Middle-school assignment within a zone depends partly on a flat's distance to the school \citep{kim2026parental}, and exam hagwons cluster around the strong general high schools whose students they serve. Proximity to such schools may be valued in its own right, for peers, the commute or the residence-based stage of high-school assignment (Section~\ref{sec:background-schools}). The high-school/exam count may therefore partly proxy for proximity to good schools. Section~\ref{sec:robust-access} controls for distance to the nearest general middle and high school and, as a proxy for school quality, to the nearest autonomous or special-purpose high school; school outcomes are not observed.

\paragraph{Timing and survivorship.} High prices attract hagwons as well as the reverse. The SEMAS count is measured in December 2018 and the registry in 2026, so neither is predetermined, and the registry lists only hagwons open in 2026. The levels are interpreted as equilibrium capitalization, and the period splits and policy windows (Section~\ref{sec:robust}) serve as evidence on dynamics. The registry's opening dates allow a time-consistent count of hagwons already open at each transaction date, and a count restricted to hagwons open before the sample began (Section~\ref{sec:robust-exposure}); both still omit hagwons that closed before 2026.

\paragraph{Threats specific to the ratio test.} Because $\theta$ is identified from differences across complexes, a determinant of the ratio that varies with exposure within a zone biases it. Three are relevant. First, lending limits that tighten with price (Section~\ref{sec:background-housing}) and holding costs that rise with value enter $\rho$, and prices are higher near hagwons. Second, unflagged renewals before June 2021 and capped renewals afterwards do not reflect current market deposits. Third, the test has power only if the ratio responds where expected appreciation is known to differ; buildings old enough for reconstruction, whose prices carry redevelopment value, provide such a positive control. Section~\ref{sec:robust-ratio} reports this control and re-estimates $\theta$ by price tier and on new contracts only.

\paragraph{Measurement error.} Complex locations come from business addresses, a minority only approximately (Appendix~\ref{app:geocoding}), and a straight-line 500~m count is a coarse measure of access. Classical error attenuates the level premia, and specification E drops the approximate locations. The grade classification relies on text cues (Appendix~\ref{app:grades}).

\paragraph{Inference and multiple testing.} Clustering by legal dong does not allow for correlation across neighboring dongs, and the densest complexes sit in three clusters. Section~\ref{sec:robust-inference} therefore reports standard errors clustered by school zone, with a wild cluster bootstrap for the small number of zones \citep{cameron2015practitioner,webb2023reworking}, grid clusters in the spirit of spatially robust errors \citep{conley1999gmm}, and estimates that drop Daechi-dong, Mok-dong and Junggye-dong in turn. The conclusions rest on \ChkFamTests{} tests of related hypotheses, listed in Table~\ref{tab:app-family}; the table flags which pass the Bonferroni threshold, $p < \ChkFamThreshold$, and reports Benjamini--Hochberg adjusted $p$-values \citep{benjamini1995controlling}.
\section{Results}\label{sec:results}

\subsection{The hagwon premium within school zones}\label{sec:results-baseline}

Table~\ref{tab:baseline} reports the hedonic estimates of Equation~\eqref{eq:hedonic} with the SEMAS count. Column (3), specification C, is the reference: flats are compared only with flats in the same middle-school assignment zone and month. The coefficient on log hagwons within 500 m is \ResCHagSale{} for sale prices (SE \ResCHagSaleSe, \ResCHagSaleP) and \ResCHagJeonse{} for jeonse deposits (SE \ResCHagJeonseSe, \ResCHagJeonseP). Doubling the number of nearby hagwons is associated with prices \ResCHagSalePct\% higher and deposits \ResCHagJeonsePct\% higher, among flats assigned to the same schools. Hagwon access is thus priced beyond the school-assignment premium, in both the ownership and the lease market.

\begin{table}[!htbp]
\centering
\begin{threeparttable}
\caption{The hagwon premium within school zones}
\label{tab:baseline}
\small
\setlength{\tabcolsep}{4pt}
\begin{tabular}{lcccccc}
\toprule
 & (1) & (2) & (3) & (4) & (5) & (6) \\
Specification & A & B & C & D & E & F \\
\midrule
\multicolumn{7}{l}{\textit{Panel A. Log sale price per m$^2$}} \\
log(1 + hagwons) & 0.025$\,^{*}$ & 0.036$\,^{***}$ & 0.026$\,^{**}$ & 0.038$\,^{***}$ & 0.024$\,^{**}$ & 0.032$\,^{***}$ \\
 & (0.013) & (0.013) & (0.012) & (0.013) & (0.012) & (0.011) \\
Observations & 771,488 & 771,488 & 771,466 & 771,466 & 723,376 & 538,514 \\
\addlinespace
\multicolumn{7}{l}{\textit{Panel B. Log jeonse deposit per m$^2$}} \\
log(1 + hagwons) & 0.047$\,^{***}$ & 0.066$\,^{***}$ & 0.051$\,^{***}$ & 0.100$\,^{***}$ & 0.050$\,^{***}$ & 0.064$\,^{***}$ \\
 & (0.014) & (0.016) & (0.014) & (0.015) & (0.014) & (0.015) \\
Observations & 1,426,021 & 1,426,021 & 1,426,021 & 1,426,021 & 1,356,646 & 923,394 \\
\addlinespace
\multicolumn{7}{l}{\textit{Panel C. Jeonse $-$ sale (pooled, fully interacted)}} \\
log(1 + hagwons) $\times$ jeonse & 0.023$\,^{**}$ & 0.030$\,^{***}$ & 0.025$\,^{***}$ & 0.062$\,^{***}$ & 0.026$\,^{***}$ & 0.032$\,^{***}$ \\
 & (0.011) & (0.008) & (0.009) & (0.013) & (0.009) & (0.009) \\
Observations & 2,197,509 & 2,197,509 & 2,197,487 & 2,197,487 & 2,080,022 & 1,461,908 \\
\addlinespace
\midrule
Hagwon count within & dong & 500 m & 500 m & 1 km & 500 m & 500 m \\
District $\times$ month FE & Yes & Yes &  &  &  &  \\
School zone $\times$ month FE &  &  & Yes & Yes & Yes & Yes \\
Age bins, log area, floor & Yes & Yes & Yes & Yes & Yes & Yes \\
Sample & All & All & All & All & Exact & $<$25 yrs \\
\midrule
\multicolumn{7}{l}{\textit{Panel D. Count groups (specification G; base: 0--9 hagwons within 500 m)}} \\
 & \multicolumn{2}{c}{Sale} & \multicolumn{2}{c}{Jeonse} & \multicolumn{2}{c}{Jeonse $-$ sale} \\
\cmidrule(lr){2-3}\cmidrule(lr){4-5}\cmidrule(lr){6-7}
10--29 hagwons & \multicolumn{2}{c}{0.017 (0.018)} & \multicolumn{2}{c}{0.049$\,^{**}$ (0.020)} & \multicolumn{2}{c}{0.033$\,^{**}$ (0.015)} \\
 & \multicolumn{2}{c}{[1.7\%]} & \multicolumn{2}{c}{[5.1\%]} & \multicolumn{2}{c}{} \\
30--99 hagwons & \multicolumn{2}{c}{0.053$\,^{**}$ (0.025)} & \multicolumn{2}{c}{0.120$\,^{***}$ (0.023)} & \multicolumn{2}{c}{0.067$\,^{***}$ (0.021)} \\
 & \multicolumn{2}{c}{[5.4\%]} & \multicolumn{2}{c}{[12.8\%]} & \multicolumn{2}{c}{} \\
100 or more & \multicolumn{2}{c}{0.142$\,^{***}$ (0.037)} & \multicolumn{2}{c}{0.223$\,^{***}$ (0.032)} & \multicolumn{2}{c}{0.081$\,^{**}$ (0.035)} \\
 & \multicolumn{2}{c}{[15.2\%]} & \multicolumn{2}{c}{[25.0\%]} & \multicolumn{2}{c}{} \\
\bottomrule
\end{tabular}
\begin{tablenotes}[flushleft]
\footnotesize
\item \textit{Notes:} Hedonic regressions of log value per m$^2$ (brokered sale prices; jeonse deposits) on the log of one plus the number of academic cram hagwons (sub-category \textit{ipsi$\cdot$gyogwa hagwon}) in the Small Business Market Service business data, December 2018, within 500 m (1 km in D) of the complex, or in its legal dong (A). Panels A and B are separate regressions; Panel C reports hagwons $\times$ jeonse from a pooled regression in which every control and fixed effect is interacted with a jeonse indicator, so it equals the difference between the two premia. Controls: log exclusive area, floor (clipped to $-1$..50) and five-year building-age bins. Exact: drops complexes placed only approximately (block tier; Section~\ref{sec:data}). Panel D replaces the log count with group indicators (specification C otherwise); brackets give $100(e^{b}-1)$. Observations exclude singleton fixed effects. Standard errors in parentheses, clustered by legal dong (district and dong name; 332 clusters in the full sample). $^{*}$\,$p<0.10$, $^{**}$\,$p<0.05$, $^{***}$\,$p<0.01$.
\end{tablenotes}
\end{threeparttable}
\end{table}

The other columns show how the estimate depends on the comparison group and the exposure measure. With the neighborhood (legal dong) count and district $\times$ month effects (column 1), the coefficients are \ResAHagSale{} and \ResAHagJeonse. Measuring exposure within 500 m but keeping district effects (column 2) raises them to \ResBHagSale{} and \ResBHagJeonse. Replacing district $\times$ month with zone $\times$ month effects (column 3) lowers both, consistent with part of the cross-district premium reflecting school assignment. Within 1 km (column 4) the coefficients are larger, \ResDHagSale{} and \ResDHagJeonse. Dropping complexes placed only approximately (column 5) leaves the estimates essentially unchanged (\ResEHagSale{} and \ResEHagJeonse), so the approximate locations do not drive them. In buildings under 25 years old (column 6), whose prices carry little redevelopment value, the premia are somewhat larger (\ResFHagSale{} and \ResFHagJeonse).

\begin{figure}[!htbp]
\centering
\includegraphics[width=0.62\textwidth]{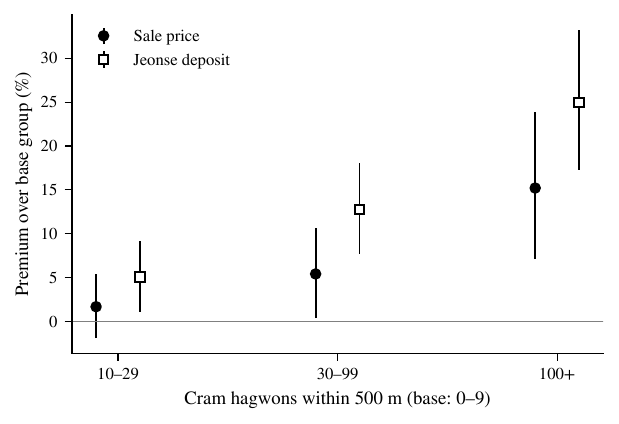}
\caption{Premium by hagwon density, sale prices and jeonse deposits}
\label{fig:bins}
\begin{minipage}{0.9\textwidth}\footnotesize
\textit{Notes:} Specification G of Table~\ref{tab:baseline}: indicators for the number of academic cram hagwons within 500 m (SEMAS data, December 2018), with 0--9 as the base, school zone $\times$ month and building-age fixed effects, log area and floor. Points show $100(e^{b}-1)$ and bars 95\% confidence intervals from standard errors clustered by legal dong.
\end{minipage}
\end{figure}

Panel D and Figure~\ref{fig:bins} replace the log count with groups. The premium rises steadily with hagwon density. Relative to complexes with 0--9 hagwons within 500 m, complexes with 10--29 sell at a premium of \ResGGTenSalePct\% (not significant) and rent at \ResGGTenJeonsePct\% more; with 30--99, the premia are \ResGGThirtySalePct\% and \ResGGThirtyJeonsePct\%; and with 100 or more, \ResGGHundredSalePct\% and \ResGGHundredJeonsePct\%. The top group is small (\DatMapBinHigh{} located complexes), and its premium is the largest, but the premium is not confined to the extreme clusters: it is already present in deposits at 10--29 hagwons and in both markets at 30--99.

Panel C shows the starting puzzle. In every column the jeonse coefficient exceeds the sale coefficient. In the reference specification the jeonse $-$ sale gap is \ResCHagGap{} (SE \ResCHagGapSe, \ResCHagGapP), and it is present with every exposure measure, in exactly located complexes and in newer buildings. Taken at face value, renters pay more for hagwon access than buyers, which in the terms of Section~\ref{sec:framework} would place hagwons in the third case: an amenity valued more by renters. The next three subsections show that this reading does not survive.

\subsection{Hagwons or convenience?}\label{sec:results-convenience}

Hagwon districts are also busy commercial districts. Table~\ref{tab:convenience} adds controls for other businesses within 500 m.

\begin{table}[!htbp]
\centering
\begin{threeparttable}
\caption{Hagwons or convenience? Controls for nearby businesses}
\label{tab:convenience}
\small
\setlength{\tabcolsep}{4pt}
\begin{tabular}{lccccc}
\toprule
 & (1) & (2) & (3) & (4) & (5) \\
Specification & C & H & I & J & K \\
\midrule
\multicolumn{6}{l}{\textit{Panel A. Log sale price per m$^2$}} \\
log(1 + hagwons) & 0.026$\,^{**}$ & 0.053$\,^{***}$ & 0.067$\,^{***}$ & 0.061$\,^{***}$ & 0.054$\,^{***}$ \\
 & (0.012) & (0.011) & (0.015) & (0.011) & (0.011) \\
log(1 + control businesses) &  & $-$0.064$\,^{***}$ & $-$0.077$\,^{***}$ & $-$0.080$\,^{***}$ & $-$0.072$\,^{***}$ \\
 &  & (0.013) & (0.021) & (0.016) & (0.014) \\
\addlinespace
\multicolumn{6}{l}{\textit{Panel B. Log jeonse deposit per m$^2$}} \\
log(1 + hagwons) & 0.051$\,^{***}$ & 0.031$\,^{**}$ & 0.027$\,^{**}$ & 0.025$\,^{**}$ & 0.034$\,^{***}$ \\
 & (0.014) & (0.012) & (0.014) & (0.012) & (0.012) \\
log(1 + control businesses) &  & 0.043$\,^{***}$ & 0.042$\,^{**}$ & 0.053$\,^{***}$ & 0.040$\,^{**}$ \\
 &  & (0.015) & (0.019) & (0.018) & (0.016) \\
\addlinespace
\multicolumn{6}{l}{\textit{Panel C. Jeonse $-$ sale}} \\
log(1 + hagwons) & 0.025$\,^{***}$ & $-$0.022$\,^{***}$ & $-$0.040$\,^{***}$ & $-$0.036$\,^{***}$ & $-$0.020$\,^{**}$ \\
 & (0.009) & (0.008) & (0.011) & (0.009) & (0.008) \\
log(1 + control businesses) &  & 0.108$\,^{***}$ & 0.118$\,^{***}$ & 0.132$\,^{***}$ & 0.112$\,^{***}$ \\
 &  & (0.013) & (0.017) & (0.015) & (0.013) \\
\addlinespace
\midrule
Control businesses & None & Non-educ. & Other educ. & Non-student & Non-food \\
FE and controls as in C & Yes & Yes & Yes & Yes & Yes \\
Obs.: sale / jeonse / pooled & \multicolumn{5}{c}{771,466 / 1,426,021 / 2,197,487 in each column} \\
\bottomrule
\end{tabular}
\begin{tablenotes}[flushleft]
\footnotesize
\item \textit{Notes:} Specification C of Table~\ref{tab:baseline} plus the log of one plus the number of other businesses within 500 m (SEMAS commercial-district data, December 2018; Seoul only, except other education, which also counts Gyeonggi businesses near the city edge). Non-educ.: all businesses outside education. Other educ.: education businesses other than academic cram hagwons (music, art, sports and martial arts, language and other academies). Non-student: estate agents (\textit{budongsan}), health care (\textit{bogeon-uiryo}) and repairs and personal services (\textit{suri-gaein}), types not expected to follow students. Non-food: all businesses except education, food (\textit{eumsik}), retail (\textit{somae}) and arts, sports and leisure (\textit{yesul-seupocheu}). Panel C: coefficients on each variable $\times$ jeonse in the pooled, fully interacted regression. Standard errors in parentheses, clustered by legal dong (district and dong name; 332 clusters in the full sample). $^{*}$\,$p<0.10$, $^{**}$\,$p<0.05$, $^{***}$\,$p<0.01$.
\end{tablenotes}
\end{threeparttable}
\end{table}

Column (2), specification H, adds all non-education businesses. The hagwon coefficient rises to \ResHHagSale{} for sale prices and falls to \ResHHagJeonse{} for jeonse deposits, and the gap reverses sign, to \ResHHagGap{} (SE \ResHHagGapSe, \ResHHagGapP). The control itself carries opposite signs in the two markets: \ResHShopSale{} for prices and \ResHShopJeonse{} for deposits. Column (3), specification I, controls instead for education businesses other than academic hagwons, such as music, art and sports academies. The pattern is the same: the hagwon coefficients are \ResIHagSale{} and \ResIHagJeonse{}, and other education carries \ResIOthEduSale{} in prices and \ResIOthEduJeonse{} in deposits.

Two readings fit these results. Under a \textit{convenience} reading, renters value nearby shops and services more than buyers do, and hagwon districts happen to be convenient; the raw jeonse gap is then a convenience gap. Under a \textit{bad-control} reading, caf\'es, snack bars and study rooms locate near hagwons \textit{because} students are there, so controlling for them absorbs part of the hagwon premium in jeonse.

\begin{figure}[!htbp]
\centering
\includegraphics[width=\textwidth]{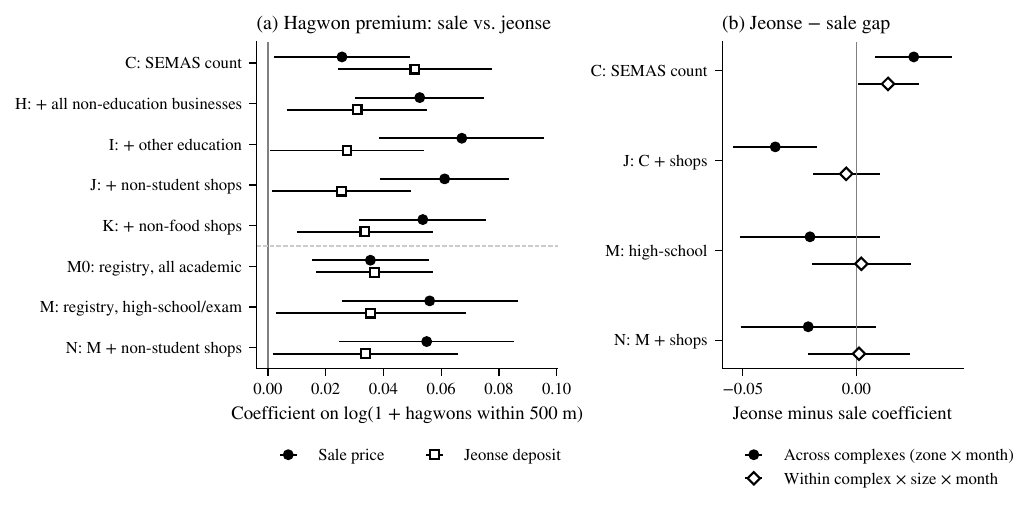}
\caption{Hagwon premia in sale prices and jeonse deposits across specifications}
\label{fig:coefplot}
\begin{minipage}{0.95\textwidth}\footnotesize
\textit{Notes:} Panel (a): coefficients on log(1 + hagwons within 500 m) from Tables~\ref{tab:baseline}--\ref{tab:grades}; C--K use the SEMAS count of academic cram hagwons (December 2018), M0 the registry total, M and N the registry high-school/exam count. Panel (b): jeonse $-$ sale gaps from the pooled cross-complex model (filled circles; Tables~\ref{tab:baseline}--\ref{tab:grades}) and from the within-complex, flat size and month comparison (open diamonds; Table~\ref{tab:withincell}, Panel A). Bars are 95\% confidence intervals, clustered by legal dong.
\end{minipage}
\end{figure}

Columns (4) and (5) discriminate between the two. Specification J controls only for business types that should not follow students: estate agents, health care, and repairs and personal services. The reversal is at least as strong as in column (2). The hagwon coefficients are \ResJHagSale{} (SE \ResJHagSaleSe) for prices and \ResJHagJeonse{} (SE \ResJHagJeonseSe) for deposits, a gap of \ResJHagGap{} (\ResJHagGapP). These non-student shops are priced with opposite signs: doubling their number is associated with changes of \ResJShopSalePct\% in prices and \ResJShopJeonsePct\% in deposits. Specification K, which drops food, retail and leisure businesses from the control, gives the same result (gap \ResKHagGap; shops \ResKShopSale{} and \ResKShopJeonse). Panel (a) of Figure~\ref{fig:coefplot} summarizes the pattern: once any measure of local commercial density is held fixed, the hagwon premium is at least as large in prices as in deposits.

The convenience premium is of the same order as the hagwon premium. In specification J, doubling nearby hagwons is associated with prices \ResJHagSalePct\% and deposits \ResJHagJeonsePct\% higher, while doubling non-student shops is associated with changes of \ResJShopSalePct\% in prices and \ResJShopJeonsePct\% in deposits. Because the two counts are positively correlated, leaving the shops out loads part of their opposite-signed effects onto hagwons: it lowers the hagwon coefficient in prices, where shops carry a negative sign, and raises it in deposits, where they carry a positive one. That is exactly the pattern of column (1) relative to columns (2)--(5).

Splitting the non-student control into its three categories locates the pattern (Table~\ref{tab:app-exposure}, Panel B). Estate agents and clinics carry premia in both markets, of similar size: \ChkExpNSplitBrkSale{} in prices and \ChkExpNSplitBrkJeonse{} in deposits for estate agents, \ChkExpNSplitHealthSale{} and \ChkExpNSplitHealthJeonse{} for clinics. The gap comes from repair and personal services, such as hair salons, laundries and repair shops, which carry \ChkExpNSplitRepairSale{} in prices and \ChkExpNSplitRepairJeonse{} in deposits, a gap of \ChkExpNSplitRepairGap{} (\ChkExpNSplitRepairGapP). None of these categories has an obvious reason to follow students, which supports the convenience reading, but what buyers discount relative to renters is the presence of everyday services rather than of shops in general. The gap is also not an artefact of the city boundary, where shop counts are incomplete: away from it the non-student jeonse coefficient is only \ChkExpNNoEdgeShopJeonse, but the jeonse $-$ sale gap remains \ChkExpNNoEdgeShopGap{} (\ChkExpNNoEdgeShopGapP). The convenience reading does not rule out bad control entirely. Non-student shops still cluster near hagwons (their log counts are correlated at \DatCorrCramNonstudent{} across complexes), and clinics often share buildings with hagwons. Why buyers pay less near shops cannot be established here. One possibility is that owners, with longer horizons, weigh congestion and noise, while renters, who are more mobile, weigh day-to-day access. Section~\ref{sec:results-withincell} returns to this question with the within-complex comparison.

\subsection{Which hagwons carry the premium?}\label{sec:results-grades}

\begin{table}[!htbp]
\centering
\begin{threeparttable}
\caption{Which hagwons carry the premium? Registry counts by grade level}
\label{tab:grades}
\footnotesize
\setlength{\tabcolsep}{4pt}
\begin{tabular}{lccc}
\toprule
 & (1) & (2) & (3) \\
Specification & M0 & M & N \\
\midrule
\multicolumn{4}{l}{\textit{Panel A. Log sale price per m$^2$}} \\
All academic hagwons & 0.036$\,^{***}$ &  &  \\
 & (0.010) &  &  \\
High-school/exam hagwons &  & 0.056$\,^{***}$ & 0.055$\,^{***}$ \\
 &  & (0.016) & (0.016) \\
Other academic hagwons &  & $-$0.016 & 0.007 \\
 &  & (0.017) & (0.017) \\
Non-student shops &  &  & $-$0.070$\,^{***}$ \\
 &  &  & (0.016) \\
6 $\to$ 50 high-school hagwons (\%) &  & 11.8 & 11.5 \\
\addlinespace
\multicolumn{4}{l}{\textit{Panel B. Log jeonse deposit per m$^2$}} \\
All academic hagwons & 0.037$\,^{***}$ &  &  \\
 & (0.010) &  &  \\
High-school/exam hagwons &  & 0.036$\,^{**}$ & 0.034$\,^{**}$ \\
 &  & (0.017) & (0.016) \\
Other academic hagwons &  & 0.005 & $-$0.015 \\
 &  & (0.018) & (0.017) \\
Non-student shops &  &  & 0.061$\,^{***}$ \\
 &  &  & (0.018) \\
6 $\to$ 50 high-school hagwons (\%) &  & 7.3 & 6.9 \\
\addlinespace
\multicolumn{4}{l}{\textit{Panel C. Jeonse $-$ sale}} \\
All academic hagwons & 0.001 &  &  \\
 & (0.008) &  &  \\
High-school/exam hagwons &  & $-$0.020 & $-$0.021 \\
 &  & (0.016) & (0.015) \\
Other academic hagwons &  & 0.021 & $-$0.021 \\
 &  & (0.015) & (0.015) \\
Non-student shops &  &  & 0.131$\,^{***}$ \\
 &  &  & (0.014) \\
\addlinespace
\midrule
Hagwon measure (registry, 2026) & Total & Split & Split \\
Non-student shop control &  &  & Yes \\
Zone $\times$ month FE, age bins, area, floor & Yes & Yes & Yes \\
Obs.: sale / jeonse / pooled & \multicolumn{3}{c}{771,466 / 1,426,021 / 2,197,487} \\
\bottomrule
\end{tabular}
\begin{tablenotes}[flushleft]
\footnotesize
\item \textit{Notes:} Same sample and model as Table~\ref{tab:baseline}, specification C, but hagwon exposure comes from the NEIS hagwon registry (snapshot of October 2026): hagwons in the academic field (entrance and qualification exams and supplementary lessons) within 500 m of the complex. M0 uses all of them. M and N split them into high-school/exam hagwons (name, course list or fees mention high school, the College Scholastic Ability Test, admissions or exam preparation) and all other academic hagwons (elementary, middle-school or unlabeled; Appendix~\ref{app:grades}). N adds non-student shops (Table~\ref{tab:convenience}). The implied-magnitude rows move the high-school count from the median complex (6) to 50: $100(e^{b\,[\ln 51-\ln 7]}-1)$. The correlation between log high-school and log other counts across complexes is 0.85. Standard errors in parentheses, clustered by legal dong (district and dong name; 332 clusters in the full sample). $^{*}$\,$p<0.10$, $^{**}$\,$p<0.05$, $^{***}$\,$p<0.01$.
\end{tablenotes}
\end{threeparttable}
\end{table}

Table~\ref{tab:grades} replaces the SEMAS count with the NEIS hagwon registry. With all registry academic hagwons (column 1, M0), the coefficients are \ResMZeroRegSale{} for prices (SE \ResMZeroRegSaleSe) and \ResMZeroRegJeonse{} for deposits (SE \ResMZeroRegJeonseSe). Their difference is \ResMZeroRegGap{} (SE \ResMZeroRegGapSe, \ResMZeroRegGapP). With the registry count, then, there is no jeonse $-$ sale gap even without any shop control: the gap in Table~\ref{tab:baseline} is specific to the SEMAS measure, which, like the shop counts, tracks commercial activity.

Column (2), specification M, splits the registry count by grade. The premium is carried by high-school and exam-preparation hagwons. Their coefficient is \ResMHsSale{} for prices (SE \ResMHsSaleSe, \ResMHsSaleP) and \ResMHsJeonse{} for deposits (SE \ResMHsJeonseSe, \ResMHsJeonseP). Holding them fixed, other academic hagwons, which teach elementary and middle-school pupils or carry no grade label, add nothing detectable: \ResMOthSale{} (SE \ResMOthSaleSe) for prices and \ResMOthJeonse{} (SE \ResMOthJeonseSe) for deposits.

In magnitude, moving a complex from the median of \DatMedNeisHs{} high-school/exam hagwons within 500 m to \ResHsTarget{} is associated with sale prices \ResMHsSalePct\% higher (95\% interval \ResMHsSalePctLo\% to \ResMHsSalePctHi\%) and jeonse deposits \ResMHsJeonsePct\% higher (\ResMHsJeonsePctLo\% to \ResMHsJeonsePctHi\%). A move across the interquartile range, from \ResHsPTwentyFive{} to \ResHsPSeventyFive{} such hagwons, is associated with \ResMHsSalePctIqr\% and \ResMHsJeonsePctIqr\%. Exam hagwons cluster near large complexes, which sell at a premium; holding complex size fixed roughly halves the sale premium, to \ChkAccSizeMHsSalePct\%, and leaves the deposit premium at \ChkAccSizeMHsJeonsePct\% (Section~\ref{sec:robust-access}). Adding non-student shops (column 3, N) leaves the high-school coefficients almost unchanged (\ResNHsSale{} and \ResNHsJeonse), while the shops again carry opposite signs (\ResNShopSale{} and \ResNShopJeonse). The other-hagwon coefficients move slightly with the shop control (to \ResNOthSale{} in prices and \ResNOthJeonse{} in deposits) but remain small and insignificant.

In money terms, consider a flat at the sample's median sale price per m$^2$ (KRW \ResMedSalePriceM{} million) and median size (\ResMedSaleArea{} m$^2$). Moving from \DatMedNeisHs{} to \ResHsTarget{} high-school hagwons corresponds to about KRW \ResMHsSaleKRW{} million in price. At the median jeonse deposit per m$^2$ (KRW \ResMedJeonsePriceM{} million) and median leased size (\ResMedJeonseArea{} m$^2$), the same move corresponds to about KRW \ResMHsJeonseKRW{} million in deposit. The deposit is returned at the end of the lease, so the tenant's annual cost is the return forgone on that amount, not the amount itself.

The gap for high-school hagwons is \ResMHsGap{} (SE \ResMHsGapSe, \ResMHsGapP). It is not statistically different from zero; if anything, the premium is larger in prices. With complex size held fixed, the gap is \ChkAccSizeMHsGap{} (Section~\ref{sec:robust-access}).

Two caveats apply. First, the high-school and other counts are correlated at \DatCorrHsOther{} across complexes. The collinearity raises the standard errors (\ResMHsSaleSe{} for the high-school sale coefficient against \ResMZeroRegSaleSe{} for the registry total), but they remain moderate. For prices, the 95\% interval for other hagwons, \ResMOthSaleLo{} to \ResMOthSaleHi, excludes the high-school estimate. For deposits it does not (\ResMOthJeonseLo{} to \ResMOthJeonseHi), so the grade contrast is sharp in prices and only suggestive in deposits. Second, of the coefficients in this table, the high-school sale premium, the registry totals and the shop coefficients pass the Bonferroni threshold for the \ChkFamTests{} tests of Table~\ref{tab:app-family} ($p < \ChkFamThreshold$). The high-school jeonse premium (\ResMHsJeonseP) does not, and with standard errors clustered by school zone its $p$-value rises to \ChkInfMHsJeonsePZone{} (Section~\ref{sec:robust-inference}), although it is positive in every period, sample and exposure measure in Section~\ref{sec:robust}.

The grade split speaks to what households pay for. Hagwons for younger children serve the same families and would capture any generic ``good neighborhood'' factor that attracts hagwons in general. Yet they carry no detectable premium once high-school hagwons are held fixed, clearly so in prices. The premium attaches to the services that prepare students for the university entrance examination, which points to competition for university places rather than to childcare or general convenience.

\subsection{Same flat, two prices: the within-complex jeonse-to-price ratio}\label{sec:results-withincell}

The level premia compare different complexes. Table~\ref{tab:withincell} instead compares the deposit and the price of the same flat type, in the same complex and month (Equation~\eqref{eq:withincell}). There are \ResCellObsCells{} cells with both a sale and a jeonse contract, in \ResCellObsComplexes{} complexes. Within a cell, the deposit averages \ResCellRatioPct\% of the sale price per m$^2$ (a log gap of \ResCellAvgGap).

\begin{table}[!htbp]
\centering
\begin{threeparttable}
\caption{Same flat, two prices: hagwons and the jeonse-to-price ratio within complexes}
\label{tab:withincell}
\small
\setlength{\tabcolsep}{4pt}
\begin{tabular}{lcccc}
\toprule
 & (1) & (2) & (3) & (4) \\
Specification & C & J & M & N \\
\midrule
\multicolumn{5}{l}{\textit{Panel A. Same complex, flat size and month: gap regression across cells}} \\
Cram hagwons (SEMAS) & 0.014$\,^{**}$ & $-$0.004 &  &  \\
 & (0.007) & (0.008) &  &  \\
High-school/exam hagwons (registry) &  &  & 0.002 & 0.001 \\
 &  &  & (0.011) & (0.011) \\
Other academic hagwons (registry) &  &  & 0.004 & $-$0.010 \\
 &  &  & (0.009) & (0.011) \\
Non-student shops &  & 0.040$\,^{***}$ &  & 0.042$\,^{***}$ \\
 &  & (0.009) &  & (0.010) \\
Cells & 222,317 & 222,317 & 222,317 & 222,317 \\
\addlinespace
\multicolumn{5}{l}{\textit{Panel B. Same rows, school zone $\times$ month comparison only (no cell comparison)}} \\
Cram hagwons (SEMAS) & 0.010 & $-$0.012$\,^{*}$ &  &  \\
 & (0.007) & (0.007) &  &  \\
High-school/exam hagwons (registry) &  &  & $-$0.004 & $-$0.006 \\
 &  &  & (0.013) & (0.013) \\
Other academic hagwons (registry) &  &  & 0.006 & $-$0.011 \\
 &  &  & (0.011) & (0.013) \\
Non-student shops &  & 0.048$\,^{***}$ &  & 0.050$\,^{***}$ \\
 &  & (0.011) &  & (0.012) \\
Transactions & 1,205,795 & 1,205,795 & 1,205,795 & 1,205,795 \\
\addlinespace
\midrule
Mean within-cell gap, log points & \multicolumn{4}{c}{$-$0.579 (0.016): jeonse = 56\% of the sale price per m$^2$} \\
Cells / complexes / zones & \multicolumn{4}{c}{222,400 / 5,123 / 46} \\
Zone $\times$ month FE, age bins & Yes & Yes & Yes & Yes \\
\bottomrule
\end{tabular}
\begin{tablenotes}[flushleft]
\footnotesize
\item \textit{Notes:} A cell is a complex $\times$ exclusive area (rounded to 1 m$^2$) $\times$ contract month with at least one brokered sale and one jeonse contract (2015--2026). Panel A regresses the cell's gap, mean log jeonse deposit per m$^2$ minus mean log sale price per m$^2$, on the complex's exposure, the within-cell differences in floor and log area, and school zone $\times$ month and building-age-bin fixed effects, weighting each cell by $n_J n_S/(n_J+n_S)$. A coefficient is the change in the log jeonse-to-price ratio of the same flat type. Panel B fits the pooled, fully interacted model of Table~\ref{tab:baseline} to the transactions in the same cells, without the cell comparison. Exposure measures as in Tables~\ref{tab:baseline}--\ref{tab:grades}; C and J use the SEMAS count, M and N the registry split; J and N add non-student shops. Observations exclude singleton fixed effects. Standard errors in parentheses, clustered by legal dong. $^{*}$\,$p<0.10$, $^{**}$\,$p<0.05$, $^{***}$\,$p<0.01$.
\end{tablenotes}
\end{threeparttable}
\end{table}

High-school hagwons do not detectably move this ratio. In column (3), specification M, the coefficient is \ResCellAllMHs{} (SE \ResCellAllMHsSe, \ResCellAllMHsP), with a 95\% interval of \ResCellAllMHsLo{} to \ResCellAllMHsHi. Moving from \DatMedNeisHs{} to \ResHsTarget{} high-school hagwons changes the ratio by \ResCellAllMHsPct\% (interval \ResCellAllMHsPctLo\% to \ResCellAllMHsPctHi\%), against level premia of \ResMHsSalePct\% and \ResMHsJeonsePct\%. Adding shops (column 4) gives \ResCellAllNHs{} (SE \ResCellAllNHsSe). Non-student shops, by contrast, raise the ratio: their coefficient is \ResCellAllNShop{} (SE \ResCellAllNShopSe, \ResCellAllNShopP), so doubling them raises the deposit relative to the price of the same flat by \ResCellAllNShopPct\% (\ResCellAllNShopPctLo\% to \ResCellAllNShopPctHi\%). With the SEMAS count, a small positive gap remains without shop controls (column 1: \ResCellAllCHag, SE \ResCellAllCHagSe, \ResCellAllCHagP), and it disappears once non-student shops are held fixed (column 2: \ResCellAllJHag, SE \ResCellAllJHagSe), while the shops carry \ResCellAllJShop{} (SE \ResCellAllJShopSe). Panel (b) of Figure~\ref{fig:coefplot} sets these within-cell gaps beside the cross-complex gaps. The design is close in spirit to \citet{zheng2016much}, who use rental differentials between paired resale and rental observations to net out unobserved neighborhood traits; here the pairing is by complex, flat size and month.

Panel B of Table~\ref{tab:withincell} fits the cross-complex model of Section~\ref{sec:strategy-hedonic} to the same transactions without the cell comparison. The high-school gap is \ResRowsAllMHs{} (SE \ResRowsAllMHsSe), already close to zero. The insignificant cross-complex gap of \ResMHsGap{} in the full sample therefore moves toward zero because the sample is restricted to frequently traded complexes, not because of the cell comparison itself. The shop gap, \ResRowsAllNShop{} without and \ResCellAllNShop{} with the cell comparison, survives both. The within-cell results are stable over time and location precision (Table~\ref{tab:app-withincell}): the high-school coefficient is \ResCellPreBanMHs{} before December 2019 and \ResCellFromBanMHs{} after, and the shop coefficient \ResCellPreBanNShop{} and \ResCellFromBanNShop.

In the terms of Section~\ref{sec:framework}, hagwon access is capitalized in proportion into prices and deposits. This is what one would expect of a current-flow amenity that both clienteles value alike and that carries no extra expected appreciation. The estimate rules out the polar alternatives. If hagwon access were valued only through expected appreciation, the ratio would fall by roughly the full sale premium (\ResMHsSale); the within-cell estimate lies \ResCellAllMHsMinusSaleZ{} of its own standard errors above that value. If only renters valued it, the ratio would rise by roughly the full jeonse premium (\ResMHsJeonse), which lies outside the 95\% interval. Intermediate combinations of an appreciation component and a renter-valued component that offset each other cannot be excluded. If both clienteles value the flow alike, the lower end of the interval allows expected appreciation to account for at most about \ResCellAllMHsMaxApprShare\% of the cross-complex sale premium. The test has the power to detect such a component where one is known to exist: in the same cells, it shows ratios about \ChkCellPosAgeThirtyMPctAbs\% lower in buildings of 30 years or more, whose prices carry reconstruction value (Section~\ref{sec:robust-ratio}).

Shops behave like the third case of the framework: something near shops raises deposits relative to the prices of the same flats. This matches the opposite-signed level coefficients in Table~\ref{tab:convenience} and survives the comparison that removes everything common to the two prices of a flat. A renter clientele is one candidate, but it can move the ratio only where tenure is segmented. A lower conversion rate, if deposits are perceived as safer there, or differences in liquidity and turnover, which search models link to the rent--price ratio \citep{huang2018accounts}, would have the same effect. The data cannot separate these channels. The contrast with school-quality capitalization elsewhere is also worth noting: \citet{gabe2022relationship} find that US apartment renters value primary-school quality more than secondary, whereas in Seoul jeonse tenants pay for access to high-school exam preparation in the same proportion as buyers.

The within-cell sample covers the more frequently traded complexes, \ResCellComplexShare\% of those in the hedonic sample. The result describes how the two markets price the same flats in those complexes; it is not an estimate for thinly traded complexes.

\subsection{Is there an entry fee for new tenants?}\label{sec:results-entryfee}

If incumbent tenants near hagwons hold below-market deposits, for instance because the 2020 renewal right capped their increases, newcomers should pay an extra premium there. Table~\ref{tab:entryfee} tests this. The sample is \ResEFLocated{} jeonse contracts signed from June 2021 on located complexes (\ResEFNew{} new contracts, \ResEFCapped{} capped renewals and \ResEFVoluntary{} voluntary renewals). There are \ResEFAllFullCells{} cells with both a new contract and a renewal, containing \ResEFAllFullContracts{} contracts.

\begin{table}[!htbp]
\centering
\begin{threeparttable}
\caption{Is there an entry fee? New versus renewal jeonse deposits in the same complex and month}
\label{tab:entryfee}
\footnotesize
\setlength{\tabcolsep}{3.5pt}
\begin{tabular}{lcccccc}
\toprule
 & (1) & (2) & (3) & (4) & (5) & (6) \\
 & Average & C: hagwons & N: high-school & N: shops & N: 6$\to$50 & Cells \\
 & markup & $\times$ new & $\times$ new & $\times$ new & (\%) &  \\
\midrule
\multicolumn{7}{l}{\textit{Panel A. New contracts vs all renewals}} \\
2021-06 to 2026-10 & 0.086$\,^{***}$ & 0.0005 & 0.0049$\,^{*}$ & $-$0.0030 & 1.0 & 74,443 \\
 & (0.002) & (0.0012) & (0.0027) & (0.0022) &  &  \\
2021-06 to 2022-12 & 0.181$\,^{***}$ & 0.0026 & 0.0107$\,^{**}$ & $-$0.0078$\,^{*}$ & 2.1 & 20,164 \\
 & (0.004) & (0.0024) & (0.0053) & (0.0041) &  &  \\
2023-01 to 2026-10 & 0.053$\,^{***}$ & $-$0.0002 & 0.0025 & $-$0.0013 & 0.5 & 54,279 \\
 & (0.002) & (0.0011) & (0.0027) & (0.0021) &  &  \\
\addlinespace
\multicolumn{7}{l}{\textit{Panel B. New contracts vs capped renewals (renewal right used, $\leq$5\%)}} \\
2021-06 to 2026-10 & 0.112$\,^{***}$ & 0.0022$\,^{*}$ & 0.0048 & $-$0.0025 & 1.0 & 49,020 \\
 & (0.002) & (0.0013) & (0.0031) & (0.0030) &  &  \\
2021-06 to 2022-12 & 0.212$\,^{***}$ & 0.0021 & 0.0131$\,^{**}$ & $-$0.0112$\,^{**}$ & 2.6 & 16,143 \\
 & (0.004) & (0.0030) & (0.0058) & (0.0045) &  &  \\
2023-01 to 2026-10 & 0.063$\,^{***}$ & 0.0018 & 0.0005 & 0.0012 & 0.1 & 32,877 \\
 & (0.002) & (0.0011) & (0.0025) & (0.0028) &  &  \\
\addlinespace
\multicolumn{7}{l}{\textit{Panel C. New contracts vs voluntary renewals (renewal right not used)}} \\
2021-06 to 2026-10 & 0.058$\,^{***}$ & $-$0.0007 & 0.0048 & $-$0.0024 & 1.0 & 45,730 \\
 & (0.002) & (0.0014) & (0.0032) & (0.0024) &  &  \\
2021-06 to 2022-12 & 0.114$\,^{***}$ & 0.0040 & 0.0092$\,^{*}$ & 0.0010 & 1.8 & 9,464 \\
 & (0.004) & (0.0029) & (0.0054) & (0.0048) &  &  \\
2023-01 to 2026-10 & 0.046$\,^{***}$ & $-$0.0016 & 0.0035 & $-$0.0028 & 0.7 & 36,266 \\
 & (0.002) & (0.0014) & (0.0035) & (0.0025) &  &  \\
\addlinespace
\bottomrule
\end{tabular}
\begin{tablenotes}[flushleft]
\footnotesize
\item \textit{Notes:} Seoul apartment jeonse contracts from June 2021 (when leases began to be reported as new or renewal) on located complexes. A cell is a complex $\times$ exclusive area (1 m$^2$) $\times$ month with at least one new contract and one renewal of the stated kind. Column (1): mean log deposit per m$^2$ of new contracts minus renewals in the same cell (log points; regression with cell fixed effects). Columns (2)--(4): regressions of the cell's markup on exposure, within-cell floor and log-area differences, and school zone $\times$ month and age-bin fixed effects, weighted by $n_{new}n_{ren}/(n_{new}+n_{ren})$. C: SEMAS cram count; N: registry high-school/exam and other academic hagwons plus non-student shops (other terms in Table~\ref{tab:app-entryfee}). Column (5) scales column (3) by the move from 6 to 50 high-school hagwons. A capped renewal used the renewal right of the 2020 lease law, which limits the increase to 5\%. Standard errors in parentheses, clustered by legal dong. $^{*}$\,$p<0.10$, $^{**}$\,$p<0.05$, $^{***}$\,$p<0.01$.
\end{tablenotes}
\end{threeparttable}
\end{table}

New contracts carry a sizeable average markup over renewals in the same complex, flat size and month: \ResEFAllFullMarkup{} log points (\ResEFAllFullMarkupPct\%) over the whole window. The markup is \ResEFAllEarlyMarkup{} log points (\ResEFAllEarlyMarkupPct\%) in June 2021 to December 2022, when renewals were anchored to deposits set in 2019--20, and \ResEFAllLateMarkup{} (\ResEFAllLateMarkupPct\%) from 2023. Against capped renewals in 2021--22 it reaches \ResEFCapEarlyMarkup{} (\ResEFCapEarlyMarkupPct\%). These averages are not themselves an entry fee. They mostly reflect market rent growth over the two years since the renewing tenant's previous contract, which was rapid in 2019--21 and which the cap prevented renewals from following. The question is whether the markup is larger near high-school hagwons.

The markup is barely larger near high-school hagwons. Over the whole window, with the shop control (specification N), the interaction is \ResEFAllFullNHs{} (SE \ResEFAllFullNHsSe, \ResEFAllFullNHsP). Moving from \DatMedNeisHs{} to \ResHsTarget{} high-school hagwons raises the markup by \ResEFAllFullNHsPct\% (\ResEFAllFullNHsPctMed\% from the entry-fee sample's own median of \ResEFAllFullMedHs). The extra markup is concentrated in 2021--22, where it is \ResEFAllEarlyNHs{} (\ResEFAllEarlyNHsP; \ResEFAllEarlyNHsPct\%), and in the comparison with capped renewals in those years, \ResEFCapEarlyNHs{} (\ResEFCapEarlyNHsP; \ResEFCapEarlyNHsPct\%). From 2023 it is \ResEFAllLateNHs{} (\ResEFAllLateNHsP), and against capped renewals \ResEFCapLateNHs{} (\ResEFCapLateNHsP). The comparison with voluntary renewals is the cleaner test of a newcomer premium, because those renewals were negotiated freely and are not mechanically held below the market: there the interaction is \ResEFVolFullNHs{} (\ResEFVolFullNHsP) over the whole window and \ResEFVolLateNHs{} (\ResEFVolLateNHsP) from 2023. The SEMAS count shows nothing (\ResEFAllFullCHag, SE \ResEFAllFullCHagSe), and non-student shops do not change the markup (\ResEFAllFullNShop, SE \ResEFAllFullNShopSe).

These interactions are small and fragile. The smallest $p$-value among the \ResEFBHTests{} high-school interactions in Table~\ref{tab:entryfee} is \ResEFMinP; after a Benjamini--Hochberg adjustment across them, the smallest adjusted $p$-value is \ResEFBHMinQ. There is no lasting entry fee for newcomers to hagwon districts. The modest 2021--22 premium against capped renewals fits a catch-up: those renewals were anchored to 2019--20 deposits, and if rents near hagwons rose faster in 2019--21 and the cap held renewals back, new contracts there would stand further above renewals. The event study in Section~\ref{sec:robust-events} points the same way: the hagwon gradient in new-contract deposits was higher in the second half of 2021 than in early 2023 (\ResEvNewJCoefTwentyOneHTwo, SE \ResEvNewJCoefTwentyOneHTwoSe). Once renewals were anchored to the 2021--22 peak, the extra markup vanished.
\section{Robustness and extensions}\label{sec:robust}

\subsection{Housing-rule periods and permit-zone dongs}\label{sec:robust-periods}

Two of Seoul's housing rules held down sale prices in the most expensive hagwon districts without directly constraining jeonse deposits: the ban on mortgages for homes above KRW 1.5 billion and the land-transaction permit zones, which require buyers in designated dongs to occupy the flat themselves (Table~\ref{tab:policy}; Section~\ref{sec:background}). Either could distort a comparison of sale and jeonse premia. Panel A of Table~\ref{tab:periods} re-estimates the main specifications by period; Table~\ref{tab:app-periods} reports every coefficient.

\begin{table}[!htbp]
\centering
\begin{threeparttable}
\caption{Stability across housing-rule periods, permit zones and contract types}
\label{tab:periods}
\footnotesize
\setlength{\tabcolsep}{3pt}
\begin{tabular}{lcccccc}
\toprule
 & (1) & (2) & (3) & (4) & (5) & (6) \\
 & Full & Before & Before ban & Ban & After ban & No permit \\
 &  & Aug.\ 2017 & and zones & years &  & zone dongs \\
 & 2015--26 & 2015--17:07 & 2015--19:11 & 2019:12--22:11 & 2022:12--26 & 2015--26 \\
\midrule
\multicolumn{7}{l}{\textit{Panel A. Housing-rule periods and permit-zone dongs (all jeonse contracts)}} \\
M: high-school, sale & 0.056$\,^{***}$ & 0.038$\,^{***}$ & 0.046$\,^{***}$ & 0.058$\,^{***}$ & 0.064$\,^{***}$ & 0.064$\,^{***}$ \\
 & (0.016) & (0.013) & (0.014) & (0.019) & (0.019) & (0.017) \\
M: high-school, jeonse & 0.036$\,^{**}$ & 0.042$\,^{**}$ & 0.040$\,^{**}$ & 0.029$\,^{*}$ & 0.039$\,^{**}$ & 0.034$\,^{**}$ \\
 & (0.017) & (0.017) & (0.016) & (0.017) & (0.018) & (0.017) \\
M: jeonse $-$ sale & $-$0.020 & 0.005 & $-$0.006 & $-$0.029 & $-$0.025 & $-$0.030$\,^{**}$ \\
 & (0.016) & (0.016) & (0.015) & (0.020) & (0.017) & (0.015) \\
N: jeonse $-$ sale & $-$0.021 & 0.005 & $-$0.005 & $-$0.026 & $-$0.031$\,^{*}$ & $-$0.027$\,^{**}$ \\
 & (0.015) & (0.016) & (0.015) & (0.019) & (0.016) & (0.013) \\
C: jeonse $-$ sale & 0.025$\,^{***}$ & 0.026$\,^{**}$ & 0.022$\,^{**}$ & 0.016 & 0.025$\,^{***}$ & 0.019$\,^{**}$ \\
 & (0.009) & (0.012) & (0.009) & (0.012) & (0.008) & (0.010) \\
J: jeonse $-$ sale & $-$0.036$\,^{***}$ & $-$0.008 & $-$0.020$\,^{**}$ & $-$0.059$\,^{***}$ & $-$0.043$\,^{***}$ & $-$0.048$\,^{***}$ \\
 & (0.009) & (0.011) & (0.010) & (0.013) & (0.009) & (0.010) \\
Observations: sale & 771,466 & 276,375 & 445,123 & 131,086 & 195,257 & 710,932 \\
Observations: jeonse & 1,426,021 & 248,784 & 507,349 & 391,792 & 526,880 & 1,270,627 \\
\midrule
\multicolumn{7}{l}{\textit{Panel B. 2021-06 to 2026-10: sales against different jeonse samples}} \\
 & All & New & Capped & All, & New, &  \\
 & jeonse & only & only & $<$25 yrs & $<$25 yrs &  \\
\cmidrule(lr){2-6}
M: high-school, sale & 0.066$\,^{***}$ & 0.066$\,^{***}$ & 0.066$\,^{***}$ & 0.086$\,^{***}$ & 0.086$\,^{***}$ & \\
 & (0.019) & (0.019) & (0.019) & (0.022) & (0.022) & \\
M: high-school, jeonse & 0.037$\,^{**}$ & 0.036$\,^{**}$ & 0.040$\,^{***}$ & 0.040$\,^{*}$ & 0.039$\,^{*}$ & \\
 & (0.017) & (0.017) & (0.015) & (0.021) & (0.022) & \\
M: jeonse $-$ sale & $-$0.029$\,^{*}$ & $-$0.030$\,^{*}$ & $-$0.025 & $-$0.046$\,^{**}$ & $-$0.047$\,^{**}$ & \\
 & (0.016) & (0.016) & (0.016) & (0.020) & (0.020) & \\
N: jeonse $-$ sale & $-$0.033$\,^{**}$ & $-$0.034$\,^{**}$ & $-$0.027$\,^{*}$ & $-$0.057$\,^{***}$ & $-$0.058$\,^{***}$ & \\
 & (0.015) & (0.015) & (0.015) & (0.016) & (0.016) & \\
C: jeonse $-$ sale & 0.024$\,^{***}$ & 0.026$\,^{***}$ & 0.003 & 0.034$\,^{***}$ & 0.037$\,^{***}$ & \\
 & (0.008) & (0.008) & (0.007) & (0.009) & (0.008) & \\
J: jeonse $-$ sale & $-$0.041$\,^{***}$ & $-$0.039$\,^{***}$ & $-$0.034$\,^{***}$ & $-$0.051$\,^{***}$ & $-$0.049$\,^{***}$ & \\
 & (0.010) & (0.010) & (0.009) & (0.011) & (0.011) & \\
Observations: sale & 222,209 & 222,209 & 222,209 & 130,150 & 130,150 &  \\
Observations: jeonse & 644,100 & 374,739 & 141,753 & 391,194 & 232,643 &  \\
\bottomrule
\end{tabular}
\begin{tablenotes}[flushleft]
\footnotesize
\item \textit{Notes:} Coefficients on the hagwon measure from the models of Tables~\ref{tab:baseline}--\ref{tab:grades} (school zone $\times$ month FE, age bins, log area, floor), re-estimated by period or sample. M and N: registry high-school/exam hagwons within 500 m (N adds non-student shops); C and J: business-data cram count (J adds non-student shops). Rows `jeonse $-$ sale' are from the pooled, fully interacted model. Panel A periods (rules and dates in Table~\ref{tab:policy}): before the measures of August 2017; before the mortgage ban on homes above KRW 1.5 billion and the land-transaction permit zones; the ban years; after the ban. Column (6) drops the ten legal dongs designated as permit zones in 2020--21, whole. Panel B (from June 2021, when contracts are flagged new or renewal): jeonse contracts on located complexes that carry the flag; capped renewals used the renewal right; $<$25 yrs keeps buildings under 25 years old in both markets. Full results with shop coefficients: Tables~\ref{tab:app-periods} and~\ref{tab:app-recent}. Standard errors in parentheses, clustered by legal dong (district and dong name; 332 clusters in the full sample). $^{*}$\,$p<0.10$, $^{**}$\,$p<0.05$, $^{***}$\,$p<0.01$.
\end{tablenotes}
\end{threeparttable}
\end{table}

The high-school premium is present in prices and deposits in every period. Before the measures of August 2017 (column 2), the sale and jeonse coefficients are \ResPerPreCurbsMHsSale{} and \ResPerPreCurbsMHsJeonse, with a gap of \ResPerPreCurbsMHsGap{} (\ResPerPreCurbsMHsGapP). Before both the mortgage ban and the permit zones (column 3, January 2015 to November 2019), they are \ResPerPreBanMHsSale{} and \ResPerPreBanMHsJeonse: moving from \DatMedNeisHs{} to \ResHsTarget{} high-school hagwons is associated with prices \ResPerPreBanMHsSalePct\% and deposits \ResPerPreBanMHsJeonsePct\% higher, and the gap is \ResPerPreBanMHsGap{} (\ResPerPreBanMHsGapP). Before the rules, then, the two markets priced high-school hagwons about equally.

A sale premium larger than the jeonse premium appears only from December 2019. The gap is \ResPerBanMHsGap{} during the ban years (\ResPerBanMHsGapP) and \ResPerPostBanMHsGap{} after the ban was lifted (\ResPerPostBanMHsGapP). Sale premia rose rather than fell while the ban was in force (\ResPerBanMHsSale{} against \ResPerPreBanMHsSale{} before), so the ban does not appear to have suppressed the price premium of hagwon districts. Dropping the ten permit-zone dongs entirely (column 6), which removes \ResPermitSaleShare\% of sales and \ResPermitJeonseShare\% of jeonse contracts, gives a sale coefficient of \ResPerNoPermitMHsSale{} and a gap of \ResPerNoPermitMHsGap{} (\ResPerNoPermitMHsGapP). These post-2019 gaps are imprecise, and Section~\ref{sec:results-withincell} shows that they do not appear when the same flat type is compared within a complex, before or after December 2019.

The rules also do not explain the raw jeonse $-$ sale gap of the SEMAS measure. It was already present before August 2017, at \ResPerPreCurbsCHagGap{} (\ResPerPreCurbsCHagGapP), before any of the rules took effect. With non-student shops controlled (specification J), the gap is \ResPerPreCurbsJHagGap{} before August 2017 and widens to \ResPerBanJHagGap{} in the ban years. If the rules restrained sale prices in hagwon districts, any bias would work against a larger sale premium; that sale premia instead rose during the ban suggests the rules did not bind on the hagwon premium.

\subsection{New contracts only and newer buildings}\label{sec:robust-newcontracts}

Once the 2020 renewal right took effect (Section~\ref{sec:background-jeonse}), the jeonse sample includes renewals whose increases are capped at 5\%. Such deposits cannot follow local demand and could flatten the jeonse premium. Contracts are flagged as new or renewal from June 2021, so Panel B of Table~\ref{tab:periods} compares sales in June 2021 to October 2026 with different jeonse samples (Table~\ref{tab:app-recent} gives every coefficient). The high-school gap is \ResRecentAllJMHsGap{} with all jeonse contracts (\ResRecentAllJMHsGapP), \ResRecentNewJMHsGap{} with new contracts only (\ResRecentNewJMHsGapP) and \ResRecentCapJMHsGap{} with capped renewals only (\ResRecentCapJMHsGapP). The cap is therefore not what separates the two premia after 2019. Capped renewals differ in one respect. With the SEMAS count their gap to sales is only \ResRecentCapJCHagGap{} (\ResRecentCapJCHagGapP), against \ResRecentNewJCHagGap{} for new contracts, and their non-student shop coefficient is \ResRecentCapJNShopJeonse, against \ResRecentNewJNShopJeonse{} for new contracts (Table~\ref{tab:app-recent}). The convenience component of deposits is thus visible in contracts set on market terms but hardly in capped renewals; this difference is not interpreted further.

Restricting both markets to buildings under 25 years, whose prices carry little redevelopment value, the sale coefficient rises to \ResRecentAllJYoungMHsSale{} and the gap widens to \ResRecentAllJYoungMHsGap{} (\ResRecentAllJYoungMHsGapP; \ResRecentNewJYoungMHsGap{} with new contracts only, \ResRecentNewJYoungMHsGapP). Redevelopment hopes in old complexes, such as the reconstruction candidates in Daechi-dong and Mok-dong, therefore do not explain the post-2019 sale premium. With non-student shops controlled (specification N), the gaps in these buildings are larger still: \ResRecentAllJYoungNHsGap{} with all jeonse contracts and \ResRecentNewJYoungNHsGap{} with new contracts only (\ResRecentAllJYoungNHsGapP{} and \ResRecentNewJYoungNHsGapP), and both pass the Bonferroni threshold of Section~\ref{sec:robust-multiple}. In newer buildings after mid-2021, then, the cross-complex sale premium for high-school hagwons clearly exceeds the jeonse premium.

Within complexes the pattern is weaker. For the same flat size in the same complex and month, in buildings under 25 years from June 2021 (\ChkCellYoungRecentNCells{} cells), the high-school coefficient on the jeonse-to-price ratio is \ChkCellYoungRecentN{} with shops controlled (SE \ChkCellYoungRecentNSe, \ChkCellYoungRecentNP) and \ChkCellYoungRecentM{} without (\ChkCellYoungRecentMP), against \ChkCellYoungEarlyN{} in the same buildings before June 2021 and \ChkCellOldRecentN{} in older buildings after it (Table~\ref{tab:app-cellvariants}, Panel D). The cross-complex model fitted to the same transactions gives \ChkCellYoungRecentRowsN, so most of the shrinkage from \ResRecentAllJYoungNHsGap{} reflects which complexes trade often enough to form cells, not the cell comparison itself. The difference from older buildings is not significant (\ChkCellYoungRecentDiffNP), and the within-complex coefficient does not pass the multiple-testing correction (Benjamini--Hochberg $q = \ChkFamQCellYoungN$). Part of the cross-complex gap may also reflect complex size, which buyers pay for more than tenants (Section~\ref{sec:robust-access}). An appreciation component in newer buildings after 2021, the second case of Section~\ref{sec:framework}, is therefore possible but not established.

\subsection{Schools, transport and complex size}\label{sec:robust-access}

Hagwon clusters may stand in for other things that families value: nearby schools, subway access, or large complexes built by major firms. Table~\ref{tab:access} adds these controls (Section~\ref{sec:data-density} describes the data).

\begin{table}[!htbp]
\centering
\begin{threeparttable}
\caption{Controls for schools, transport and complex size}
\label{tab:access}
\footnotesize
\setlength{\tabcolsep}{4pt}
\begin{tabular}{lcccc}
\toprule
 & (1) & (2) & (3) & (4) \\
 & Baseline & + Schools, & + Complex size, & + Elite high school, \\
 & & station & builder & stations within 1 km \\
\midrule
\multicolumn{5}{l}{\textit{Panel A. High-school/exam hagwons, specification M}} \\
Log sale price per m$^2$ & 0.056$\,^{***}$ & 0.054$\,^{***}$ & 0.027$\,^{**}$ & 0.025$\,^{**}$ \\
 & (0.016) & (0.015) & (0.012) & (0.012) \\
Log jeonse deposit per m$^2$ & 0.036$\,^{**}$ & 0.036$\,^{**}$ & 0.029$\,^{*}$ & 0.028$\,^{*}$ \\
 & (0.017) & (0.016) & (0.016) & (0.015) \\
Jeonse $-$ sale & $-$0.020 & $-$0.018 & 0.002 & 0.003 \\
 & (0.016) & (0.015) & (0.016) & (0.015) \\
Within complex, flat size and month & 0.002 & 0.002 & 0.009 & 0.010 \\
 & (0.011) & (0.011) & (0.010) & (0.011) \\
6 $\to$ 50 hagwons, sale (\%) & 11.8 & 11.3 & 5.6 & 5.1 \\
6 $\to$ 50 hagwons, jeonse (\%) & 7.3 & 7.4 & 5.9 & 5.7 \\
\addlinespace
\multicolumn{5}{l}{\textit{Panel B. High-school/exam hagwons with non-student shops, specification N}} \\
Log sale price per m$^2$ & 0.055$\,^{***}$ & 0.051$\,^{***}$ & 0.028$\,^{**}$ & 0.026$\,^{**}$ \\
 & (0.016) & (0.015) & (0.012) & (0.012) \\
Log jeonse deposit per m$^2$ & 0.034$\,^{**}$ & 0.035$\,^{**}$ & 0.025$\,^{*}$ & 0.024 \\
 & (0.016) & (0.016) & (0.015) & (0.015) \\
Jeonse $-$ sale & $-$0.021 & $-$0.016 & $-$0.002 & $-$0.002 \\
 & (0.015) & (0.015) & (0.015) & (0.015) \\
Within complex, flat size and month & 0.001 & 0.002 & 0.008 & 0.009 \\
 & (0.011) & (0.011) & (0.011) & (0.011) \\
\addlinespace
\multicolumn{5}{l}{\textit{Panel C. Academic cram hagwons (SEMAS), jeonse $-$ sale}} \\
Specification C & 0.025$\,^{***}$ & 0.024$\,^{***}$ & 0.024$\,^{***}$ & 0.021$\,^{**}$ \\
 & (0.009) & (0.008) & (0.008) & (0.008) \\
Specification J (non-student shops) & $-$0.036$\,^{***}$ & $-$0.034$\,^{***}$ & $-$0.020$\,^{**}$ & $-$0.021$\,^{**}$ \\
 & (0.009) & (0.010) & (0.010) & (0.010) \\
\bottomrule
\end{tabular}
\begin{tablenotes}[flushleft]
\footnotesize
\item \textit{Notes:} Coefficients on log(1 + hagwons within 500 m) from Equation~\eqref{eq:hedonic} (Panels A--C) and Equation~\eqref{eq:withincell} (``within complex'' rows). Column (1) reproduces Tables~\ref{tab:baseline}--\ref{tab:withincell}. Column (2) adds log(1 + metres) to the nearest general middle school, the nearest general high school and the nearest urban-rail station. Column (3) adds log households in the complex and an indicator for the ten largest builders, from the K-apt register (matched for \ChkKaptTxnShare\% of transactions; unmatched complexes enter with a missing-value indicator). Column (4) adds log(1 + metres) to the nearest autonomous or special-purpose high school and log(1 + stations within 1 km). Percent rows: moving from the median complex (\DatMedNeisHs) to 50 high-school/exam hagwons. Standard errors clustered by legal dong in parentheses. * $p<0.10$, ** $p<0.05$, *** $p<0.01$.
\end{tablenotes}
\end{threeparttable}
\end{table}

Distances to schools and stations barely move the hagwon coefficients (column 2). Exam hagwons cluster only moderately near schools and are unrelated to station distance, although station access is itself priced, twice as strongly in deposits as in prices (\ChkAccSizeMDStJeonse{} against \ChkAccSizeMDStSale{} per log meter, column 3). Complex size matters more. With log households and the builder indicator (column 3), the high-school sale coefficient falls by about half, to \ChkAccSizeMHsSale{} (SE \ChkAccSizeMHsSaleSe, \ChkAccSizeMHsSaleP), or \ChkAccSizeMHsSalePct\% for the move from \DatMedNeisHs{} to \ResHsTarget{} high-school hagwons, while the jeonse coefficient changes little, to \ChkAccSizeMHsJeonse{} (\ChkAccSizeMHsJeonseP; \ChkAccSizeMHsJeonsePct\%). Households, not the builder, account for the change: with households alone the sale coefficient is \ChkDecHhMHsSale, with the builder alone \ChkDecBuilderMHsSale. Within a zone and month, log households rises by \ChkAuxHhOnHs{} (SE \ChkAuxHhOnHsSe) per log point of high-school hagwons, and large complexes sell at a premium (\ChkAccSizeMHhSale{} per log household in prices, against \ChkAccSizeMHhJeonse{} in deposits). The halving also occurs among complexes matched to the register (\ChkAccSizeKaptMHsSale{} against \ChkAccBaseKaptMHsSale), so it does not come from the missing-value indicator, and in buildings under 25 years (\ChkDecSizeYoungMHsSale{} against \ChkDecBaseYoungMHsSale), so it does not come from the redevelopment value of large old estates.

Whether complex size should be held fixed depends on why the two go together. If hagwons locate where large complexes supply many pupils, size is a confounder and column (3) is the better estimate. If hagwon districts drew large developments, size absorbs part of the premium. Either way, the premium does not depend on proximity to schools or stations, and with size held fixed it is about equal in the two markets: the jeonse $-$ sale gap is \ChkAccSizeMHsGap{} (SE \ChkAccSizeMHsGapSe). The within-complex coefficient, which already holds the complex fixed, moves only from \ResCellAllMHs{} to \ChkAccSizeCellMHs. Size does not remove the convenience pattern of Section~\ref{sec:results-convenience}: the jeonse $-$ sale gap of the SEMAS count stays at \ChkAccSizeCHagGap{} in specification C and \ChkAccSizeJHagGap{} in specification J (Panel C). Adding distance to the nearest autonomous or special-purpose high school, the only available proxy for school quality, and the number of stations within 1~km changes little (column 4). One by-product bears on the asset-pricing question. Within cells, larger complexes have lower jeonse-to-price ratios (\ChkAccSizeCellMHh{} per log household, SE \ChkAccSizeCellMHhSe): buyers pay for something in large complexes that tenants do not. This is left for future work.

\subsection{Measuring exposure}\label{sec:robust-location}\label{sec:robust-exposure}

Complexes placed only approximately (the block tier of the geocoding; Appendix~\ref{app:geocoding}) can be dropped from every design. The level premia are unchanged (Table~\ref{tab:baseline}, column 5). In the within-complex comparison the high-school coefficient is \ResCellExactMHs{} and the shop coefficient \ResCellExactNShop{} (Table~\ref{tab:app-withincell}), and in the entry-fee comparison the high-school interaction is \ResEFAllFullExactNHs{} (\ResEFAllFullExactNHsP; Table~\ref{tab:app-entryfee}).

Table~\ref{tab:app-exposure} varies the registry measure itself. Three variants address timing and survival. The registry lists only hagwons open in 2026, many of which opened after the contracts in the sample. Counting only hagwons already open on 1 July of each contract's year gives coefficients of \ChkExpMTimeHsSale{} for prices and \ChkExpMTimeHsJeonse{} for deposits, against \ResMHsSale{} and \ResMHsJeonse{} with the 2026 count. Counting only hagwons that opened before 2015, and so predate every contract, gives \ChkExpMPreHsSale{} and \ChkExpMPreHsJeonse, and \ChkExpMPreEarlyHsSale{} and \ChkExpMPreEarlyHsJeonse{} on contracts of 2015--19. Hagwons that followed high prices into a neighborhood therefore do not drive the premium, although every version still omits hagwons that closed before 2026. Weighting hagwons by enrolment capacity rather than counting them gives \ChkExpMCapHsSale{} and \ChkExpMCapHsJeonse{} (log hundreds of places), so the count is not driven by small tutoring rooms. Within 1~km the coefficients are larger and less precise (\ChkExpMKmHsSale{} and \ChkExpMKmHsJeonse), and the high-school and other counts are hard to separate at that radius. With shops also counted within 1~km, the deposit coefficient exceeds the price coefficient (\ChkExpNKmHsJeonse{} against \ChkExpNKmHsSale), but the gap is not significant (\ChkExpNKmHsGapP). This is consistent with a wider radius picking up the general commercial density that renters value more than buyers (Section~\ref{sec:results-convenience}), so the 500~m results are treated as the main evidence. Finally, the registry and shop counts stop at the city boundary. Dropping the complexes within 1~km of it leaves the premia unchanged (\ChkExpMNoEdgeHsSale{} and \ChkExpMNoEdgeHsJeonse). Across these \ChkExpSpecs{} specifications, the high-school sale coefficient is significant at 5\% in every one, and the jeonse coefficient in \ChkExpJeonseNSig.

\subsection{Inference}\label{sec:robust-inference}

The paper clusters standard errors by legal dong. Exposure crosses dong lines, and the densest complexes sit in a few dongs. Table~\ref{tab:inference} reports three alternatives: clustering by school zone, with only \DatSampleZones{} clusters and a wild cluster bootstrap \citep{cameron2015practitioner,webb2023reworking}; square grid clusters of 2 and 4~km, a coarse version of spatially robust errors \citep{conley1999gmm}; and estimates without the complexes of Daechi-dong, the Mok-dong area and Junggye-dong.

\begin{table}[!htbp]
\centering
\begin{threeparttable}
\caption{Inference with fewer, larger clusters and without the densest areas}
\label{tab:inference}
\footnotesize
\setlength{\tabcolsep}{3.5pt}
\begin{tabular}{lccccccc}
\toprule
 & & \multicolumn{4}{c}{$p$-value, clustered by} & \multicolumn{2}{c}{Estimate without} \\
\cmidrule(lr){3-6}\cmidrule(lr){7-8}
 & Estimate & Dong & Zone & Zone, & Grid & Daechi- & Three \\
 & & (332) & (46) & wild & (max) & dong & areas \\
\midrule
High-school/exam, sale (M) & 0.056 & $<$0.001 & 0.003 & 0.005 & 0.006 & 0.057$\,^{***}$ & 0.053$\,^{***}$ \\
High-school/exam, jeonse (M) & 0.036 & 0.034 & 0.071 & 0.095 & 0.046 & 0.028 & 0.022 \\
High-school/exam, jeonse $-$ sale (M) & $-$0.020 & 0.19 & 0.33 & 0.37 & 0.22 & $-$0.029 & $-$0.031 \\
High-school/exam, sale (N) & 0.055 & $<$0.001 & 0.005 & 0.006 & 0.005 & 0.056$\,^{***}$ & 0.052$\,^{***}$ \\
High-school/exam, jeonse (N) & 0.034 & 0.040 & 0.054 & 0.081 & 0.054 & 0.026 & 0.019 \\
Cram hagwons, jeonse (C) & 0.051 & $<$0.001 & 0.004 & 0.018 & 0.001 & 0.047$\,^{***}$ & 0.041$\,^{**}$ \\
Cram hagwons, jeonse $-$ sale (C) & 0.025 & 0.004 & 0.022 & 0.032 & 0.022 & 0.021$\,^{**}$ & 0.024$\,^{**}$ \\
Cram hagwons, jeonse $-$ sale (J) & $-$0.036 & $<$0.001 & 0.001 & 0.004 & 0.001 &  &  \\
Within complex, high-school/exam (M) & 0.002 & 0.85 & 0.88 & 0.89 & 0.87 & $-$0.006 & $-$0.005 \\
Within complex, cram hagwons (C) & 0.014 & 0.043 & 0.075 & 0.12 & 0.099 & 0.010 & 0.014$\,^{**}$ \\
\bottomrule
\end{tabular}
\begin{tablenotes}[flushleft]
\footnotesize
\item \textit{Notes:} Estimates from Tables~\ref{tab:baseline}--\ref{tab:withincell}. $p$-values from $t$ tests with $G-1$ degrees of freedom, clustering by legal dong (332 clusters, as in the paper), by middle-school zone (46) and, in the ``wild'' column, from a restricted wild cluster bootstrap at the zone level (Webb weights, \ChkWildDraws{} draws). ``Grid'' is the largest $p$-value across square grids of 2 km and 4 km cells (about \ChkGridTwoCells{} and \ChkGridFourCells{} occupied cells), each laid out twice, with origins half a cell apart. The last two columns re-estimate without the complexes of Daechi-dong and without those of Daechi-dong, the Mok-dong area and Junggye-dong; stars use zone-clustered $p$-values: * $p<0.10$, ** $p<0.05$, *** $p<0.01$.
\end{tablenotes}
\end{threeparttable}
\end{table}

The high-school sale premium survives all of them. Its bootstrap $p$-value with zone clusters is \ChkInfMHsSalePWild{}, no grid gives more than \ChkInfMHsSalePGridMax, and without the three districts the coefficient is \ChkLooThreeMHsSale. The deposit premium is weaker. Clustering by zone raises its $p$-value to \ChkInfMHsJeonsePZone{} (\ChkInfMHsJeonsePWild{} with the bootstrap), and without Daechi-dong the coefficient falls to \ChkLooDaechiMHsJeonse{} (\ChkLooThreeMHsJeonse{} without all three districts), still positive but no longer significant. Because it is positive in every specification, sample and exposure measure, the evidence is best read as a deposit premium of similar or smaller size than the sale premium, estimated less precisely and concentrated partly in Daechi-dong. The convenience results hold: the SEMAS jeonse $-$ sale gap has a bootstrap $p$-value of \ChkInfCHagGapPWild{}, and its reversal once shops are held fixed (specification J) one of \ChkInfJHagGapPWild. The within-complex high-school coefficient remains a precise zero (bootstrap \ChkInfMHsCellPWild). The small positive within-complex coefficient of the SEMAS count (Table~\ref{tab:withincell}, column 1) does not survive zone clustering (\ChkInfCHagCellPWild).

To gauge selection on unobservables, the degree of proportional selection $\delta$ at which the high-school coefficient would be zero is computed \citep{oster2019unobservable}, treating the flat controls, age bins and other academic hagwons as observed controls and measuring $R^2$ within zone $\times$ month cells. With the bound $R_{\max} = 1.3\tilde{R}$ proposed by \citet{oster2019unobservable}, $\delta$ is \ChkOsterMSaleLow{} for prices and \ChkOsterMJeonseLow{} for deposits: selection on unobservables would have to be \ChkOsterMSaleLow{} times as strong as selection on the observed controls to remove the sale premium, and about as strong to remove the deposit premium. With a stricter bound of $R_{\max} = 2.2\tilde{R}$, $\delta$ falls to \ChkOsterMSaleHigh{} and \ChkOsterMJeonseHigh. The bounds are weak partly because the count of other hagwons, an observed control, absorbs most of the variation in the high-school count. They do not settle the sorting question, but by the usual standard of $\delta > 1$ the sale premium is not fragile.

\begin{figure}[!htbp]
\centering
\includegraphics[width=\textwidth]{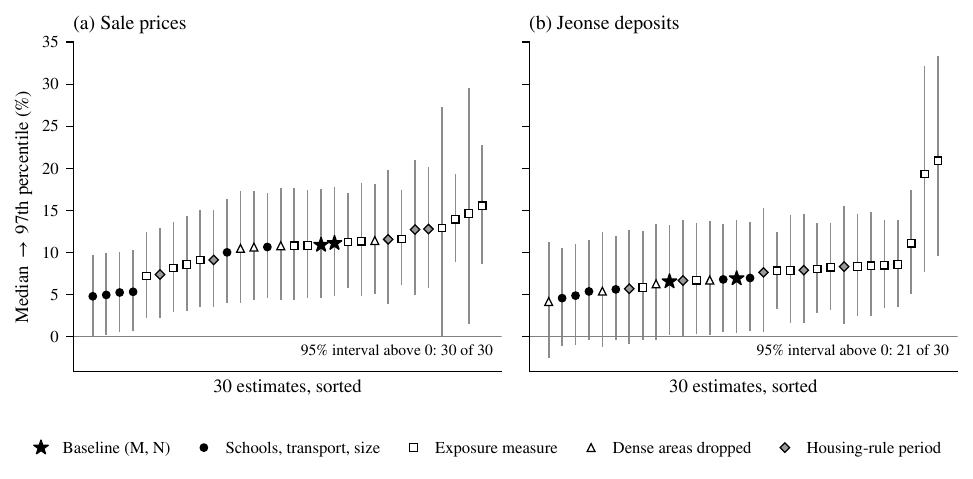}
\caption{The high-school premium across robustness checks}
\label{fig:speccurve}
\begin{minipage}{0.95\textwidth}\footnotesize
\textit{Notes:} Each point is one estimate of the coefficient on a high-school/exam hagwon measure (specifications M or N), sorted within each panel, with its 95\% interval (standard errors clustered by legal dong). Estimates are expressed as the percent change for moving the measure from its median to its 97th percentile across located complexes; for the baseline 500~m registry count this is a move from \FigSpecBaseMove{} hagwons. Stars: baseline M and N (Table~\ref{tab:grades}). Filled circles: controls for schools, stations, complex size and builder (Table~\ref{tab:access}). Open squares: alternative exposure measures (Table~\ref{tab:app-exposure}). Triangles: without Daechi-dong, the Mok-dong area, Junggye-dong or all three (Table~\ref{tab:inference}). Gray diamonds: housing-rule periods and the sample without permit-zone dongs (Table~\ref{tab:periods}).
\end{minipage}
\end{figure}

Figure~\ref{fig:speccurve} collects the high-school premium across the checks of this section: the control sets of Table~\ref{tab:access}, the exposure measures of Table~\ref{tab:app-exposure}, the estimates without the densest areas and the housing-rule periods of Table~\ref{tab:periods}. To make different measures comparable, each estimate is scaled to the move from the median to the 97th percentile of its own measure. The sale premium ranges from \FigSpecSaleMin\% to \FigSpecSaleMax\%, and its 95\% interval excludes zero in all \FigSpecN{} estimates. The deposit premium ranges from \FigSpecJeonseMin\% to \FigSpecJeonseMax\%, and its interval excludes zero in \FigSpecJeonseAbove{} of them; the lowest estimates come without the dense hagwon districts and with complex-size controls.

\subsection{Power and stability of the ratio test}\label{sec:robust-ratio}

\begin{figure}[!htbp]
\centering
\includegraphics[width=0.8\textwidth]{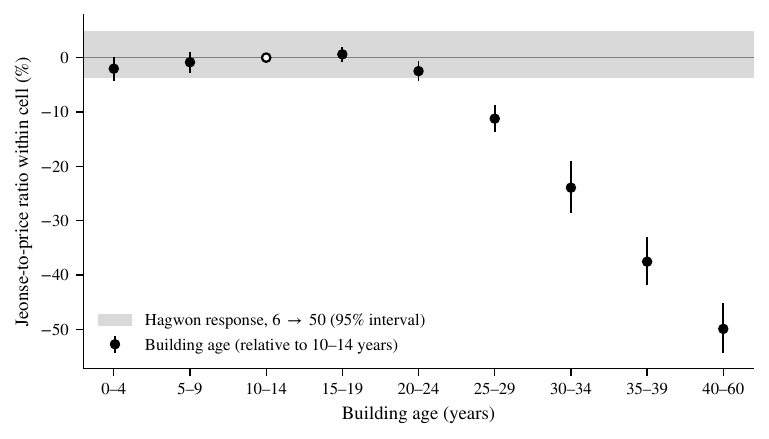}
\caption{The ratio test detects reconstruction value but no hagwon response}
\label{fig:positive-control}
\begin{minipage}{0.8\textwidth}\footnotesize
\textit{Notes:} Black dots: coefficients on building-age indicators in a version of Equation~\eqref{eq:withincell} that replaces the exposure measure with age bins (specification M, reference 10--14 years), shown as $100(e^{b}-1)$ with 95\% intervals clustered by legal dong. The gray band is the 95\% interval for the change in the within-cell jeonse-to-price ratio implied by moving from \DatMedNeisHs{} to \ResHsTarget{} high-school/exam hagwons (Table~\ref{tab:withincell}, column 3).
\end{minipage}
\end{figure}

A null response of the within-complex ratio is informative only if the test can detect an appreciation component where one is known to exist. Old buildings provide such a case: their prices carry the value of reconstruction, which the current lease does not deliver. In the same cells, the jeonse-to-price ratio is \ChkCellPosAgeThirtyMPctAbs\% lower in buildings of 30 years or more than in younger ones (coefficient \ChkCellPosAgeThirtyM, SE \ChkCellPosAgeThirtyMSe), and it falls steadily with age from 25 years on (Figure~\ref{fig:positive-control}; Table~\ref{tab:app-cellvariants}, Panel E). This effect is \ChkCellPosSeRatio{} times the standard error of $\theta$. The 95\% interval for the hagwon response, \ResCellAllMHsPctLo\% to \ResCellAllMHsPctHi\% for the move from \DatMedNeisHs{} to \ResHsTarget{} high-school hagwons, is at most \ChkCellPosPowerShare\% of it. The hagwon coefficient itself is near zero in old and younger buildings alike (\ChkCellPosHsOldM{} and \ChkCellPosHsYoungM).

$\theta$ is also stable across the samples that Section~\ref{sec:strategy-identification} flags (Table~\ref{tab:app-cellvariants}). It is \ChkCellRegOneM{} before the mortgage ban, \ChkCellRegTwoM{} from December 2019 to May 2021 and \ChkCellRegThreeM{} from June 2021 (equality \ChkCellRegEqualMP). The 24 half-year estimates differ (\ChkCellHalfEqualMP), mainly because of two thin half-years in 2022, when few flats were sold: \ChkCellHalfTwentyTwoHOneM{} and \ChkCellHalfTwentyTwoHTwoM{} (without them, \ChkCellHalfEqualNoTwentyTwoMP; Figure~\ref{fig:cell-halfyears}). With new contracts only, from June 2021, $\theta$ is \ChkCellRecentNewM{} (SE \ChkCellRecentNewMSe), so unflagged and capped renewals do not hide a response. By the sale price of the flat, it is \ChkCellTierLowM, \ChkCellTierMidM{} and \ChkCellTierHighM{} for flats up to KRW 0.9 billion, from 0.9 to 1.5 billion and above 1.5 billion (equality \ChkCellTierEqualMP). Lending limits and holding costs that rise with price therefore do not create a ratio response near hagwons, although the top tier is imprecise (SE \ChkCellTierHighMSe). Finally, hagwon access does not change the share of leases written as jeonse rather than monthly rent (\ChkCellShareRecentM, SE \ChkCellShareRecentMSe, from June 2021), so the ratio is not measured on a selected set of leases near hagwons.

\subsection{Education-policy shocks}\label{sec:robust-events}

\begin{figure}[!htbp]
\centering
\includegraphics[width=\textwidth]{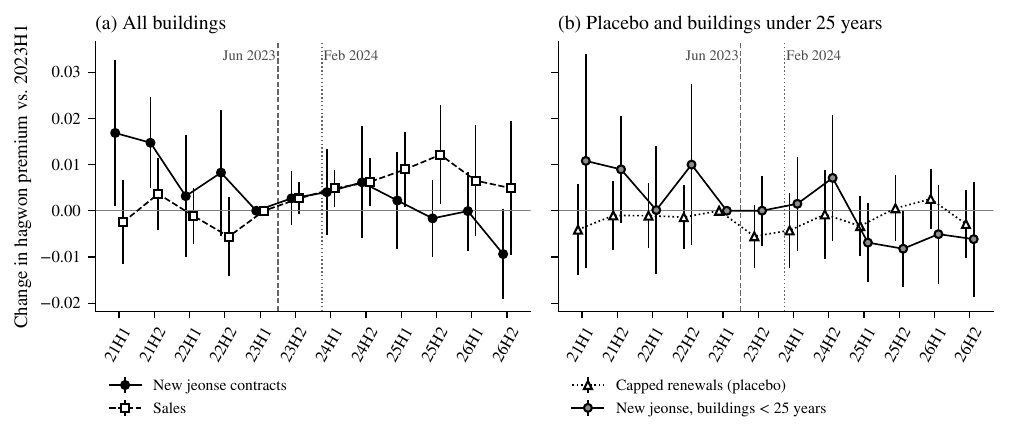}
\caption{The hagwon premium around the 2023--24 education-policy windows}
\label{fig:events}
\begin{minipage}{0.95\textwidth}\footnotesize
\textit{Notes:} Coefficients $\lambda_{\tau}$ from Equation~\eqref{eq:events}: the change in the coefficient on log(1 + academic cram hagwons in the legal dong) relative to 2023H1, for 60--85 m$^2$ flats, with complex and district $\times$ month fixed effects. Exposure: coverage-adjusted, time-varying dong count from the SEMAS data. Bars are 95\% confidence intervals, clustered by legal dong. Dashed line: June 2023 policy window; dotted line: February 2024 policy window (Section~\ref{sec:background-shocks}). 21H1 contains June 2021 only; 26H2 runs to mid-October 2026.
\end{minipage}
\end{figure}

If the premium reflects the stakes of the university entrance race, it might respond to policies that change those stakes. Each of the three windows bundles several pieces of exam-policy news, some of which should raise and others lower the demand for exam preparation (Section~\ref{sec:background-shocks}), so the tests below measure net responses. Table~\ref{tab:events} and Figure~\ref{fig:events} report the event study of Equation~\eqref{eq:events}. Recall that this design uses the dong-level, coverage-adjusted SEMAS count with complex and district $\times$ month effects, not the 500 m registry measure.

\begin{table}[!htbp]
\centering
\begin{threeparttable}
\caption{Education-policy shocks: did the hagwon premium jump?}
\label{tab:events}
\footnotesize
\setlength{\tabcolsep}{3pt}
\begin{tabular}{lcccccc}
\toprule
 & (1) & (2) & (3) & (4) & (5) & (6) \\
 & New & Sales & New + & Capped & New jeonse, & Sales, \\
 & jeonse &  & vol. ren. & renewals & $<$25 yrs & $<$25 yrs \\
\midrule
\multicolumn{7}{l}{\textit{Panel A. New-contract window, June 2021 to October 2026 (reference: 2023H1)}} \\
Jump at Jun.\ 2023 & 0.0027 & 0.0027 & 0.0031 & $-$0.0055 & 0.0000 & 0.0012 \\
 & (0.0030) & (0.0018) & (0.0029) & (0.0035) & (0.0039) & (0.0021) \\
\quad Daechi vs median dong (\%) & 1.2 & 1.2 & 1.4 & $-$2.4 & 0.0 & 0.5 \\
\quad 95\% interval (\%) & [$-$1.4, 3.8] & [$-$0.4, 2.7] & [$-$1.1, 3.9] & [$-$5.3, 0.6] & [$-$3.3, 3.4] & [$-$1.3, 2.3] \\
Jump at Feb.\ 2024 & 0.0013 & 0.0022 & 0.0001 & 0.0013 & 0.0015 & 0.0029$\,^{*}$ \\
 & (0.0043) & (0.0013) & (0.0038) & (0.0032) & (0.0061) & (0.0016) \\
\quad Daechi vs median dong (\%) & 0.6 & 1.0 & 0.1 & 0.6 & 0.7 & 1.3 \\
\quad 95\% interval (\%) & [$-$3.1, 4.4] & [$-$0.2, 2.1] & [$-$3.1, 3.4] & [$-$2.1, 3.3] & [$-$4.5, 6.1] & [$-$0.1, 2.7] \\
Post halves = 0: $F$ & 1.20 & 1.34 & 0.96 & 1.62 & 1.51 & 1.14 \\
\quad $p$-value & 0.300 & 0.231 & 0.458 & 0.131 & 0.163 & 0.337 \\
Pre halves = 0: $F$ & 3.81 & 1.36 & 2.51 & 0.21 & 1.35 & 1.45 \\
\quad $p$-value & 0.005 & 0.248 & 0.042 & 0.934 & 0.251 & 0.217 \\
Hagwons $\times$ winter & $-$0.0018 & 0.0002 & $-$0.0029 & 0.0002 & $-$0.0018 & $-$0.0004 \\
 & (0.0019) & (0.0009) & (0.0018) & (0.0012) & (0.0026) & (0.0011) \\
Observations & 163,576 & 99,455 & 224,244 & 66,586 & 107,253 & 62,990 \\
Dongs (clusters) & 300 & 301 & 303 & 279 & 280 & 280 \\
\midrule
\multicolumn{7}{l}{\textit{Panel B. All contracts, 2015--2026 (reference: 2019H1; November 2019 policy window)}} \\
 & \multicolumn{3}{c}{Sales} & \multicolumn{3}{c}{Jeonse (all contracts)} \\
\cmidrule(lr){2-4}\cmidrule(lr){5-7}
2019H2 $-$ 2019H1 & \multicolumn{3}{c}{0.0048$\,^{**}$ (0.0024) [2.1\%]} & \multicolumn{3}{c}{0.0031 (0.0028) [1.4\%]} \\
2020H1 $-$ 2019H1 & \multicolumn{3}{c}{0.0022 (0.0033) [1.0\%]} & \multicolumn{3}{c}{0.0030 (0.0035) [1.3\%]} \\
\bottomrule
\end{tabular}
\begin{tablenotes}[flushleft]
\footnotesize
\item \textit{Notes:} Panel A: event-study regressions of log value per m$^2$ for 60--85 m$^2$ flats on the hagwon measure interacted with half-year indicators, the measure itself, log area and floor, with complex and district $\times$ month fixed effects. The hagwon measure is the log of one plus the number of academic cram hagwons in the flat's legal dong, from the SEMAS data, adjusted for the database's growing coverage and interpolated between snapshots; it is not the 500 m or registry measure of Tables~\ref{tab:baseline}--\ref{tab:withincell}. Samples: (1) new jeonse contracts; (2) brokered sales; (3) new contracts plus renewals that did not use the renewal right; (4) capped renewals (placebo); (5)--(6) buildings under 25 years. Jumps: 2023H2 $-$ 2023H1 (June 2023 window) and 2024H1 $-$ 2023H2 (February 2024 window; components of each window in Section~\ref{sec:background-shocks}); $F$-tests are joint Wald tests that all half-year coefficients after (post) or before (pre) 2023H1 are zero; winter is December to February. Percent rows scale a coefficient by the gap in exposure between Daechi-dong (733 cram hagwons in December 2018) and the median apartment dong (8), $\ln 734 - \ln 9 = 4.40$. Panel B: the earlier specification on all sales and all jeonse contracts (all sizes) with the fixed December 2018 dong count, complex and district $\times$ month fixed effects; coefficients relative to 2019H1, with percent rows scaled as in Panel A. Standard errors in parentheses, clustered by legal dong. $^{*}$\,$p<0.10$, $^{**}$\,$p<0.05$, $^{***}$\,$p<0.01$.
\end{tablenotes}
\end{threeparttable}
\end{table}

New jeonse contracts show no jump. The change at June 2023 is \ResEvNewJJumpJun{} (SE \ResEvNewJJumpJunSe), which scales to \ResEvNewJJumpJunPct\% for Daechi-dong relative to the median apartment dong (95\% interval \ResEvNewJJumpJunPctLo\% to \ResEvNewJJumpJunPctHi\%). The change at February 2024 is \ResEvNewJJumpFeb{} (SE \ResEvNewJJumpFebSe), or \ResEvNewJJumpFebPct\% (\ResEvNewJJumpFebPctLo\% to \ResEvNewJJumpFebPctHi\%). The post-2023H1 coefficients are jointly insignificant ($F = \ResEvNewJPostF$, \ResEvNewJPostP). The data therefore rule out jumps in new-contract deposits larger than \ResEvNewJJumpJunPctHi\% (June 2023) and \ResEvNewJJumpFebPctHi\% (February 2024) for Daechi-dong relative to a median dong. The fixed December 2018 and December 2022 counts give similar jumps (Table~\ref{tab:app-events}). There is no winter moving-season effect (\ResEvNewJWinter, SE \ResEvNewJWinterSe).

Sales show no significant jump at either date in the full sample (\ResEvSalesJumpJun, \ResEvSalesJumpJunP; \ResEvSalesJumpFeb, \ResEvSalesJumpFebP; in buildings under 25 years the February 2024 jump is \ResEvSalesYoungJumpFeb, \ResEvSalesYoungJumpFebP), but the sale premium drifted up through 2024--25, reaching \ResEvSalesCoefTwentyFiveHTwo{} in the second half of 2025 (SE \ResEvSalesCoefTwentyFiveHTwoSe). The post-period coefficients are jointly insignificant (\ResEvSalesPostP), and the drift halves in buildings under 25 years (\ResEvSalesYoungCoefTwentyFiveHTwo), so it is at most a gradual change, partly in older complexes, rather than a response to either announcement. Before 2023, the new-contract premium was higher in 2021 than in early 2023 (pre-period $F = \ResEvNewJPreF$, \ResEvNewJPreP). This pre-trend comes from buildings 25 years and older: in younger buildings it disappears (\ResEvNewJYoungPreP). It matches the 2021--22 entry-fee pattern in Section~\ref{sec:results-entryfee}. Capped renewals, which cannot respond, are flat before (\ResEvCappedPreP) and after (\ResEvCappedPostP), so the placebo passes.

The November 2019 window predates the new-contract sample. In the earlier specification on all contracts (Table~\ref{tab:events}, Panel B), the jeonse gradient in the first half of 2020 was \ResOldJeonseTwentyHOne{} above its level in the first half of 2019 (SE \ResOldJeonseTwentyHOneSe; \ResOldJeonseTwentyHOnePct\% for Daechi-dong against the median dong), and the sale gradient \ResOldSaleTwentyHOne{} (SE \ResOldSaleTwentyHOneSe). The sale coefficient for the second half of 2019 is \ResOldSaleNineteenHTwo{} (\ResOldSaleNineteenHTwoP), but that half-year lies mostly before the November announcements and the change does not persist into 2020. Because this specification mixes renewals with new contracts, it is a weaker test than the 2023--24 one.

The coarse dong-level measure could hide a response near exam hagwons in particular. The event study is therefore re-estimated with the 500~m registry count of high-school hagwons, complex effects and school zone $\times$ month effects (Figure~\ref{fig:events-registry}). New-contract deposits again show no jump: \ChkEvRegNewJJumpJunPct\% at June 2023 (95\% interval \ChkEvRegNewJJumpJunPctLo\% to \ChkEvRegNewJJumpJunPctHi\%) and \ChkEvRegNewJJumpFebPct\% at February 2024 (\ChkEvRegNewJJumpFebPctLo\% to \ChkEvRegNewJJumpFebPctHi\%) for the move from \DatMedNeisHs{} to \ResHsTarget{} high-school hagwons; with zone clusters the upper bounds are \ChkEvRegNewJJumpJunZonePctHi\% and \ChkEvRegNewJJumpFebZonePctHi\%. Capped renewals show no jump at either date (\ChkEvRegCappedJumpJunP{} and \ChkEvRegCappedJumpFebP). Sales show small positive changes at June 2023 (\ChkEvRegSalesJumpJunPct\%, \ChkEvRegSalesJumpJunP), February 2024 (\ChkEvRegSalesJumpFebPct\%, \ChkEvRegSalesJumpFebP) and November 2019 (\ChkEvRegAllSaleJumpNovPct\%, \ChkEvRegAllSaleJumpNovP). But they sit on a steady rise in the sales gradient that began in 2017 (pre-2019 coefficients jointly \ChkEvRegAllSalePreP), so they are not read as responses to the announcements. The registry count is a 2026 snapshot, measured after the 2019 and 2023 windows.

Taken together, the policy windows did not move the deposit premium measurably, and any change in the sale premium was gradual. Because each window mixes news that pushes demand for exam preparation in opposite directions, a null is consistent either with a premium driven by long-run demand that these policies did not change or with offsetting responses; what the data rule out is a large net jump.

\subsection{Multiple testing}\label{sec:robust-multiple}

The paper reports many tests of related hypotheses. Table~\ref{tab:app-family} lists the \ChkFamTests{} tests on which Sections~\ref{sec:results} and~\ref{sec:robust} base a conclusion, each in the specification the text reports; re-estimates of the same hypothesis under other controls, measures, samples or clusters are not counted again. Of these tests, \ChkFamPassFive{} have $p < 0.05$, \ChkFamPassBonf{} pass the Bonferroni threshold $p < \ChkFamThreshold$, and \ChkFamPassBH{} have Benjamini--Hochberg adjusted $p$-values below 0.05 \citep{benjamini1995controlling}. Combining these corrections with the checks above, the results fall into three groups.

\textit{Robust:} the high-school sale premium (\ResMHsSaleP; zone bootstrap \ChkInfMHsSalePWild), the registry-total premia in both markets, the shop coefficients in the sale equations, the jeonse $-$ sale shop gaps, the reversal of the SEMAS gap once shops are held fixed (\ResJHagGapP), the within-complex shop coefficients (\ResCellAllNShopP), and the shop-controlled post-2021 gaps across complexes in buildings under 25 years (\ResRecentNewJYoungNHsGapP). The shop coefficients in the jeonse equations pass the threshold in specification N but not in J (\ResJShopJeonseP).

\textit{Precise nulls:} the within-complex response of the jeonse-to-price ratio to high-school hagwons (95\% interval \ResCellAllMHsLo{} to \ResCellAllMHsHi), the registry-total jeonse $-$ sale gap (\ResMZeroRegGapP) and the policy jumps in new-contract deposits (upper bounds of \ResEvNewJJumpJunPctHi\% and \ResEvNewJJumpFebPctHi\% for Daechi-dong against a median dong, and of \ChkEvRegNewJJumpJunPctHi\% and \ChkEvRegNewJJumpFebPctHi\% with the registry measure).

\textit{Suggestive:} the high-school jeonse premium (\ResMHsJeonseP; $q = \ChkFamQMHsJeonse$; zone bootstrap \ChkInfMHsJeonsePWild), the SEMAS sale premium (\ResCHagSaleP), the post-2019 cross-complex gaps without shop controls (from \ResRecentNewJYoungMHsGapP{} to \ResPerPostBanMHsGapP), the within-complex gap in newer buildings after 2021 (\ChkCellYoungRecentNP; $q = \ChkFamQCellYoungN$) and the 2021--22 entry-fee interactions (smallest $q$ = \ChkFamQEntryMin). The conclusion that deposits, too, price high-school hagwons does not rest on the full-sample $p$-value alone: the coefficient is positive in every period and contract sample of Table~\ref{tab:periods} and every exposure measure of Table~\ref{tab:app-exposure}, and the registry-total jeonse premium is robust.
\section{Discussion and conclusion}\label{sec:conclusion}

Seoul families pay to live near hagwons, even among flats assigned to the same public schools. The premium attaches to hagwons that prepare high-school students for the university entrance exam and, at least in sale prices, not to those that teach younger children. About half of the sale premium is shared with the size of the complexes near which exam hagwons cluster; with size held fixed, the move from a median complex to one with \ResHsTarget{} such hagwons is associated with prices \ChkAccSizeMHsSalePct\% and deposits \ChkAccSizeMHsJeonsePct\% higher. It is paid by buyers and by jeonse tenants alike, in the same proportion for the same flat, and it was stable across Seoul's housing-rule regimes and through three windows of education-policy news. An apparently larger premium in the lease market belongs to everyday services nearby, which buyers value less than renters, rather than to hagwons.

\paragraph{What the premium buys.} The grade split locates the premium in the competition for university places. That is the margin on which models of status competition in education place the costs of child-rearing in Korea \citep{anderson2013education,kim2024status}, and the premium adds a housing component to those costs that measures of tutoring spending leave out. Because the premium is capitalized, its economic cost to a family is not the premium itself but the return forgone on it while the family lives there, plus the risk of not recovering it on exit. For a jeonse tenant this is explicit: the extra deposit near exam hagwons is returned, and the tenant gives up the return on it. That the two markets price hagwon access in proportion, on average, suggests that owners are not paying for expected gains on top of the current service, although newer buildings after 2021 may be an exception (Section~\ref{sec:robust-newcontracts}), so the housing cost of exam preparation is borne similarly by the families who buy and the families who rent.

\paragraph{Policy.} Restrictions on hagwon hours have not reduced tutoring \citep{choi2016regulating}, and in these data neither the 2023 removal of ``killer questions'' and the accompanying crackdown nor the 2019 and 2024 announcements moved the deposit premium measurably; the sale premium changed only gradually. Each window bundled measures that pushed demand for exam preparation in opposite directions, so the evidence is of a stable net premium rather than of a policy that failed. It is consistent with a premium that reflects persistent demand for exam preparation, which these announcements did not change. China's tutoring restrictions moved competition into school-district housing \citep{jia2025unintended}; in Seoul, where tutoring is legal and dense, the housing market prices access to tutoring itself. Policies aimed at the supply of tutoring may therefore leave the underlying competition, and its housing cost, in place. The paper does not test policies that change the stakes of university entrance.

\paragraph{Implications for real estate research.} Jeonse offers a second capitalized price of the same flat, observed in the same complex and month, and so a direct test of whether an amenity carries an asset-market component beyond its current service. The comparison shows how a cross-complex difference between rent and price capitalization can arise from correlated amenities and from the composition of what is sold and what is rented, and how it can disappear once the same flat type is compared within a complex. Studies that compare rent and price capitalization across locations, in Korea or elsewhere, may face the same confounds \citep{zheng2016much,gabe2022relationship}. This appears to be the first evidence that deposits are higher relative to prices near shops and services within the same complex and month. Across complexes the gap comes from repair and personal services; why buyers discount them, and why buyers pay more than tenants for large complexes, are questions left open.

\paragraph{Limitations and next steps.} The level premia are conditional associations, not causal effects of hagwon supply. By the usual standard the sale premium is not fragile to selection on unobservables \citep{oster2019unobservable}, but the bounds are weak under stricter assumptions, and holding complex size fixed halves it. Three extensions would strengthen the estimates: comparisons within elementary-school catchments, which are smaller than middle-school zones; measures of school quality, proxied here only by distance to selective high schools; and estimates from the opening and closing of hagwons near the same complex over time, which require records of closed hagwons that the registry does not keep and that the quarterly business data provide only with changing coverage. The registry omits hagwons that have closed, although counts restricted to hagwons open at each contract date, or before 2015, give similar premia. Finally, the paper measures a cost of the education race but not its consequences. Linking hagwon access and its housing cost to births across Seoul's neighborhoods is the natural next step.

\section*{Statements and declarations}
\paragraph{Acknowledgments} The author thanks Korebap and other Korean YouTube creators who make videos for English-speaking audiences. Their videos introduced the author to jeonse leases and hagwons.

\paragraph{Funding} The author did not receive any funding from any source.

\paragraph{Competing interests} The author declares no conflict of interest.

\paragraph{Ethics} Not applicable: the study uses public administrative and business records and involves no human participants.

\paragraph{Data availability} The transaction records, business-location data, hagwon registry and school-zone boundaries are public Korean government data; Section~\ref{sec:data} and Appendix~\ref{app:geocoding} give their sources. Code that reproduces every table and figure from the raw files will be deposited in a public repository on publication and is available to the editor on request during review.

\clearpage
\singlespacing

\clearpage
\appendix
\counterwithin{table}{section}
\counterwithin{figure}{section}
\renewcommand{\thetable}{\Alph{section}\arabic{table}}
\renewcommand{\thefigure}{\Alph{section}\arabic{figure}}
\section{Geocoding apartment complexes}\label{app:geocoding}

This appendix describes how the apartment complexes in the MOLIT records are located, how accurate the locations are, and where the method falls short. Table~\ref{tab:geocode} summarizes the tiers and the checks.

\subsection{Why a business-address geocoder}\label{app:geocoding-why}

MOLIT identifies a complex by its district, legal dong, lot number (\textit{jibeon}), road address and name, but gives no coordinates. Three official sources would provide them: the location database of the national road-address service, the national spatial-data portal (VWorld), and Seoul's apartment-complex dataset \citep{seoul2026apartment}. Each provides its files only after Korean identity verification by mobile phone or digital certificate, which could not be completed for this study. The geocoder therefore uses the business data of the Small Enterprise and Market Service \citep{semas2026commercial}. Shops, clinics and tutoring rooms in the commercial buildings of apartment complexes are registered at the complex's own lot number and road address, and every record carries longitude and latitude.

\subsection{Business points and complexes}\label{app:geocoding-points}

The geocoder pools the two complete snapshots of Seoul businesses (December 2018 and June 2026) with the Seoul records of the \DatSnapshots{} education snapshots (December 2015 to June 2026), \DatGeoPointFiles{} files in all. Pooling across years adds businesses that opened or closed between snapshots and so raises the number of addresses with at least one business. After records with coordinates outside Seoul and records on mountain lots (\textit{san}) are dropped, the pool has \DatGeoPointRows{} records of \DatGeoPointBiz{} distinct businesses. Each record carries the lot address (district, legal dong, main and sub lot number), the road address (road name, main and sub building number), the building name and the coordinates. The coordinates are projected to EPSG:5186 (Korea 2000, Central Belt 2010), in meters, with pyproj \citep{snow2026pyproj}.

For each address key, a business counts once, however many snapshots list it. The key's position is the coordinate-wise median of its businesses, and its spread is the median distance of those businesses from that position. The median resists a few misplaced records, and the spread flags keys whose businesses are scattered.

A complex is a unique combination of district, legal dong, lot number and name in the MOLIT records. There are \DatGeoComplexes{} complexes with at least one sale or lease between \DatYearFirst{} and \DatYearLast{}.

\subsection{Match tiers}\label{app:geocoding-tiers}

Each complex is matched to the business keys in four tiers (Table~\ref{tab:geocode}, Panel~A):
\begin{enumerate}
  \item \textit{Lot}: same district, legal dong, main lot number and sub-number.
  \item \textit{Road}: same district, road name and building number.
  \item \textit{Name}: same district and legal dong, and a building name equal to the complex name after removing spaces, punctuation, bracketed text and the word ``apartment'' in Korean or English. The match is accepted only if the businesses lie within \DatGeoNameSpread{}~m of their median (spread).
  \item \textit{Block}: same district, legal dong and main lot number, with any sub-number. The match is accepted only if the spread is at most \DatGeoBlockSpread{}~m. A main lot number can cover a large redeveloped block divided into several sub-lots, so this position is approximate.
\end{enumerate}
When several tiers match, the higher one is used. The lot tier places \DatGeoLotN{} complexes (\DatGeoLotPct\% of complexes and \DatGeoLotTxPct\% of transactions). The road tier adds \DatGeoRoadN{}, the name tier \DatGeoNameN{} and the block tier \DatGeoBlockN{} (\DatGeoBlockTxPct\% of transactions). In total, \DatGeoLocated{} complexes (\DatGeoLocatedPct\%) with \DatGeoLocatedTxPct\% of transactions are located. Block-tier complexes are many but small: they hold \DatGeoBlockPct\% of complexes and only \DatGeoBlockTxPct\% of transactions. A lot-tier position rests on a median of \DatGeoLotPointsMedian{} businesses. Rebuilding the tiers from the raw files reproduces the stored locations exactly (\DatGeoRebuildAgreePct\% agreement).

\subsection{Accuracy checks}\label{app:geocoding-checks}

Without official coordinates there is no full benchmark, so accuracy is assessed with four checks (Table~\ref{tab:geocode}, Panel~B).

\paragraph{Lot versus road.} \DatValLotRoadN{} complexes match on both their lot number and their road address. The two positions are identical (less than 1~m apart) for \DatValLotRoadSamePct\% of them, and only \DatValLotRoadFarPct\% are more than 300~m apart. Both keys often select the same businesses, so this check mainly detects inconsistent address fields rather than errors in the coordinates themselves.

\paragraph{Name versus lot.} For the \DatValLotNameN{} complexes that match on both name and lot, the positions are identical for \DatValLotNameSamePct\%, and the 90th-percentile gap is \DatValLotNamePNinety{}~m.

\paragraph{Block re-location test.} The block tier is the least precise, so it is tested on complexes whose true position is known precisely. Take a lot-matched complex whose main lot also has businesses on other sub-lots (\DatValBlockAllN{} complexes). The test discards the businesses on the complex's own sub-lot and places the complex as the block tier would: at the median of the businesses on the other sub-lots, provided their spread is within \DatGeoBlockSpread{}~m (\DatValBlockN{} complexes). The distance from this position to the lot position has a median of \DatValBlockMedian{}~m and a 90th percentile of \DatValBlockPNinety{}~m; \DatValBlockFarPct\% of complexes are off by more than 300~m. A more conservative version uses each other sub-lot on its own, as if only one neighboring sub-lot had businesses. Over \DatValPairN{} complex and sub-lot pairs from \DatValPairComplexes{} complexes, the median distance is \DatValPairMedian{}~m and the 90th percentile \DatValPairPNinety{}~m. Errors of this size are small relative to a 1~km radius but not relative to 500~m. At the 90th-percentile error, the 500~m disk around the estimated position shares \DatValBlockOverlapPct\% of its area with the disk around the true position (\DatValPairOverlapPct\% in the conservative version). Block-tier locations make up \DatGeoBlockTxPct\% of all transactions and \DatSampleBlockTxPct\% of the analysis sample. Section~\ref{sec:robust} reports the main estimates without them.

\paragraph{Seoul apartment-complex data.} Seoul's open-data portal publishes a public sample of \DatValSeoulSampleRows{} complexes from its apartment-complex dataset, with coordinates \citep{seoul2026apartment}. Of these, \DatValSeoulSampleN{} match one of the located complexes by road address, both in the lot tier. The geocoded positions lie \DatValSeoulSampleDistOne{}~m and \DatValSeoulSampleDistTwo{}~m from Seoul's. So few complexes cannot measure accuracy, but they are the only independent benchmark available. A full comparison would require the complete dataset, which is issued only after identity verification.

\subsection{Coverage}\label{app:geocoding-coverage}

The \DatGeoUnmatchedN{} complexes that cannot be located (\DatGeoUnmatchedPct\% of complexes, \DatGeoUnmatchedTxPct\% of transactions) have no business registered at their address in any snapshot. They are smaller on average than located complexes (a mean of \DatGeoUnmatchedMeanTx{} against \DatGeoLocatedMeanTx{} transactions), although the medians are similar (\DatGeoUnmatchedMedianTx{} against \DatGeoLocatedMedianTx{}). They are not newer: \DatGeoUnmatchedNewPct\% of unmatched and \DatGeoLocatedNewPct\% of located complexes were built in 2018 or later. The share located is similar for the smallest and largest quartiles of complexes by transactions (\DatGeoLocatedSmallPct\% and \DatGeoLocatedLargePct\%). It is lowest in the old city center: \DatGeoLocatedJungguPct\% in Jung-gu and \DatGeoLocatedJongnoPct\% in Jongno-gu, against up to \DatGeoLocatedGuMaxPct\% elsewhere. Because the analysis compares flats within school zones, this geographic gap in coverage mainly reduces the number of comparisons in central zones.

\subsection{Limitations}\label{app:geocoding-limits}

\begin{itemize}
  \item \textit{One point per complex.} The position is the median of the complex's businesses, usually a commercial building inside or beside the complex. Large complexes extend over several hundred meters, so a 500~m count measured from that point is a noisy measure of access for flats on the far side. The records give no building or unit locations that would allow finer placement.
  \item \textit{Inherited coordinates.} Business coordinates are geocoded by the data provider, and the geocoder inherits its errors. The same provider supplies the hagwon points. Errors that are common to nearby records therefore shift a complex and its hagwons together and partly cancel in the counts; independent errors do not.
  \item \textit{Approximate block tier.} Block-tier positions carry the errors measured above.
  \item \textit{Changing complexes.} Pooling snapshots from 2015 to 2026 assumes that a complex keeps its address and footprint. A complex rebuilt during the period may keep its lot number but change its footprint and name.
  \item \textit{Uneven coverage.} Unmatched complexes are concentrated in the city center, so the sample under-represents its small complexes.
  \item \textit{Limited external validation.} The only independent benchmark is the \DatValSeoulSampleN{}-complex public sample. The internal checks show that the method is consistent, not that it is unbiased.
\end{itemize}

\begin{table}[!htbp]
\centering
\begin{threeparttable}
\caption{Locating apartment complexes from business addresses: tiers and accuracy}
\label{tab:geocode}
\footnotesize
\begin{tabular}{@{}l p{6.3cm} rrr@{}}
\toprule
\multicolumn{5}{@{}l}{\textit{Panel A. Match tiers, in order of preference}} \\
 & & & \multicolumn{2}{c}{Share (\%)} \\
\cmidrule(l){4-5}
Tier & Rule & Complexes & Complexes & Transactions \\
\midrule
Lot & \raggedright Same district, legal dong and lot number (main and sub-number) & 4,688 & 46.3 & 86.6 \tabularnewline
Road & \raggedright Same district, road name and building number & 148 & 1.5 & 0.8 \tabularnewline
Name & \raggedright Same district and legal dong; building name equals the normalized complex name; points within 100~m of their median & 206 & 2.0 & 0.9 \tabularnewline
Block & \raggedright Same district, legal dong and main lot number, any sub-number; points within 150~m of their median (approximate) & 3,687 & 36.4 & 5.0 \tabularnewline
Unmatched & \raggedright No business registered at the complex's address & 1,387 & 13.7 & 6.7 \tabularnewline
\addlinespace
Located & \raggedright Lot, road, name or block & 8,729 & 86.3 & 93.3 \tabularnewline
All complexes & & 10,116 & 100.0 & 100.0 \\
\midrule
\multicolumn{5}{@{}l}{\textit{Panel B. Accuracy checks}} \\
Check & Comparison & Complexes & \multicolumn{2}{l}{Result} \\
\midrule
Lot vs road & \raggedright Positions from the lot-number and road-address keys, complexes matched both ways & 4,580 & \multicolumn{2}{p{3.3cm}}{\raggedright 90\% identical ($<$1~m); 0.5\% over 300~m apart} \tabularnewline
Name vs lot & \raggedright Positions from the building-name and lot-number keys & 3,253 & \multicolumn{2}{p{3.3cm}}{\raggedright 90\% identical ($<$1~m); 90th percentile 1~m} \tabularnewline
Block re-location & \raggedright Lot-matched complexes placed only from the other sub-lots of their main lot, as the block tier would place them & 2,458 & \multicolumn{2}{p{3.3cm}}{\raggedright Median 66~m; 90th percentile 195~m} \tabularnewline
Block, one sub-lot & \raggedright As above, but from each other sub-lot alone (25,336 pairs) & 2,503 & \multicolumn{2}{p{3.3cm}}{\raggedright Median 87~m; 90th percentile 302~m} \tabularnewline
Seoul public sample & \raggedright The 5 complexes in the public sample of Seoul's apartment-complex dataset, matched by road address & 2 & \multicolumn{2}{p{3.3cm}}{\raggedright 19~m and 103~m} \tabularnewline
\bottomrule
\end{tabular}
\begin{tablenotes}[flushleft]\footnotesize
\item \textit{Notes:} Complexes are MOLIT apartment complexes (district, legal dong, lot number and name) with at least one sale or lease, 2015--2026. Each complex takes the median position of the businesses registered at its address in 16 Small Enterprise and Market Service snapshots (two complete Seoul snapshots and 14 education snapshots; each business counted once per address). When several tiers match, the higher tier is used. Transaction shares count all sales and leases. Distances are straight-line distances in EPSG:5186. Appendix~\ref{app:geocoding} describes each check.
\end{tablenotes}
\end{threeparttable}
\end{table}
\section{Classifying hagwons by grade}\label{app:grades}

The grade split in Section~\ref{sec:results} separates hagwons that prepare students for high school and university entrance from those that teach younger children. The business data cannot make this split, because business names rarely state the grade level. The classification therefore uses the hagwons in the official registry. This appendix sets out the rules, the resulting classes and their limitations. Table~\ref{tab:gradeclass} gives the counts.

\subsection{Registry records}\label{app:grades-registry}

The NEIS registry \citep{neis2026hagwon} lists every registered hagwon and tutoring room with its name, opening date, capacity, field, course list, posted fees and road address. The Seoul snapshot of 12 October 2026 has \DatNeisSeoulAll{} records, all with the status ``open''. The classification keeps the \DatNeisAcad{} records (\DatNeisAcadPct\%) in the academic field, exam preparation and supplementary study: \DatNeisAcadHagwon{} hagwons and \DatNeisAcadRoom{} tutoring rooms. The other fields cover, among others, the arts, foreign languages, vocational training and reading rooms.

The registry has no coordinates. The road address of \DatNeisKeyedPct\% of academic records is standardized into a key of district, road name and building number, the same key as the road tier of the geocoder (Appendix~\ref{app:geocoding}). Each hagwon is placed at the median position of the business-data records with that key. This places \DatNeisPlacedPct\% of academic hagwons (\DatNeisPlacedN{}); the lowest rate in any district is \DatNeisPlacedMinGuPct\%. Business records at the same address almost always share coordinates (median spread \DatNeisPlacedSpread{}~m). The placement rate is nearly the same across grade classes (Table~\ref{tab:gradeclass}), so placement does not tilt the split.

\subsection{Classification rules}\label{app:grades-rules}

A hagwon's class comes from regular-expression cues found in three text fields: its name, its course list and its posted fees. The fees field often lists classes by grade with a price for each. There are three kinds of cue (the full lists are in the notes to Table~\ref{tab:gradeclass}):
\begin{itemize}
  \item \textit{High school or exam}: high school, grades 10 to 12, the College Scholastic Ability Test (CSAT), admissions, repeat test-takers, essay tests, pre-high-school courses, school grades, mock exams and medical school.
  \item \textit{Middle school}: middle school, grades 7 to 9 and pre-middle-school courses.
  \item \textit{Elementary}: elementary school, grades 1 to 6, preschool, children, kids and junior.
\end{itemize}
A hagwon can carry several cues. The analysis uses two classes. \textit{High-school/exam} hagwons carry at least one high-school or exam cue: \DatGradeHsN{} hagwons, or \DatGradeHsPct\%. \textit{Other} hagwons carry no such cue: \DatGradeOtherN{} hagwons, or \DatGradeOtherPct\%. Middle-school cues appear in \DatGradeMidPct\% of records and elementary cues in \DatGradeElemPct\%. Of the high-school/exam hagwons, \DatGradeHsAlsoLowerPct\% also carry a middle-school or elementary cue; many teach both middle- and high-school students, and they stay in the high-school class. The rule thus asks whether a hagwon serves the exam years at all, not whether it serves them only.

\subsection{Reliance on text cues}\label{app:grades-cues}

The classification depends on text that many records lack. Only \DatGradeAnyCuePct\% of academic records carry any grade cue. The remaining \DatGradeNoCuePct\% (\DatGradeNoCueN{} records) carry none and fall into the ``other'' class by default. The name alone carries a cue in only \DatGradeNameCuePct\% of records; for \DatGradeTextOnlyPct\% the cue appears only in the course list or the fees. The course list is filled in for \DatNeisCourseFilledPct\% of records and the fees for only \DatNeisFeeFilledPct\%. Fees contribute most of the middle-school and elementary cues (Table~\ref{tab:gradeclass}).

The ``other'' class is therefore a mix of hagwons for younger children and unlabeled hagwons, some of which probably teach high-school students. Misclassification of this kind moves part of the high-school count into the ``other'' count. It narrows the contrast between the two classes rather than creating it. A premium carried by high-school hagwons alone, with no premium for ``other'' hagwons, is thus unlikely to be an artifact of the cues. The two counts are also strongly correlated across complexes (\DatCorrHsOther{} in logs, within 500~m), which leaves limited independent variation and widens the standard errors of the split.

\subsection{Agreement with the business data}\label{app:grades-agreement}

The registry and the business data measure hagwon access similarly. The registry lists \DatNeisAcad{} academic hagwons and tutoring rooms in Seoul; the business data list \DatSbizCramTwentyFive{} cram hagwons in December 2025 and \DatSbizCramTwentySix{} in June 2026. Across complexes, log registry counts within 500~m correlate at \DatCorrNeisSbiz{} with business-data counts in December 2025 and at \DatCorrNeisSbizEighteen{} with those in December 2018. The high-school/exam count alone correlates at \DatCorrHsSbizEighteen{} with the December 2018 business count. The registry's grade split thus refines, rather than replaces, the location pattern in the main measure.

\subsection{Survivorship}\label{app:grades-survivorship}

The registry is a snapshot: it lists only hagwons open on 12 October 2026 and records no closures. Opening dates are available for \DatNeisOpenedKnownPct\% of academic records. Of these hagwons, \DatNeisOpenedAfterEighteenPct\% opened after December 2018 (\DatNeisHsOpenedAfterEighteenPct\% of high-school/exam hagwons), and only \DatNeisOpenedBeforeFifteenPct\% opened before 2015; the median opening year is \DatNeisOpenedMedianYear{}. Applying the 2026 counts to prices from 2015 onward therefore includes hagwons that did not yet exist and omits hagwons that have since closed. If hagwons entered where housing demand later rose, the registry measure partly reflects that later demand, and early-period estimates are the most exposed to this. The period splits in Section~\ref{sec:robust} find the high-school premium already in 2015--2019. A history rebuilt from opening dates would still contain only survivors. Regular snapshots of the registry from 2026 onward would build a proper panel of openings and closures.

\begin{table}[!htbp]
\centering
\begin{threeparttable}
\caption{Registry hagwons by the grade level they teach}
\label{tab:gradeclass}
\small
\begin{tabular}{@{}l rr rrr r@{}}
\toprule
 & & & \multicolumn{3}{c}{Cue found in} & \\
\cmidrule(lr){4-6}
Grade cue & Hagwons & \% & Name & Course list & Fees & Placed (\%) \\
\midrule
\multicolumn{7}{@{}l}{\textit{Panel A. Cues (a hagwon can carry several)}} \\
\quad High school / exam & 5,278 & 37 & 1,557 & 3,052 & 2,281 & 98.8 \\
\quad Middle school & 2,938 & 21 & 84 & 532 & 2,787 & 98.7 \\
\quad Elementary & 2,905 & 21 & 97 & 2,310 & 2,559 & 98.7 \\
\quad Any grade cue & 6,551 & 46 & 1,731 & \multicolumn{2}{c}{4,820 only here} & 98.8 \\
\quad No grade cue & 7,595 & 54 & & & & 99.2 \\
\addlinespace
\multicolumn{7}{@{}l}{\textit{Panel B. Classes used in the analysis}} \\
\quad High-school/exam & 5,278 & 37 & & & & 98.8 \\
\quad Other (no high-school cue) & 8,868 & 63 & & & & 99.1 \\
\midrule
All academic hagwons & 14,146 & 100 & & & & 99.0 \\
\bottomrule
\end{tabular}
\begin{tablenotes}[flushleft]\footnotesize
\item \textit{Notes:} Seoul records in the NEIS hagwon registry (snapshot of 12 October 2026) whose field is academic (exam preparation, qualification and supplementary study); the count includes 8,477 hagwons and 5,669 tutoring rooms (\textit{gyoseupso}). A cue is a regular-expression match in the hagwon's name, its course list or its posted fees, which often name classes. Cues are written here in romanization. High school / exam: \textit{godeung} (high school), \textit{go1}--\textit{go3} (grades 10--12), \textit{suneung} (College Scholastic Ability Test), \textit{ipsi} (admissions), \textit{jaesu} and \textit{N-su} (repeat test-takers), \textit{nonsul} (essay test), \textit{yebi-go} (pre-high school), \textit{naesin} (school grades), \textit{moui gosa} (mock exams), \textit{uidae} (medical school). Middle school: \textit{jungdeung}, \textit{junghak}, \textit{jung1}--\textit{jung3}, \textit{yebi-jung} (pre-middle school). Elementary: \textit{chodeung}, \textit{cho1}--\textit{cho6}, \textit{yua} (preschool), \textit{eorini} (children), \textit{kijeu} (kids), \textit{junieo} (junior). The three cue columns count matches in each field and overlap; ``only here'' counts hagwons whose cue appears in the course list or fees but not in the name. Placed: share located at the median position of the businesses with the same road address.
\end{tablenotes}
\end{threeparttable}
\end{table}
\section{Additional results}\label{app:additional}

This appendix documents the business-density controls and reports the full versions of the robustness tables in Sections~\ref{sec:results} and~\ref{sec:robust}. All tables are generated by the same script as the main tables, from the same estimates.

\subsection{Business-density controls}\label{app:additional-shops}

Table~\ref{tab:app-shopdefs} defines the controls used in specifications H--K and N. All come from the December 2018 full snapshot of the SEMAS commercial-district data, counted within 500 m of each complex, and enter as $\ln(1 + \text{count})$. The non-student control of specifications J and N keeps three major categories whose location should not depend on students: estate agents, health care, and repairs and personal services. The non-food control of specification K keeps all businesses except education, food, retail and arts, sports and leisure; the last category contains PC rooms, karaoke rooms and study caf\'es, which plausibly serve students. Because the full snapshots cover Seoul only, complexes near the city boundary have some neighbors missing from these counts; the hagwon and other-education counts include Gyeonggi businesses.

\begin{table}[!htbp]
\centering
\begin{threeparttable}
\caption{Definitions of the business-density controls}
\label{tab:app-shopdefs}
\footnotesize
\setlength{\tabcolsep}{4pt}
\begin{tabular}{p{3.3cm}p{1.1cm}p{8.2cm}}
\toprule
Measure & Used in & Business categories counted within 500 m \\
\midrule
Cram hagwons (SEMAS) & A--K & Education businesses whose sub-category contains \textit{ipsi} (entrance exam), i.e.\ \textit{ipsi$\cdot$gyogwa hagwon}, Dec.\ 2018 \\
Other education & I & All other education businesses: music, art, yoga and pilates, taekwondo and martial arts, vocational, language, education support \\
Non-education businesses & H & All businesses outside the education major category \\
Non-student shops & J, N & Estate agents (\textit{budongsan}), health care (\textit{bogeon-uiryo}), repairs and personal services (\textit{suri-gaein}) \\
Non-food shops & K & All except education, food (\textit{eumsik}), retail (\textit{somae}) and arts, sports and leisure (\textit{yesul-seupocheu}), which includes PC rooms, karaoke and study caf\'es \\
Registry academic hagwons & M0--N & NEIS hagwon registry, academic field (entrance and qualification exams, supplementary lessons), Oct.\ 2026; split into high-school/exam and other (Appendix~\ref{app:grades}) \\
\bottomrule
\end{tabular}
\begin{tablenotes}[flushleft]
\footnotesize
\item \textit{Notes:} Business counts are from the December 2018 full snapshot of the SEMAS commercial-district data (Seoul only); cram and other-education counts also include Gyeonggi businesses near the city edge. Major categories as named (in romanized form) in the source.
\end{tablenotes}
\end{threeparttable}
\end{table}

\subsection{Period split and contract types: all coefficients}\label{app:additional-periods}

Table~\ref{tab:app-periods} reports, for each period and for the sample without permit-zone dongs, the hagwon and shop coefficients of specifications C, J, M and N in both markets, together with their jeonse $-$ sale differences. Table~\ref{tab:app-recent} does the same for June 2021 to October 2026 with different jeonse samples.

The sign pattern of the shop coefficients holds in every period: negative in prices and positive in deposits. The deposit coefficient is smaller before 2020 and not always significant on its own, but the jeonse $-$ sale difference is significant in every period. With the registry split (specification N), the shop coefficient is \ResPerPreCurbsNShopSale{} in prices and \ResPerPreCurbsNShopJeonse{} in deposits before August 2017, and \ResPerBanNShopSale{} and \ResPerBanNShopJeonse{} during the ban years. The difference is largest during the ban years (\ResPerBanNShopGap) and smallest before August 2017 (\ResPerPreCurbsNShopGap). In Table~\ref{tab:app-recent}, the shop coefficient is \ResRecentNewJNShopJeonse{} for new jeonse contracts but only \ResRecentCapJNShopJeonse{} for capped renewals. Capped deposits, which can rise by at most 5\%, show almost no convenience premium, just as they show no response to the policy shocks in Section~\ref{sec:robust-events}.

\begin{table}[!htbp]
\centering
\begin{threeparttable}
\caption{Period split and permit-zone dongs: all specifications}
\label{tab:app-periods}
\scriptsize
\setlength{\tabcolsep}{3pt}
\begin{tabular}{lcccccc}
\toprule
 & \multicolumn{3}{c}{Hagwon measure} & \multicolumn{3}{c}{Non-student shops} \\
\cmidrule(lr){2-4}\cmidrule(lr){5-7}
 & Sale & Jeonse & J $-$ S & Sale & Jeonse & J $-$ S \\
\midrule
\multicolumn{7}{l}{\textit{Specification C: SEMAS cram count}} \\
Full, 2015-01 to 2026-10 & 0.026$\,^{**}$ & 0.051$\,^{***}$ & 0.025$\,^{***}$ &  &  &  \\
 & (0.012) & (0.014) & (0.009) &  &  &  \\
2015-01 to 2017-07 & 0.020$\,^{*}$ & 0.046$\,^{***}$ & 0.026$\,^{**}$ &  &  &  \\
 & (0.011) & (0.014) & (0.012) &  &  &  \\
2015-01 to 2019-11 & 0.022$\,^{*}$ & 0.045$\,^{***}$ & 0.022$\,^{**}$ &  &  &  \\
 & (0.012) & (0.014) & (0.009) &  &  &  \\
2019-12 to 2022-11 & 0.036$\,^{***}$ & 0.052$\,^{***}$ & 0.016 &  &  &  \\
 & (0.012) & (0.013) & (0.012) &  &  &  \\
2022-12 to 2026-10 & 0.030$\,^{**}$ & 0.055$\,^{***}$ & 0.025$\,^{***}$ &  &  &  \\
 & (0.013) & (0.013) & (0.008) &  &  &  \\
Permit-zone dongs dropped & 0.035$\,^{***}$ & 0.054$\,^{***}$ & 0.019$\,^{**}$ &  &  &  \\
 & (0.012) & (0.014) & (0.010) &  &  &  \\
\addlinespace
\multicolumn{7}{l}{\textit{Specification J: C + non-student shops}} \\
Full, 2015-01 to 2026-10 & 0.061$\,^{***}$ & 0.025$\,^{**}$ & $-$0.036$\,^{***}$ & $-$0.080$\,^{***}$ & 0.053$\,^{***}$ & 0.132$\,^{***}$ \\
 & (0.011) & (0.012) & (0.009) & (0.016) & (0.018) & (0.015) \\
2015-01 to 2017-07 & 0.046$\,^{***}$ & 0.038$\,^{***}$ & $-$0.008 & $-$0.062$\,^{***}$ & 0.018 & 0.080$\,^{***}$ \\
 & (0.011) & (0.012) & (0.011) & (0.014) & (0.018) & (0.015) \\
2015-01 to 2019-11 & 0.052$\,^{***}$ & 0.032$\,^{***}$ & $-$0.020$\,^{**}$ & $-$0.069$\,^{***}$ & 0.028 & 0.097$\,^{***}$ \\
 & (0.011) & (0.012) & (0.010) & (0.015) & (0.019) & (0.015) \\
2019-12 to 2022-11 & 0.083$\,^{***}$ & 0.023$\,^{*}$ & $-$0.059$\,^{***}$ & $-$0.105$\,^{***}$ & 0.060$\,^{***}$ & 0.166$\,^{***}$ \\
 & (0.012) & (0.012) & (0.013) & (0.016) & (0.016) & (0.017) \\
2022-12 to 2026-10 & 0.065$\,^{***}$ & 0.023$\,^{*}$ & $-$0.043$\,^{***}$ & $-$0.077$\,^{***}$ & 0.065$\,^{***}$ & 0.142$\,^{***}$ \\
 & (0.013) & (0.012) & (0.009) & (0.019) & (0.017) & (0.015) \\
Permit-zone dongs dropped & 0.062$\,^{***}$ & 0.014 & $-$0.048$\,^{***}$ & $-$0.061$\,^{***}$ & 0.079$\,^{***}$ & 0.140$\,^{***}$ \\
 & (0.012) & (0.013) & (0.010) & (0.016) & (0.018) & (0.015) \\
\addlinespace
\multicolumn{7}{l}{\textit{Specification M: registry high-school/exam (other academic also included)}} \\
Full, 2015-01 to 2026-10 & 0.056$\,^{***}$ & 0.036$\,^{**}$ & $-$0.020 &  &  &  \\
 & (0.016) & (0.017) & (0.016) &  &  &  \\
2015-01 to 2017-07 & 0.038$\,^{***}$ & 0.042$\,^{**}$ & 0.005 &  &  &  \\
 & (0.013) & (0.017) & (0.016) &  &  &  \\
2015-01 to 2019-11 & 0.046$\,^{***}$ & 0.040$\,^{**}$ & $-$0.006 &  &  &  \\
 & (0.014) & (0.016) & (0.015) &  &  &  \\
2019-12 to 2022-11 & 0.058$\,^{***}$ & 0.029$\,^{*}$ & $-$0.029 &  &  &  \\
 & (0.019) & (0.017) & (0.020) &  &  &  \\
2022-12 to 2026-10 & 0.064$\,^{***}$ & 0.039$\,^{**}$ & $-$0.025 &  &  &  \\
 & (0.019) & (0.018) & (0.017) &  &  &  \\
Permit-zone dongs dropped & 0.064$\,^{***}$ & 0.034$\,^{**}$ & $-$0.030$\,^{**}$ &  &  &  \\
 & (0.017) & (0.017) & (0.015) &  &  &  \\
\addlinespace
\multicolumn{7}{l}{\textit{Specification N: M + non-student shops}} \\
Full, 2015-01 to 2026-10 & 0.055$\,^{***}$ & 0.034$\,^{**}$ & $-$0.021 & $-$0.070$\,^{***}$ & 0.061$\,^{***}$ & 0.131$\,^{***}$ \\
 & (0.016) & (0.016) & (0.015) & (0.016) & (0.018) & (0.014) \\
2015-01 to 2017-07 & 0.037$\,^{***}$ & 0.042$\,^{**}$ & 0.005 & $-$0.054$\,^{***}$ & 0.028 & 0.081$\,^{***}$ \\
 & (0.013) & (0.017) & (0.016) & (0.014) & (0.018) & (0.015) \\
2015-01 to 2019-11 & 0.045$\,^{***}$ & 0.040$\,^{**}$ & $-$0.005 & $-$0.059$\,^{***}$ & 0.038$\,^{**}$ & 0.097$\,^{***}$ \\
 & (0.014) & (0.016) & (0.015) & (0.015) & (0.019) & (0.015) \\
2019-12 to 2022-11 & 0.055$\,^{***}$ & 0.028$\,^{*}$ & $-$0.026 & $-$0.090$\,^{***}$ & 0.070$\,^{***}$ & 0.161$\,^{***}$ \\
 & (0.019) & (0.016) & (0.019) & (0.016) & (0.017) & (0.017) \\
2022-12 to 2026-10 & 0.065$\,^{***}$ & 0.034$\,^{**}$ & $-$0.031$\,^{*}$ & $-$0.065$\,^{***}$ & 0.071$\,^{***}$ & 0.136$\,^{***}$ \\
 & (0.019) & (0.017) & (0.016) & (0.019) & (0.016) & (0.014) \\
Permit-zone dongs dropped & 0.061$\,^{***}$ & 0.034$\,^{**}$ & $-$0.027$\,^{**}$ & $-$0.050$\,^{***}$ & 0.084$\,^{***}$ & 0.134$\,^{***}$ \\
 & (0.017) & (0.016) & (0.013) & (0.016) & (0.017) & (0.014) \\
\addlinespace
\bottomrule
\end{tabular}
\begin{tablenotes}[flushleft]
\footnotesize
\item \textit{Notes:} Full version of Table~\ref{tab:periods}, Panel A, with the non-student shop coefficients. J $-$ S: jeonse minus sale from the pooled, fully interacted model. Standard errors in parentheses, clustered by legal dong (district and dong name; 332 clusters in the full sample). $^{*}$\,$p<0.10$, $^{**}$\,$p<0.05$, $^{***}$\,$p<0.01$.
\end{tablenotes}
\end{threeparttable}
\end{table}
\begin{table}[!htbp]
\centering
\begin{threeparttable}
\caption{June 2021 to October 2026: sales against new, renewal and younger-building jeonse}
\label{tab:app-recent}
\scriptsize
\setlength{\tabcolsep}{3pt}
\begin{tabular}{lcccccc}
\toprule
 & \multicolumn{3}{c}{Hagwon measure} & \multicolumn{3}{c}{Non-student shops} \\
\cmidrule(lr){2-4}\cmidrule(lr){5-7}
 & Sale & Jeonse & J $-$ S & Sale & Jeonse & J $-$ S \\
\midrule
\multicolumn{7}{l}{\textit{Specification C: SEMAS cram count}} \\
All jeonse & 0.029$\,^{**}$ & 0.052$\,^{***}$ & 0.024$\,^{***}$ &  &  &  \\
 & (0.013) & (0.013) & (0.008) &  &  &  \\
New contracts only & 0.029$\,^{**}$ & 0.055$\,^{***}$ & 0.026$\,^{***}$ &  &  &  \\
 & (0.013) & (0.012) & (0.008) &  &  &  \\
Capped renewals only & 0.029$\,^{**}$ & 0.032$\,^{***}$ & 0.003 &  &  &  \\
 & (0.013) & (0.011) & (0.007) &  &  &  \\
All jeonse, $<$25 yrs & 0.031$\,^{*}$ & 0.065$\,^{***}$ & 0.034$\,^{***}$ &  &  &  \\
 & (0.016) & (0.016) & (0.009) &  &  &  \\
New only, $<$25 yrs & 0.031$\,^{*}$ & 0.067$\,^{***}$ & 0.037$\,^{***}$ &  &  &  \\
 & (0.016) & (0.015) & (0.008) &  &  &  \\
\addlinespace
\multicolumn{7}{l}{\textit{Specification J: C + non-student shops}} \\
All jeonse & 0.069$\,^{***}$ & 0.028$\,^{**}$ & $-$0.041$\,^{***}$ & $-$0.086$\,^{***}$ & 0.050$\,^{***}$ & 0.136$\,^{***}$ \\
 & (0.013) & (0.012) & (0.010) & (0.019) & (0.016) & (0.014) \\
New contracts only & 0.069$\,^{***}$ & 0.030$\,^{**}$ & $-$0.039$\,^{***}$ & $-$0.086$\,^{***}$ & 0.051$\,^{***}$ & 0.137$\,^{***}$ \\
 & (0.013) & (0.012) & (0.010) & (0.019) & (0.015) & (0.014) \\
Capped renewals only & 0.069$\,^{***}$ & 0.035$\,^{***}$ & $-$0.034$\,^{***}$ & $-$0.086$\,^{***}$ & $-$0.007 & 0.079$\,^{***}$ \\
 & (0.013) & (0.009) & (0.009) & (0.019) & (0.014) & (0.012) \\
All jeonse, $<$25 yrs & 0.072$\,^{***}$ & 0.020 & $-$0.051$\,^{***}$ & $-$0.079$\,^{***}$ & 0.079$\,^{***}$ & 0.157$\,^{***}$ \\
 & (0.016) & (0.015) & (0.011) & (0.023) & (0.017) & (0.015) \\
New only, $<$25 yrs & 0.072$\,^{***}$ & 0.023 & $-$0.049$\,^{***}$ & $-$0.079$\,^{***}$ & 0.077$\,^{***}$ & 0.156$\,^{***}$ \\
 & (0.016) & (0.014) & (0.011) & (0.023) & (0.017) & (0.016) \\
\addlinespace
\multicolumn{7}{l}{\textit{Specification M: registry high-school/exam (other academic also included)}} \\
All jeonse & 0.066$\,^{***}$ & 0.037$\,^{**}$ & $-$0.029$\,^{*}$ &  &  &  \\
 & (0.019) & (0.017) & (0.016) &  &  &  \\
New contracts only & 0.066$\,^{***}$ & 0.036$\,^{**}$ & $-$0.030$\,^{*}$ &  &  &  \\
 & (0.019) & (0.017) & (0.016) &  &  &  \\
Capped renewals only & 0.066$\,^{***}$ & 0.040$\,^{***}$ & $-$0.025 &  &  &  \\
 & (0.019) & (0.015) & (0.016) &  &  &  \\
All jeonse, $<$25 yrs & 0.086$\,^{***}$ & 0.040$\,^{*}$ & $-$0.046$\,^{**}$ &  &  &  \\
 & (0.022) & (0.021) & (0.020) &  &  &  \\
New only, $<$25 yrs & 0.086$\,^{***}$ & 0.039$\,^{*}$ & $-$0.047$\,^{**}$ &  &  &  \\
 & (0.022) & (0.022) & (0.020) &  &  &  \\
\addlinespace
\multicolumn{7}{l}{\textit{Specification N: M + non-student shops}} \\
All jeonse & 0.067$\,^{***}$ & 0.034$\,^{**}$ & $-$0.033$\,^{**}$ & $-$0.074$\,^{***}$ & 0.058$\,^{***}$ & 0.132$\,^{***}$ \\
 & (0.019) & (0.016) & (0.015) & (0.018) & (0.016) & (0.013) \\
New contracts only & 0.067$\,^{***}$ & 0.034$\,^{**}$ & $-$0.034$\,^{**}$ & $-$0.074$\,^{***}$ & 0.060$\,^{***}$ & 0.134$\,^{***}$ \\
 & (0.019) & (0.016) & (0.015) & (0.018) & (0.015) & (0.013) \\
Capped renewals only & 0.067$\,^{***}$ & 0.040$\,^{***}$ & $-$0.027$\,^{*}$ & $-$0.074$\,^{***}$ & 0.003 & 0.078$\,^{***}$ \\
 & (0.019) & (0.014) & (0.015) & (0.018) & (0.014) & (0.011) \\
All jeonse, $<$25 yrs & 0.089$\,^{***}$ & 0.032$\,^{*}$ & $-$0.057$\,^{***}$ & $-$0.064$\,^{***}$ & 0.084$\,^{***}$ & 0.148$\,^{***}$ \\
 & (0.021) & (0.020) & (0.016) & (0.022) & (0.017) & (0.014) \\
New only, $<$25 yrs & 0.089$\,^{***}$ & 0.031 & $-$0.058$\,^{***}$ & $-$0.064$\,^{***}$ & 0.083$\,^{***}$ & 0.148$\,^{***}$ \\
 & (0.021) & (0.019) & (0.016) & (0.022) & (0.015) & (0.015) \\
\addlinespace
\bottomrule
\end{tabular}
\begin{tablenotes}[flushleft]
\footnotesize
\item \textit{Notes:} Full version of Table~\ref{tab:periods}, Panel B. Sales are the same in every row except the $<$25 yrs rows, which restrict both markets to buildings under 25 years. Standard errors in parentheses, clustered by legal dong (district and dong name; 332 clusters in the full sample). $^{*}$\,$p<0.10$, $^{**}$\,$p<0.05$, $^{***}$\,$p<0.01$.
\end{tablenotes}
\end{threeparttable}
\end{table}

\subsection{Within-complex gap by period and location precision}\label{app:additional-withincell}

Table~\ref{tab:app-withincell} repeats Panel A of Table~\ref{tab:withincell} for cells before December 2019, for cells from December 2019, and for exactly located complexes. High-school hagwons produce no detectable change in the jeonse-to-price ratio in any subsample (\ResCellPreBanMHs, \ResCellFromBanMHs{} and \ResCellExactMHs), while non-student shops raise it (\ResCellPreBanNShop, \ResCellFromBanNShop{} and \ResCellExactNShop). With the SEMAS count, the small positive gap without shop controls is concentrated before December 2019 (\ResCellPreBanCHag, \ResCellPreBanCHagP) and is absent afterwards (\ResCellFromBanCHag).

\begin{table}[!htbp]
\centering
\begin{threeparttable}
\caption{Within-complex jeonse-to-price gap by period and location precision}
\label{tab:app-withincell}
\footnotesize
\setlength{\tabcolsep}{4pt}
\begin{tabular}{lcccc}
\toprule
 & C & J & M & N \\
\midrule
\multicolumn{5}{l}{\textit{All cells}} \\
Cram hagwons (SEMAS) & 0.014$\,^{**}$ & $-$0.004 &  &  \\
 & (0.007) & (0.008) &  &  \\
High-school/exam (registry) &  &  & 0.002 & 0.001 \\
 &  &  & (0.011) & (0.011) \\
Other academic (registry) &  &  & 0.004 & $-$0.010 \\
 &  &  & (0.009) & (0.011) \\
Non-student shops &  & 0.040$\,^{***}$ &  & 0.042$\,^{***}$ \\
 &  & (0.009) &  & (0.010) \\
Cells & 222,317 & 222,317 & 222,317 & 222,317 \\
\addlinespace
\multicolumn{5}{l}{\textit{2015-01 to 2019-11}} \\
Cram hagwons (SEMAS) & 0.022$\,^{**}$ & 0.005 &  &  \\
 & (0.010) & (0.010) &  &  \\
High-school/exam (registry) &  &  & 0.009 & 0.009 \\
 &  &  & (0.013) & (0.013) \\
Other academic (registry) &  &  & 0.009 & $-$0.005 \\
 &  &  & (0.008) & (0.009) \\
Non-student shops &  & 0.038$\,^{***}$ &  & 0.039$\,^{***}$ \\
 &  & (0.010) &  & (0.010) \\
Cells & 111,361 & 111,361 & 111,361 & 111,361 \\
\addlinespace
\multicolumn{5}{l}{\textit{2019-12 to 2026-10}} \\
Cram hagwons (SEMAS) & 0.008 & $-$0.014$\,^{*}$ &  &  \\
 & (0.006) & (0.008) &  &  \\
High-school/exam (registry) &  &  & 0.002 & 0.000 \\
 &  &  & (0.012) & (0.012) \\
Other academic (registry) &  &  & $-$0.005 & $-$0.020 \\
 &  &  & (0.012) & (0.013) \\
Non-student shops &  & 0.046$\,^{***}$ &  & 0.048$\,^{***}$ \\
 &  & (0.011) &  & (0.011) \\
Cells & 110,956 & 110,956 & 110,956 & 110,956 \\
\addlinespace
\multicolumn{5}{l}{\textit{Exact locations}} \\
Cram hagwons (SEMAS) & 0.014$\,^{**}$ & $-$0.003 &  &  \\
 & (0.007) & (0.008) &  &  \\
High-school/exam (registry) &  &  & 0.003 & 0.002 \\
 &  &  & (0.011) & (0.011) \\
Other academic (registry) &  &  & 0.004 & $-$0.009 \\
 &  &  & (0.009) & (0.011) \\
Non-student shops &  & 0.036$\,^{***}$ &  & 0.038$\,^{***}$ \\
 &  & (0.009) &  & (0.010) \\
Cells & 214,908 & 214,908 & 214,908 & 214,908 \\
\addlinespace
\bottomrule
\end{tabular}
\begin{tablenotes}[flushleft]
\footnotesize
\item \textit{Notes:} Panel A of Table~\ref{tab:withincell} by period and with block-tier (approximately located) complexes dropped. Standard errors in parentheses, clustered by legal dong. $^{*}$\,$p<0.10$, $^{**}$\,$p<0.05$, $^{***}$\,$p<0.01$.
\end{tablenotes}
\end{threeparttable}
\end{table}

\subsection{Entry fee: capped versus voluntary renewals}\label{app:additional-entryfee}

Table~\ref{tab:app-entryfee} reports the remaining coefficients of the entry-fee regressions. The split by renewal type helps interpret the average markups in Table~\ref{tab:entryfee}. Over the whole window, new contracts stand \ResEFCapFullMarkup{} log points above capped renewals (\ResEFCapFullMarkupPct\%) and \ResEFVolFullMarkup{} above voluntary renewals (\ResEFVolFullMarkupPct\%). Voluntary renewals are negotiated freely, so their gap to new contracts measures what incumbents keep without the legal cap; capped renewals add the effect of the 5\% limit. The high-school interaction is similar against both (\ResEFCapFullNHs{} and \ResEFVolFullNHs{} over the whole window) and is largest in 2021--22 against both (\ResEFCapEarlyNHs, \ResEFCapEarlyNHsP; \ResEFVolEarlyNHs, \ResEFVolEarlyNHsP). From 2023 it is \ResEFCapLateNHs{} against capped and \ResEFVolLateNHs{} against voluntary renewals. Other academic hagwons do not raise the markup in any comparison (\ResEFAllFullMOth{} over the whole window, all renewals). In 2021--22, the markup over capped renewals was smaller near non-student shops (\ResEFCapEarlyNShop, \ResEFCapEarlyNShopP); this is one of many interactions tested and is not interpreted. Dropping approximately located complexes leaves the high-school interaction at \ResEFAllFullExactNHs{} (\ResEFAllFullExactNHsP).

\begin{table}[!htbp]
\centering
\begin{threeparttable}
\caption{Entry fee: remaining coefficients, by renewal type and period}
\label{tab:app-entryfee}
\footnotesize
\setlength{\tabcolsep}{3pt}
\begin{tabular}{lccccccc}
\toprule
 & M: high- & M: other & N: other & J: hagwons & J: shops & Contracts & Median \\
 & school & academic & academic &  &  & in cells & high-school \\
\midrule
\multicolumn{8}{l}{\textit{All renewals}} \\
2021-06 to 2026-10 & 0.0048$\,^{*}$ & $-$0.0035 & $-$0.0026 & 0.0020 & $-$0.0032 & 349,424 & 9 \\
 & (0.0028) & (0.0027) & (0.0029) & (0.0016) & (0.0022) & & \\
\quad exact locations & 0.0044 & $-$0.0032 & $-$0.0025 &  &  &  &  \\
 & (0.0028) & (0.0027) & (0.0029) &  &  & & \\
2021-06 to 2022-12 & 0.0104$\,^{**}$ & $-$0.0049 & $-$0.0026 & 0.0061$\,^{**}$ & $-$0.0079$\,^{*}$ & 89,121 & 9 \\
 & (0.0052) & (0.0054) & (0.0055) & (0.0030) & (0.0042) & & \\
2023-01 to 2026-10 & 0.0024 & $-$0.0027 & $-$0.0024 & 0.0006 & $-$0.0017 & 260,303 & 10 \\
 & (0.0027) & (0.0028) & (0.0030) & (0.0015) & (0.0022) & & \\
\addlinespace
\multicolumn{8}{l}{\textit{Capped renewals}} \\
2021-06 to 2026-10 & 0.0046 & $-$0.0012 & $-$0.0005 & 0.0034$\,^{*}$ & $-$0.0025 & 213,420 & 10 \\
 & (0.0030) & (0.0031) & (0.0033) & (0.0019) & (0.0031) & & \\
2021-06 to 2022-12 & 0.0124$\,^{**}$ & $-$0.0068 & $-$0.0037 & 0.0071$\,^{**}$ & $-$0.0111$\,^{**}$ & 67,503 & 9 \\
 & (0.0057) & (0.0060) & (0.0062) & (0.0035) & (0.0047) & & \\
2023-01 to 2026-10 & 0.0006 & 0.0012 & 0.0008 & 0.0013 & 0.0010 & 145,917 & 10 \\
 & (0.0025) & (0.0027) & (0.0029) & (0.0016) & (0.0029) & & \\
\addlinespace
\multicolumn{8}{l}{\textit{Voluntary renewals}} \\
2021-06 to 2026-10 & 0.0046 & $-$0.0050 & $-$0.0043 & 0.0004 & $-$0.0025 & 204,329 & 9 \\
 & (0.0032) & (0.0032) & (0.0034) & (0.0019) & (0.0026) & & \\
2021-06 to 2022-12 & 0.0092$\,^{*}$ & $-$0.0039 & $-$0.0042 & 0.0036 & 0.0008 & 37,122 & 9 \\
 & (0.0054) & (0.0064) & (0.0063) & (0.0036) & (0.0052) & & \\
2023-01 to 2026-10 & 0.0033 & $-$0.0048 & $-$0.0040 & $-$0.0002 & $-$0.0030 & 167,207 & 9 \\
 & (0.0036) & (0.0035) & (0.0037) & (0.0020) & (0.0028) & & \\
\addlinespace
\bottomrule
\end{tabular}
\begin{tablenotes}[flushleft]
\footnotesize
\item \textit{Notes:} Remaining coefficients of the entry-fee regressions in Table~\ref{tab:entryfee} (each coefficient is the exposure term $\times$ new contract). Contracts in cells: new and renewal contracts in cells with both. Median high-school: median registry high-school/exam count within 500 m among those contracts; the entry-fee CSV scales coefficients from this median to 50 rather than from the median complex. Exact locations: block-tier complexes dropped (specifications C and J not run). Standard errors in parentheses, clustered by legal dong. $^{*}$\,$p<0.10$, $^{**}$\,$p<0.05$, $^{***}$\,$p<0.01$.
\end{tablenotes}
\end{threeparttable}
\end{table}

\subsection{Event study: alternative exposure measures}\label{app:additional-events}

Table~\ref{tab:app-events} reports the jump, joint and winter tests for new jeonse contracts and sales under the three versions of the dong-level exposure: the time-varying, coverage-adjusted count used in the main text, and the counts of December 2018 and December 2022 held fixed. The June 2023 jump for new contracts is \ResEvNewJJumpJun{} with the time-varying count, \ResEvNewJFixedEighteenJumpJun{} with the 2018 count and \ResEvNewJFixedTwentyTwoJumpJun{} with the 2022 count; none is significant. The lower panel lists every half-year coefficient behind Figure~\ref{fig:events}.

\begin{table}[!htbp]
\centering
\begin{threeparttable}
\caption{Event study: alternative exposure measures and half-year coefficients}
\label{tab:app-events}
\footnotesize
\setlength{\tabcolsep}{3pt}
\begin{tabular}{lcccccc}
\toprule
 & \multicolumn{3}{c}{New jeonse contracts} & \multicolumn{3}{c}{Sales} \\
\cmidrule(lr){2-4}\cmidrule(lr){5-7}
 & Time-varying & Fixed 2018 & Fixed 2022 & Time-varying & Fixed 2018 & Fixed 2022 \\
\midrule
Jump at June 2023 & 0.0027 & 0.0012 & 0.0022 & 0.0027 & 0.0019 & 0.0020 \\
 & (0.0030) & (0.0029) & (0.0029) & (0.0018) & (0.0017) & (0.0017) \\
Jump at Feb.\ 2024 & 0.0013 & 0.0026 & 0.0011 & 0.0022 & 0.0015 & 0.0017 \\
 & (0.0043) & (0.0050) & (0.0042) & (0.0013) & (0.0013) & (0.0013) \\
Hagwons $\times$ winter & $-$0.0018 & $-$0.0030 & $-$0.0022 & 0.0002 & 0.0001 & 0.0003 \\
 & (0.0019) & (0.0022) & (0.0018) & (0.0009) & (0.0009) & (0.0009) \\
Post halves = 0: $p$ & 0.300 & 0.062 & 0.185 & 0.231 & 0.255 & 0.266 \\
Pre halves = 0: $p$ & 0.005 & 0.013 & 0.002 & 0.248 & 0.349 & 0.253 \\
\midrule
\multicolumn{7}{l}{\textit{Half-year coefficients (time-varying exposure; 2023H1 = 0)}} \\
 & New jeonse & Sales & New + vol. & Capped & New, $<$25 & Sales, $<$25 \\
2021H1 & 0.0168$\,^{**}$ & $-$0.0025 & 0.0143$\,^{*}$ & $-$0.0041 & 0.0108 & $-$0.0049 \\
 & (0.0081) & (0.0046) & (0.0073) & (0.0050) & (0.0118) & (0.0045) \\
2021H2 & 0.0147$\,^{***}$ & 0.0037 & 0.0115$\,^{**}$ & $-$0.0010 & 0.0090 & 0.0042 \\
 & (0.0050) & (0.0039) & (0.0045) & (0.0038) & (0.0059) & (0.0041) \\
2022H1 & 0.0032 & $-$0.0011 & 0.0031 & $-$0.0011 & 0.0002 & $-$0.0025 \\
 & (0.0067) & (0.0031) & (0.0056) & (0.0036) & (0.0071) & (0.0035) \\
2022H2 & 0.0083 & $-$0.0056 & 0.0060 & $-$0.0014 & 0.0100 & $-$0.0005 \\
 & (0.0070) & (0.0044) & (0.0058) & (0.0036) & (0.0088) & (0.0045) \\
2023H2 & 0.0027 & 0.0027 & 0.0031 & $-$0.0055 & 0.0000 & 0.0012 \\
 & (0.0030) & (0.0018) & (0.0029) & (0.0035) & (0.0039) & (0.0021) \\
2024H1 & 0.0040 & 0.0048$\,^{**}$ & 0.0032 & $-$0.0042 & 0.0015 & 0.0040$\,^{**}$ \\
 & (0.0048) & (0.0020) & (0.0043) & (0.0041) & (0.0052) & (0.0019) \\
2024H2 & 0.0062 & 0.0062$\,^{**}$ & 0.0034 & $-$0.0008 & 0.0071 & 0.0052$\,^{**}$ \\
 & (0.0062) & (0.0026) & (0.0061) & (0.0049) & (0.0069) & (0.0026) \\
2025H1 & 0.0022 & 0.0090$\,^{**}$ & 0.0030 & $-$0.0033 & $-$0.0069 & 0.0059 \\
 & (0.0054) & (0.0042) & (0.0048) & (0.0033) & (0.0044) & (0.0036) \\
2025H2 & $-$0.0016 & 0.0121$\,^{**}$ & 0.0007 & 0.0005 & $-$0.0082$\,^{*}$ & 0.0059 \\
 & (0.0043) & (0.0055) & (0.0042) & (0.0036) & (0.0042) & (0.0044) \\
2026H1 & $-$0.0001 & 0.0065 & 0.0021 & 0.0026 & $-$0.0051 & 0.0031 \\
 & (0.0044) & (0.0061) & (0.0056) & (0.0033) & (0.0055) & (0.0054) \\
2026H2 & $-$0.0094$\,^{*}$ & 0.0049 & $-$0.0070 & $-$0.0029 & $-$0.0061 & 0.0036 \\
 & (0.0049) & (0.0074) & (0.0069) & (0.0038) & (0.0064) & (0.0060) \\
\bottomrule
\end{tabular}
\begin{tablenotes}[flushleft]
\footnotesize
\item \textit{Notes:} Time-varying: coverage-adjusted, interpolated dong count (as in Table~\ref{tab:events}). Fixed 2018 and Fixed 2022: the dong count in December 2018 (raw) or December 2022 (coverage-adjusted), held fixed. 2021H1 contains June 2021 only; 2026H2 runs to mid-October 2026. Standard errors in parentheses, clustered by legal dong. $^{*}$\,$p<0.10$, $^{**}$\,$p<0.05$, $^{***}$\,$p<0.01$.
\end{tablenotes}
\end{threeparttable}
\end{table}

\subsection{Alternative exposure measures and shop categories}\label{app:additional-exposure}

Table~\ref{tab:app-exposure} reports every variant of the registry measure discussed in Section~\ref{sec:robust-exposure} and the separate shop categories discussed in Section~\ref{sec:results-convenience}. Of today's registry hagwons, only those open on 1 July of a contract's year enter the time-consistent count. The capacity variants use the registry's total enrolment places and the places that can be taught at one time.

\begin{table}[!htbp]
\centering
\begin{threeparttable}
\caption{Alternative exposure measures and shop categories}
\label{tab:app-exposure}
\scriptsize
\renewcommand{\arraystretch}{0.92}
\setlength{\tabcolsep}{3pt}
\begin{tabular}{lccccc}
\toprule
 & Sale & Jeonse & Jeonse $-$ sale & Sale & Jeonse \\
 & & & & \multicolumn{2}{c}{Median $\to$ p97 (\%)} \\
\midrule
\multicolumn{6}{l}{\textit{Panel A. High-school/exam hagwons}} \\
Baseline (registry, 2026, 500 m), M & 0.056$\,^{***}$ & 0.036$\,^{**}$ & $-$0.020 & 11.1 & 6.9 \\
 & (0.016) & (0.017) & (0.016) & & \\
Open by the year of the contract & 0.052$\,^{***}$ & 0.040$\,^{***}$ & $-$0.012 & 11.3 & 8.5 \\
 & (0.012) & (0.012) & (0.012) & & \\
Opened before 2015 & 0.042$\,^{***}$ & 0.040$\,^{***}$ & $-$0.002 & 8.2 & 7.8 \\
 & (0.013) & (0.011) & (0.011) & & \\
Opened before 2015, contracts 2015--19 & 0.037$\,^{***}$ & 0.042$\,^{***}$ & 0.005 & 7.2 & 8.3 \\
 & (0.013) & (0.013) & (0.013) & & \\
Baseline, contracts 2015--19 & 0.046$\,^{***}$ & 0.040$\,^{**}$ & $-$0.006 & 9.1 & 7.9 \\
 & (0.014) & (0.016) & (0.015) & & \\
Enrolment places (hundreds) & 0.045$\,^{***}$ & 0.029$\,^{***}$ & $-$0.017$\,^{**}$ & 13.9 & 8.6 \\
 & (0.008) & (0.008) & (0.008) & & \\
Places taught at one time (hundreds) & 0.062$\,^{***}$ & 0.045$\,^{***}$ & $-$0.017 & 15.6 & 11.1 \\
 & (0.013) & (0.012) & (0.013) & & \\
Within 1 km & 0.074$\,^{**}$ & 0.095$\,^{***}$ & 0.022 & 14.6 & 19.3 \\
 & (0.033) & (0.028) & (0.031) & & \\
Within 1 km, shops within 1 km (N) & 0.065$\,^{**}$ & 0.102$\,^{***}$ & 0.037 & 12.9 & 20.9 \\
 & (0.033) & (0.027) & (0.028) & & \\
Baseline with shops, N & 0.055$\,^{***}$ & 0.034$\,^{**}$ & $-$0.021 & 10.9 & 6.6 \\
 & (0.016) & (0.016) & (0.015) & & \\
Shop categories entered separately (N) & 0.044$\,^{***}$ & 0.030$\,^{*}$ & $-$0.013 & 8.6 & 5.9 \\
 & (0.014) & (0.016) & (0.015) & & \\
Shops without estate agents (N) & 0.055$\,^{***}$ & 0.035$\,^{**}$ & $-$0.020 & 10.8 & 6.7 \\
 & (0.015) & (0.016) & (0.015) & & \\
Complexes 1 km or more from the city edge, M & 0.055$\,^{***}$ & 0.043$\,^{***}$ & $-$0.012 & 10.8 & 8.4 \\
 & (0.016) & (0.015) & (0.015) & & \\
Complexes 1 km or more from the city edge, N & 0.057$\,^{***}$ & 0.043$\,^{***}$ & $-$0.014 & 11.3 & 8.3 \\
 & (0.016) & (0.015) & (0.015) & & \\
\addlinespace
\multicolumn{6}{l}{\textit{Panel B. Non-student shop terms (a doubling)}} \\
Non-student shops, N & $-$0.070$\,^{***}$ & 0.061$\,^{***}$ & 0.131$\,^{***}$ & & \\
 & (0.016) & (0.018) & (0.014) & & \\
Estate agents, N-split & 0.093$\,^{***}$ & 0.085$\,^{***}$ & $-$0.008 & & \\
 & (0.025) & (0.028) & (0.029) & & \\
Health care, N-split & 0.053$\,^{***}$ & 0.029$\,^{**}$ & $-$0.024$\,^{*}$ & & \\
 & (0.013) & (0.014) & (0.014) & & \\
Repairs and personal services, N-split & $-$0.213$\,^{***}$ & $-$0.044 & 0.169$\,^{***}$ & & \\
 & (0.022) & (0.028) & (0.028) & & \\
Health care + repairs and personal services & $-$0.077$\,^{***}$ & 0.051$\,^{***}$ & 0.128$\,^{***}$ & & \\
 & (0.015) & (0.018) & (0.014) & & \\
Non-student shops, 1 km or more from the edge & $-$0.093$\,^{***}$ & 0.006 & 0.098$\,^{***}$ & & \\
 & (0.019) & (0.019) & (0.015) & & \\
\bottomrule
\end{tabular}
\begin{tablenotes}[flushleft]
\footnotesize
\item \textit{Notes:} Each row re-estimates Equation~\eqref{eq:hedonic} (and its pooled jeonse $-$ sale version) with a different exposure measure from the NEIS registry; all other terms as in specifications M and N. ``Open by the year of the contract'' counts registry hagwons whose opening date precedes 1 July of the contract year; ``opened before 2015'' counts those that predate every contract. Both still omit hagwons that closed before 2026. Capacity rows use log(1 + places/100), with zero or missing values replaced by the median for the same kind and grade group and values capped at the 99th percentile. Panel B: shop categories are SEMAS major categories within 500 m (December 2018); the three are entered jointly in N-split. The last two columns move each measure from its median to its 97th percentile across located complexes. Standard errors clustered by legal dong. * $p<0.10$, ** $p<0.05$, *** $p<0.01$.
\end{tablenotes}
\end{threeparttable}
\end{table}

\subsection{Within-complex ratio test: further samples}\label{app:additional-cellvariants}

\begin{figure}[!htbp]
\centering
\includegraphics[width=\textwidth]{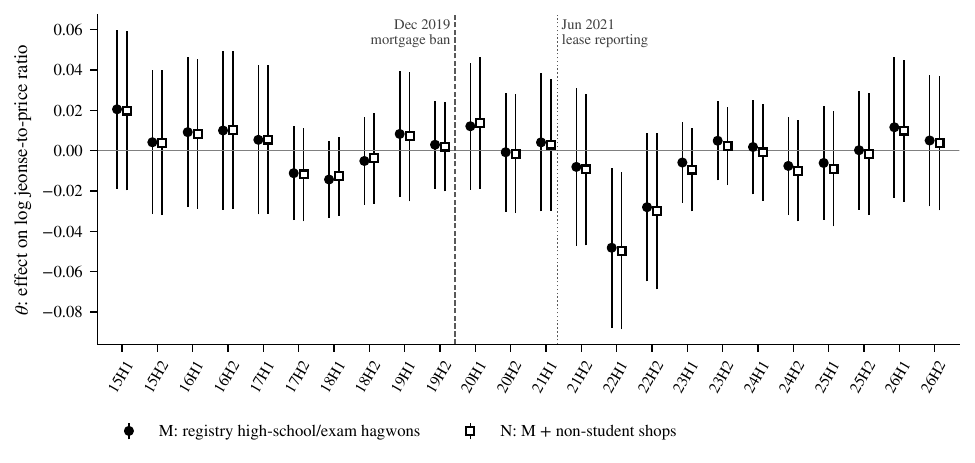}
\caption{The within-complex ratio response by half-year}
\label{fig:cell-halfyears}
\begin{minipage}{0.95\textwidth}\footnotesize
\textit{Notes:} Coefficient $\theta$ on log(1 + registry high-school/exam hagwons within 500 m) from Equation~\eqref{eq:withincell}, interacted with half-year indicators, for specifications M and N. Bars are 95\% confidence intervals, clustered by legal dong. The 2022 half-years contain \ChkCellHalfTwentyTwoHOneMCells{} and \ChkCellHalfTwentyTwoHTwoMCells{} cells, against a median of \ChkCellHalfMedianCells{} across half-years.
\end{minipage}
\end{figure}

Table~\ref{tab:app-cellvariants} reports the within-complex coefficient $\theta$ by period, contract type, price tier and building age, and the positive control of Section~\ref{sec:robust-ratio}. Figure~\ref{fig:cell-halfyears} plots $\theta$ by half-year.

\begin{table}[!htbp]
\centering
\begin{threeparttable}
\caption{Within-complex ratio test: periods, contracts, price tiers and building age}
\label{tab:app-cellvariants}
\scriptsize
\renewcommand{\arraystretch}{0.92}
\setlength{\tabcolsep}{4pt}
\begin{tabular}{lccc}
\toprule
 & M & N & Cells \\
\midrule
All cells (Table~\ref{tab:withincell}) & 0.002 & 0.001 & 222,317 \\
 & (0.011) & (0.011) & \\
\addlinespace
\multicolumn{4}{l}{\textit{Panel A. Periods}} \\
2015-01 to 2019-11 & 0.003 & 0.003 & 111,361 \\
 & (0.013) & (0.013) & \\
2019-12 to 2021-05 & 0.006 & 0.007 & 32,413 \\
 & (0.015) & (0.015) & \\
2021-06 to 2026-10 & $-$0.002 & $-$0.005 & 78,543 \\
 & (0.012) & (0.012) & \\
Equal across periods & $p=0.61$ & $p=0.42$ &  \\
Equal across 24 half-years & $p=0.009$ & $p=0.008$ &  \\
Equal across half-years, without 2022 & $p=0.063$ & $p=0.051$ &  \\
\addlinespace
\multicolumn{4}{l}{\textit{Panel B. Contracts from June 2021}} \\
All jeonse contracts & 0.002 & $-$0.001 & 78,543 \\
 & (0.012) & (0.012) & \\
New contracts only & 0.003 & 0.001 & 61,099 \\
 & (0.012) & (0.012) & \\
Capped renewals only & 0.010 & 0.008 & 33,494 \\
 & (0.013) & (0.013) & \\
\addlinespace
\multicolumn{4}{l}{\textit{Panel C. Sale price per flat}} \\
KRW 0.9 billion or less & 0.004 & 0.007 & 145,903 \\
 & (0.011) & (0.012) & \\
KRW 0.9--1.5 billion & 0.014 & 0.011 & 47,636 \\
 & (0.015) & (0.015) & \\
Above KRW 1.5 billion & 0.007 & 0.008 & 28,778 \\
 & (0.025) & (0.025) & \\
Equal across tiers & $p=0.68$ & $p=0.95$ &  \\
\addlinespace
\multicolumn{4}{l}{\textit{Panel D. Building age}} \\
Under 30 years & 0.004 & 0.002 & 177,538 \\
 & (0.011) & (0.011) & \\
30 years or more & $-$0.009 & $-$0.007 & 44,779 \\
 & (0.021) & (0.021) & \\
Under 25 years, from June 2021 & $-$0.023$\,^{**}$ & $-$0.032$\,^{***}$ & 43,528 \\
 & (0.012) & (0.011) & \\
Under 25 years, from June 2021, new contracts & $-$0.023$\,^{**}$ & $-$0.031$\,^{***}$ & 33,932 \\
 & (0.011) & (0.011) & \\
25 years or more, from June 2021 & 0.020 & 0.020 & 34,651 \\
 & (0.020) & (0.020) & \\
Under 25 years, before June 2021 & $-$0.008 & $-$0.009 & 100,751 \\
 & (0.006) & (0.006) & \\
\addlinespace
\multicolumn{4}{l}{\textit{Panel E. Positive control: buildings of 30 years or more}} \\
Age 30+ indicator & $-$0.358$\,^{***}$ & $-$0.355$\,^{***}$ & 222,317 \\
 & (0.032) & (0.032) & \\
Age 30+ indicator, price-tier effects & $-$0.355$\,^{***}$ & $-$0.353$\,^{***}$ & 222,317 \\
 & (0.031) & (0.031) & \\
\bottomrule
\end{tabular}
\begin{tablenotes}[flushleft]
\footnotesize
\item \textit{Notes:} Coefficient $\theta$ on log(1 + registry high-school/exam hagwons within 500 m) in Equation~\eqref{eq:withincell}: the log jeonse-to-price ratio within cells of the same complex, flat size and month, weighted as in Table~\ref{tab:withincell}. M and N as in Table~\ref{tab:grades}. Panels A, C and the first two rows of D interact the exposure with group indicators; Panel C adds tier effects and Panel D age-bin effects. Panel C groups cells by the cell's mean sale price. The ``from June 2021'' rows use contracts flagged as new or renewal. Panel E replaces the exposure with an indicator for buildings of 30 years or more, whose prices carry reconstruction value; the second row adds price-tier effects. Equality tests are Wald tests. Standard errors clustered by legal dong. * $p<0.10$, ** $p<0.05$, *** $p<0.01$.
\end{tablenotes}
\end{threeparttable}
\end{table}

\subsection{Event study with the registry measure}\label{app:additional-eventsregistry}

\begin{figure}[!htbp]
\centering
\includegraphics[width=\textwidth]{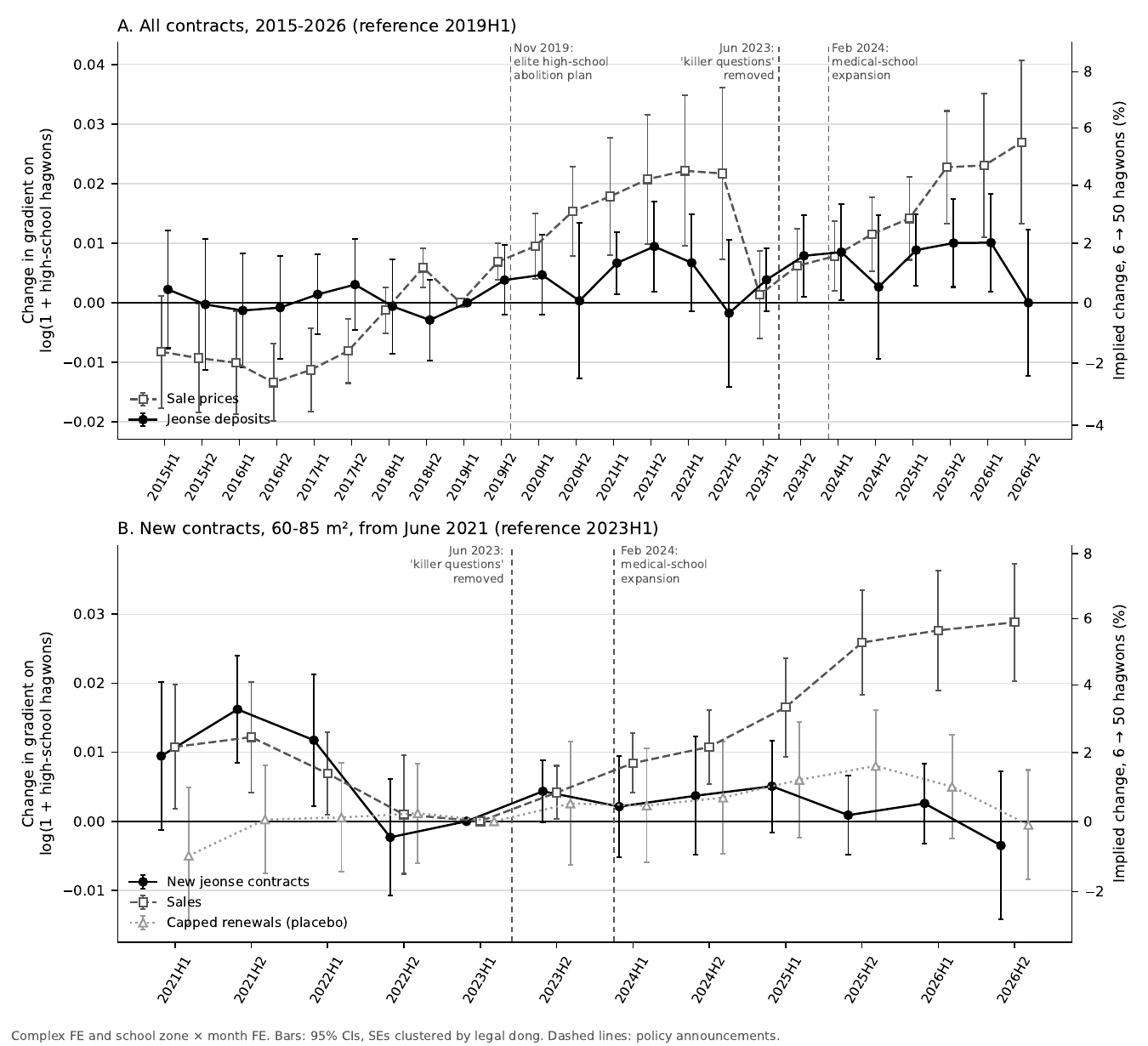}
\caption{The hagwon premium around the policy windows, registry measure}
\label{fig:events-registry}
\begin{minipage}{0.95\textwidth}\footnotesize
\textit{Notes:} Change in the coefficient on log(1 + registry high-school/exam hagwons within 500 m) relative to the reference half-year, with complex and school zone $\times$ month fixed effects. The right-hand axis expresses the change in percent for the move from \DatMedNeisHs{} to \ResHsTarget{} such hagwons. Bars are 95\% confidence intervals, clustered by legal dong. Dashed lines: policy windows (Section~\ref{sec:background-shocks}).
\end{minipage}
\end{figure}

Figure~\ref{fig:events-registry} repeats the event study of Section~\ref{sec:robust-events} with log(1 + registry high-school/exam hagwons within 500 m) as the exposure, complex fixed effects and school zone $\times$ month fixed effects. Panel A uses all contracts from 2015 (reference 2019H1); Panel B uses new contracts of 60--85 m$^2$ from June 2021 (reference 2023H1), with capped renewals as a placebo. Panel B has \ChkEvRegNewJObs{} new jeonse contracts and \ChkEvRegSalesObs{} sales.

\subsection{The multiple-testing family}\label{app:additional-family}

Table~\ref{tab:app-family} lists the tests counted in Section~\ref{sec:robust-multiple}, with their $p$-values, Benjamini--Hochberg adjusted $p$-values and whether each passes the Bonferroni threshold.

\begin{table}[!htbp]
\centering
\begin{threeparttable}
\caption{The multiple-testing family}
\label{tab:app-family}
\scriptsize
\renewcommand{\arraystretch}{0.85}
\setlength{\tabcolsep}{4pt}
\begin{tabular}{lcccc}
\toprule
 & Estimate & $p$ & BH $q$ & Bonferroni \\
\midrule
\multicolumn{5}{l}{\textit{A. Level premia, registry (Table~\ref{tab:grades})}} \\
High-school/exam, sale (M) & 0.056 & $<$0.001 & 0.002 & \checkmark \\
High-school/exam, jeonse (M) & 0.036 & 0.034 & 0.079 &  \\
High-school/exam, jeonse $-$ sale (M) & $-$0.020 & 0.19 & 0.26 &  \\
Other academic, sale (M) & $-$0.016 & 0.32 & 0.41 &  \\
Other academic, jeonse (M) & 0.0050 & 0.78 & 0.83 &  \\
All academic, sale (M0) & 0.036 & $<$0.001 & 0.002 & \checkmark \\
All academic, jeonse (M0) & 0.037 & $<$0.001 & 0.002 & \checkmark \\
All academic, jeonse $-$ sale (M0) & 0.0014 & 0.86 & 0.86 &  \\
Non-student shops, sale (N) & $-$0.070 & $<$0.001 & $<$0.001 & \checkmark \\
Non-student shops, jeonse (N) & 0.061 & $<$0.001 & 0.003 & \checkmark \\
Non-student shops, jeonse $-$ sale (N) & 0.131 & $<$0.001 & $<$0.001 & \checkmark \\
\multicolumn{5}{l}{\textit{B. Level premia, SEMAS (Tables~\ref{tab:baseline} and~\ref{tab:convenience})}} \\
Cram hagwons, sale (C) & 0.026 & 0.032 & 0.079 &  \\
Cram hagwons, jeonse (C) & 0.051 & $<$0.001 & 0.001 & \checkmark \\
Cram hagwons, jeonse $-$ sale (C) & 0.025 & 0.003 & 0.012 &  \\
Cram hagwons, sale (J) & 0.061 & $<$0.001 & $<$0.001 & \checkmark \\
Cram hagwons, jeonse (J) & 0.025 & 0.039 & 0.086 &  \\
Cram hagwons, jeonse $-$ sale (J) & $-$0.036 & $<$0.001 & 0.001 & \checkmark \\
Non-student shops, sale (J) & $-$0.080 & $<$0.001 & $<$0.001 & \checkmark \\
Non-student shops, jeonse (J) & 0.053 & 0.004 & 0.013 &  \\
Non-student shops, jeonse $-$ sale (J) & 0.132 & $<$0.001 & $<$0.001 & \checkmark \\
\multicolumn{5}{l}{\textit{C. Within complex (Table~\ref{tab:withincell})}} \\
Cram hagwons (C) & 0.014 & 0.042 & 0.086 &  \\
Cram hagwons (J) & $-$0.0045 & 0.55 & 0.63 &  \\
Non-student shops (J) & 0.040 & $<$0.001 & $<$0.001 & \checkmark \\
High-school/exam (M) & 0.0022 & 0.85 & 0.86 &  \\
Non-student shops (N) & 0.042 & $<$0.001 & $<$0.001 & \checkmark \\
\multicolumn{5}{l}{\textit{D. Entry fee, high-school/exam $\times$ new contract (Table~\ref{tab:entryfee})}} \\
Against all renewals, 2021--26 & 0.0049 & 0.068 & 0.12 &  \\
Against all renewals, 2021--22 & 0.011 & 0.042 & 0.086 &  \\
Against all renewals, 2023--26 & 0.0025 & 0.35 & 0.44 &  \\
Against capped renewals, 2021--26 & 0.0048 & 0.11 & 0.18 &  \\
Against capped renewals, 2021--22 & 0.013 & 0.025 & 0.063 &  \\
Against capped renewals, 2023--26 & 0.0005 & 0.85 & 0.86 &  \\
Against voluntary renewals, 2021--26 & 0.0048 & 0.14 & 0.20 &  \\
Against voluntary renewals, 2021--22 & 0.0092 & 0.091 & 0.15 &  \\
Against voluntary renewals, 2023--26 & 0.0035 & 0.32 & 0.41 &  \\
\multicolumn{5}{l}{\textit{E. Policy windows (Table~\ref{tab:events})}} \\
New jeonse, June 2023 & 0.0027 & 0.37 & 0.45 &  \\
New jeonse, February 2024 & 0.0013 & 0.76 & 0.83 &  \\
Sales, June 2023 & 0.0027 & 0.13 & 0.20 &  \\
Sales, February 2024 & 0.0022 & 0.10 & 0.16 &  \\
All jeonse, 2020H1 against 2019H1 & 0.0030 & 0.39 & 0.46 &  \\
All sales, 2020H1 against 2019H1 & 0.0022 & 0.50 & 0.59 &  \\
\multicolumn{5}{l}{\textit{F. Housing rules and contract samples, jeonse $-$ sale (Table~\ref{tab:periods})}} \\
Before August 2017 (M) & 0.0046 & 0.78 & 0.83 &  \\
Before December 2019 (M) & $-$0.0059 & 0.68 & 0.77 &  \\
Ban years (M) & $-$0.029 & 0.15 & 0.21 &  \\
After the ban (M) & $-$0.025 & 0.16 & 0.22 &  \\
Permit-zone dongs dropped (M) & $-$0.030 & 0.044 & 0.086 &  \\
From June 2021, all jeonse (M) & $-$0.029 & 0.073 & 0.13 &  \\
From June 2021, new contracts (M) & $-$0.030 & 0.066 & 0.12 &  \\
From June 2021, capped renewals (M) & $-$0.025 & 0.10 & 0.16 &  \\
Under 25 years, all jeonse (M) & $-$0.046 & 0.019 & 0.050 &  \\
Under 25 years, new contracts (M) & $-$0.047 & 0.017 & 0.049 &  \\
Under 25 years, all jeonse (N) & $-$0.057 & $<$0.001 & 0.002 & \checkmark \\
Under 25 years, new contracts (N) & $-$0.058 & $<$0.001 & 0.001 & \checkmark \\
\multicolumn{5}{l}{\textit{G. Within complex, under 25 years from June 2021 (Table~\ref{tab:app-cellvariants})}} \\
High-school/exam (M) & $-$0.023 & 0.044 & 0.086 &  \\
High-school/exam (N) & $-$0.032 & 0.005 & 0.014 &  \\
\bottomrule
\end{tabular}
\begin{tablenotes}[flushleft]
\footnotesize
\item \textit{Notes:} The family is the \ChkFamTests{} tests on which Sections~\ref{sec:results} and~\ref{sec:robust} base a conclusion, each in the specification the text reports. Re-estimates of the same hypothesis under other controls, exposure measures, samples or clusters (Tables~\ref{tab:access}, \ref{tab:inference}, \ref{tab:app-exposure} and \ref{tab:app-cellvariants}) are not counted as separate tests. $p$: as reported in the source table, clustered by legal dong. BH $q$: Benjamini--Hochberg adjusted $p$-value across the whole family. \checkmark: $p$ below the Bonferroni threshold $0.05/\ChkFamTests = \ChkFamThreshold$.
\end{tablenotes}
\end{threeparttable}
\end{table}

\end{document}